\documentclass{emulateapj}
\usepackage{apjfonts,rotating}
\usepackage{color}

\def\athena{{\itshape Athena\/}}
\def\surveyor{{\itshape X-ray Surveyor\/}}
\def\chandra{{\itshape Chandra\/}}

\def\hubble{{\itshape Hubble\/}}
\def\hst{{\itshape HST\/}}

\def\spitzer{{\itshape Spitzer\/}}
\def\herschel{{\itshape Herschel\/}}
\def\galex{{\itshape GALEX\/}}

\def\xmm{{\itshape XMM-Newton\/}}
\def\nustar{{\itshape NuSTAR\/}}

\def\xray{\hbox{X-ray}}

\def\cdfs{\hbox{CDF-S}}

\def\etal{{et\,al.}}

\def\ltsima{$\; \buildrel < \over \sim \;$}
\def\simlt{\lower.5ex\hbox{\ltsima}}
\def\gtsima{$\; \buildrel > \over \sim \;$}
\def\simgt{\lower.5ex\hbox{\gtsima}}
\def\kms{\ifmmode{~{\rm km~s^{-1}}}\else{~km s$^{-1}$}\fi}
\def\lsim{\lower0.3em\hbox{$\,\buildrel <\over\sim\,$}}
\def\gsim{\lower0.3em\hbox{$\,\buildrel >\over\sim\,$}}

\def\msol{$M_\odot$}
\def\h2{H$_2$}
\def\flux{erg~cm$^{-2}$~s$^{-1}$}
\def\lum{erg~s$^{-1}$}

\def\arcsec{\mbox{$^{\prime\prime}$}}
\def\arcmin{\mbox{$^\prime$}}
\def\sfr{$M_{\odot}$~yr$^{-1}$}

\def\aap{A\&A}
\def\apj{ApJ}
\def\apjl{ApJL}
\def\apjs{ApJS}
\def\aj{AJ}
\def\mnras{MNRAS}
\def\araa{ARA\&A}

\def\nstk{63}
\def\nsbs{23}
\def\nsbh{60}
\def\nhbs{11}
\def\nundet{3}
\def\nmcut{4,898}
\newcommand{\angstrom}{\mbox{\normalfont\AA}}

\begin{document}

\shortauthors{{\sc Lehmer et al.}}
\shorttitle{Evolution of X-ray Emission in the 6~Ms CDF-S}

%
\title{The Evolution of Normal Galaxy X-ray Emission Through Cosmic History: Constraints from the 6~Ms {\itshape Chandra} Deep Field-South}
%

\author{
B.~D.~Lehmer,\altaffilmark{1,2,3}
A.~R.~Basu-Zych,\altaffilmark{3,4}
S.~Mineo,\altaffilmark{5,6}
W.~N.~Brandt,\altaffilmark{7,8}
R.~T.~Eufrasio,\altaffilmark{2,3}
T.~Fragos,\altaffilmark{9}
A.~E.~Hornschemeier,\altaffilmark{3}
B.~Luo,\altaffilmark{7,8,10}
Y.~Q.~Xue,\altaffilmark{11}
F.~E.~Bauer,\altaffilmark{12,13,14}
M.~Gilfanov,\altaffilmark{5}
P.~Ranalli,\altaffilmark{15}
D.~P.~Schneider,\altaffilmark{7,8}
O.~Shemmer,\altaffilmark{16}
P.~Tozzi,\altaffilmark{17}
J.~R.~Trump,\altaffilmark{7}
C.~Vignali,\altaffilmark{18}
J.-X.~Wang,\altaffilmark{11}
M.~Yukita,\altaffilmark{2,3}
\& 
A.~Zezas\altaffilmark{19,20}
}

\altaffiltext{1}{Department of Physics, University of Arkansas, 226 Physics
Building, 835 West Dickson Street, Fayetteville, AR 72701, USA}
\altaffiltext{2}{The Johns Hopkins University, Homewood Campus, Baltimore, MD
21218, USA}
\altaffiltext{3}{NASA Goddard Space Flight Center, Code 662, Greenbelt, MD
20771, USA} 
\altaffiltext{4}{Center for Space Science and Technology, University of
Maryland Baltimore County, 1000 Hilltop Circle, Baltimore, MD 21250, USA}
\altaffiltext{5}{Max Planck Institut f\"{u}r Astrophysik,
Karl-Schwarzschild-Str. 1, 85741 Garching, Germany}
\altaffiltext{6}{XAIA Investment GmbH, Sonnenstra\ss e 19, 80331 M\"unchen,
Germany}
\altaffiltext{7}{Department of Astronomy \& Astrophysics and Institute for
Gravitation and the Cosmos, 525 Davey Lab, The Pennsylvania State University,
University Park, PA 16802, USA}
\altaffiltext{8}{Department of Physics, The Pennsylvania State University,
University Park, PA 16802, USA}
\altaffiltext{9}{Geneva Observatory, Geneva University, Chemin des Maillettes
51, 1290 Sauverny, Switzerland}
\altaffiltext{10}{School of Astronomy and Space Science, Nanjing University,
Nanjing 210093, China}
\altaffiltext{11}{CAS Key Laboratory for Researches in Galaxies and Cosmology,
Center for Astrophysics, Department of Astronomy, University of Science and
Technology of China, Chinese Academy of Sciences, Hefei, Anhui 230026, China}
\altaffiltext{12}{Instituto de Astrof\'{\i}sica, Facultad de F\'{i}sica,
Pontificia Universidad Cat\'{o}lica de Chile, Casilla 306, Santiago 22, Chile} 
\altaffiltext{13}{Millennium Institute of Astrophysics (MAS), Nuncio
Monse\~{n}or S\'{o}tero Sanz 100, Providencia, Santiago, Chile} 
\altaffiltext{14}{Space Science Institute, 4750 Walnut Street, Suite 205,
Boulder, Colorado 80301} 
\altaffiltext{15}{Institute for Astronomy, Astrophysics, Space Applications and
Remote Sensing (IAASARS), National Observatory of Athens, 15236 Penteli,
Greece}
\altaffiltext{16}{Department of Physics, University of North Texas, Denton, TX
76203}
\altaffiltext{17}{INAF - Osservatorio Astrofisico di Arcetri, Largo E. Fermi 5,
I-50125, Florence, Italy}
\altaffiltext{18}{Universit\'a di Bologna, Via Ranzani 1, Bologna, Italy}
\altaffiltext{19}{Physics Department, University of Crete, Heraklion, Greece}
\altaffiltext{20}{Harvard-Smithsonian Center for Astrophysics, 60 Garden
Street, Cambridge, MA 02138, USA}
%

\begin{abstract}
%

We present measurements of the evolution of normal-galaxy \xray\ emission from
$z \approx$~\hbox{0--7} using local galaxies and galaxy samples in the
$\approx$6~Ms \chandra\ Deep Field-South (CDF-S) survey.  The majority of the
CDF-S galaxies are observed at rest-frame energies above 2~keV, where the
emission is expected to be dominated by \xray\ binary (XRB) populations;
however, hot gas is expected to provide small contributions to the
observed-frame $\simlt$1~keV emission at $z \simlt 1$.  We show that a single
scaling relation between X-ray luminosity ($L_{\rm X}$) and star-formation rate
(SFR) is insufficient for characterizing the average \xray\
emission at all redshifts.  We establish that scaling relations involving not
only SFR, but also stellar mass ($M_\star$) and redshift, provide significantly
improved characterizations of the average \xray\ emission from normal galaxy
populations at $z \approx$~0--7.  We further provide the first empirical
constraints on the redshift evolution of \xray\ emission from both low-mass XRB
(LMXB) and high-mass XRB (HMXB) populations and their scalings with $M_\star$
and SFR, respectively.  We find $L_{\rm 2-10~keV}$(LMXB)/$M_\star \propto
(1+z)^{2-3}$ and $L_{\rm 2-10~keV}$(HMXB)/SFR~$\propto (1+z)$, and show that
these relations are consistent with XRB population-synthesis model predictions,
which attribute the increase in LMXB and HMXB scaling relations with redshift
as being due to declining host galaxy stellar ages and metallicities,
respectively.  We discuss how emission from XRBs could
provide an important source of heating to the intergalactic medium in the early
Universe, exceeding that of active galactic nuclei.

%
\end{abstract}
%


%
\section{Introduction}
%

\chandra\ studies of local galaxies have yielded remarkable insight into the
formation and evolution of populations of X-ray binaries (XRBs; see, e.g.,
Fabbiano \etal\ 2006 for a review).  
Expansive multiwavelength
(e.g., from \galex, \herschel, \hubble, and \spitzer) and \chandra\
observations of local star-forming and passive galaxy samples have constrained
basic scaling relations between the X-ray emission from the high-mass XRB
(HMXB) and low-mass XRB (LMXB) populations with star-formation rate (SFR) and
stellar mass ($M_\star$), respectively (see, e.g., Grimm \etal\ 2003; Ranalli
\etal\ 2003; Colbert \etal\ 2004; Gilfanov~2004; Hornschemeier \etal\ 2005;
Lehmer \etal\ 2010; Boroson \etal\ 2011; Mineo \etal\ 2012a; Zhang \etal\
2012); hereafter, we refer to these scaling relations as the $L_{\rm
X}$(HMXB)/SFR and $L_{\rm X}$(LMXB)/$M_\star$ relations, respectively.
However, the scatters in these relations are factors of \hbox{$\approx$2--5}
times larger than the expected variations due to measurement errors and
statistical noise (e.g., Gilfanov~2004; Hornschemeier \etal\ 2005; Mineo \etal\
2012a), thus indicating that real physical variations in the galaxy population
(e.g., stellar ages, metallicities, and star-formation histories) likely have a
significant influence on XRB formation.

Recently, Fragos \etal\ (2013a; hereafter, F13a) used measurements of the local
scaling relations to constrain theoretical XRB population-synthesis models.
The F13a framework was developed using jointly the {\ttfamily Millenium~II}
cosmological simulation (Guo \etal\ 2011) and the {\ttfamily Startrack} XRB
population-synthesis code (e.g., Belczynski \etal \ 2002, 2008) to track the
evolution of the stellar populations in the Universe and predict the \xray\
emission associated with the underlying XRB populations, respectively.  The
F13a models follow the evolution of XRB populations and their parent stellar
populations throughout the history of the Universe, spanning $z \approx$~20 to
the present day, and provide predictions for how the \xray\ scaling relations
(i.e., $L_{\rm X}$(HMXB)/SFR and $L_{\rm X}$(LMXB)/$M_\star$) evolve with
redshift.  From this work, F13a identified a ``best fit'' theoretical model
that simultaneously fits well both the observed $L_{\rm X}$(HMXB)/SFR and
$L_{\rm X}$(LMXB)/$M_\star$ scaling relations at $z=0$.  Subsequent
observational tests have shown that the F13a best-fit model provides reasonable
predictions for (1) the XRB luminosity functions of a sample of nearby galaxies
that had simple star-formation history estimates from the literature
(Tzanavaris \etal\ 2013); (2) the metallicity dependence of the $L_{\rm
X}$(HMXB)/SFR relation for powerful star-forming galaxies (Basu-Zych \etal\
2013a; Brorby \etal\ 2016); (3) the stellar-age dependence of the $L_{\rm
X}$(LMXB)/$M_\star$ relation for early-type galaxies (Lehmer \etal\ 2014;
however, see Boroson \etal\ 2011 and Zhang \etal\ 2012); (4) the redshift
evolution out to $z \approx 1.5$ of the normal galaxy \xray\ luminosity
functions in extragalactic \chandra\ surveys (Tremmel \etal\ 2013); and (5) the
redshift evolution of the total $L_{\rm X}$/SFR relation (i.e., using the
summed HMXB plus LMXB emission) for star-forming galaxies out to $z \approx 4$
(e.g., Basu-Zych \etal\ 2013b).

The F13a theoretical modeling framework, as well as the broad observational
testing of its predictions, represent major steps forward in our understanding
of how XRBs form and evolve along with their parent stellar populations.
Within the F13a framework, the most fundamental predictions include
prescriptions for how the $L_{\rm X}$(HMXB)/SFR and $L_{\rm X}$(LMXB)/$M_\star$
scaling relations evolve as a function of redshift (see Fig.~5 of F13a).  Due
to sensitivity and angular-resolution limitations, it is not possible to detect
complete and representative populations of cosmologically distant galaxies and
separate spatially their HMXB and LMXB contributions.  However, with deep
($\simgt$1~Ms) \chandra\ exposures and new multiwavelength databases, several
extragalactic surveys now have data sufficient to isolate large populations of
galaxies, measure their global physical properties (e.g., SFR and $M_\star$), and
study their population-averaged \xray\ emission via stacking techniques (see,
e.g., Hornschemeier \etal\ 2002; Laird \etal\ 2006; Lehmer \etal\ 2007, 2008;
Cowie \etal\ 2012; Basu-Zych \etal\ 2013b).  With these advances, we can now
conduct the most robust tests to date of the F13a model predictions by 
examining the XRB emission of galaxies dependence on SFR, $M_\star$, and redshift.

The \chandra\ Deep Field-South (CDF-S) survey is the deepest \xray\ survey
yet conducted.  In this paper, we utilize data products based on the first
$\approx$6~Ms of data, which were produced following the procedures outlined
for the $\approx$4~Ms exposure in Xue \etal\ (2011).  An additional
$\approx$1~Ms of data will be added to the CDF-S, eventually bringing the total
exposure to $\approx$7~Ms; these results will be presented in Luo \etal\ (2016,
in preparation).  In the $\approx$6~Ms exposure, 919 highly reliable \xray\
sources are detected to an ultimate \hbox{0.5--2~keV} flux limit of $\approx$$7
\times 10^{-18}$~\flux, including 650 AGN candidates, 257 normal galaxy
candidates, and 12 Galactic stars (see $\S$3 for classification details).  For
comparison, the $\approx$4~Ms CDF-S catalog contained 740 sources down to an
ultimate \hbox{0.5--2~keV} flux limit of $\approx$$10^{-17}$~\flux, of which
568, 162, and 10 were classified as AGN, normal galaxies, and Galactic stars.
In the most sensitive regions of the survey field, the \hbox{0.5--2~keV} detected normal
galaxies rival or exceed the AGN in terms of source density (see, e.g., the
Lehmer \etal\ 2012 analysis of the 4~Ms data).  

Source catalogs based on optical/near-IR imaging contain $\approx$25,000
galaxies within $\approx$7~arcmin of the CDF-S \chandra\ aimpoint (e.g., Luo
\etal\ 2011; Xue \etal\ 2012), indicating that only a small fraction
($\simlt$1\%) of the known normal galaxy population is currently detected in
\xray\ emission.  In this paper, we utilize \xray\ stacking analyses of the
galaxy populations within the CDF-S, divided into redshift intervals and
subsamples of specific SFR, sSFR~$\equiv$~SFR/$M_\star$, which is an indicator
of the ratio of HMXB-to-LMXB emission.  These measurements provide the first
accounting of both HMXBs and LMXBs at $z>0$ to simultaneously constrain the
evolution of the $L_{\rm X}$(HMXB)/SFR and $L_{\rm X}$(LMXB)/$M_\star$ scaling
relations and provide the most powerful and robust test of the F13a model
predictions.

This paper is organized as follows: In $\S$2, we define our galaxy sample and
describe our methods for measuring SFR and $M_\star$ values for the galaxies.
In $\S$3, we discuss the \xray\ properties of galaxies and scaling relations of
our sample that are individually detected in the $\approx$6~Ms images.  In
$\S$4, we detail our stacking procedure, and in $\S$5 we define galaxy
subsamples to be stacked.  Results from our stacking analyses, including
characterizations of the evolution of \xray\ scaling relations, are presented
in $\S$6.  Finally, in $\S$7, we interpret our results in the context of XRB
population-synthesis models, construct models of the evolution of the \xray\
emissivity of the Universe, and estimate the cosmic \xray\ background
contributions from normal galaxies.

Throughout this paper, we make use of the main point-source catalog and data
products for the $\approx$6~Ms CDF-S as will be outlined in Luo \etal\ (2016,
in-preparation).  The Luo \etal\ (2016, in-preparation) procedure is identical
in nature to that presented for the $\approx$4~Ms CDF-S in Xue \etal\ (2011).
The Galactic column density for the \cdfs\ is $8.8 \times 10^{19}$~cm$^{-2}$
(Stark \etal\ 1992).  All of the \hbox{X-ray} fluxes and luminosities quoted
throughout this paper have been corrected for Galactic absorption.  Estimates
of $M_\star$ and SFR presented throughout this paper have been derived assuming
a Kroupa~(2001) initial mass function (IMF); when making comparisons with other
studies, we have adjusted all values to correspond to our adopted IMF.  Values
of $H_0$ = 70~\hbox{km s$^{-1}$ Mpc$^{-1}$}, $\Omega_{\rm M}$ = 0.3, and
$\Omega_{\Lambda}$ = 0.7 are adopted throughout this paper.

%
\section{Galaxy Sample and Physical Properties}
%

We began with an initial sample of 32,508 galaxies in the Great Observatories
Origins Deep Survey South (GOODS-S) footprint as presented in $\S$2 of Xue
\etal\ (2012; hereafter X12; see also Luo \etal\ 2011).  The X12 galaxy sample
was selected using the \hst\ F850LP photometric data from Dahlen \etal\ (2010),
and contains objects down to a 5$\sigma$ limiting magnitude of $z_{850} \approx
28.1$.  We cut our initial sample to the 24,941 objects that were within
7\arcmin\ of the mean $\approx$6~Ms CDF-S aimpoint, a region where the
\chandra\ point-spread function (PSF) is sharpest and the corresponding \xray\
sensitivity is highest.  Hereafter, we refer to the resulting sample as our
base sample.  Of the 24,941 objects in our base sample, 1,124 (4.5\%) have
secure spectroscopic redshifts.  To measure redshifts for the full base sample,
X12 used 12--15 band PSF-matched photometric data (including upper limits)
covering 0.3--8~$\mu$m, and performed spectral energy distribution (SED)
fitting using the Zurich Extragalactic Bayesian Redshift Analyzer (ZEBRA;
Feldmann \etal\ 2006).  The full redshift range of the sample spans $z
=$~0.01--7.  A variety of tests indicated that the resulting photometric
redshift estimates are of high quality (normalized median absolute deviation
$\sigma_{\rm NMAD} = 0.043$) over the range of $z_{850} \approx$~\hbox{16--26},
with a low outlier fraction (fraction of sources with $\vert \Delta z
\vert/(1+z_{\rm spec}) > 0.15$) of $\approx$7\%.

For each of the 24,941 galaxies in the base sample, X12 adopted the
best-available redshift and best-fitting SED for that redshift to estimate
galaxy absolute magnitudes.  In this procedure secure spectroscopic redshifts
were used when available and photometric redshifts were used otherwise.  For a
given galaxy, absolute magnitudes were computed for rest-frame $B$, $V$, $R$,
$I$, $J$, $H$, and $K$ bands using the best-fit SED.  Stellar masses,
$M_\star$, were calculated following the prescription in Zibetti \etal\ (2009),
using rest-frame $B - V$ colors and $K$-band luminosities, $L_K$, along with
the following equation:
\begin{equation}
\log M_\star = \log L_K + 1.176 (B - V) - 1.39,
\end{equation}
where $M_\star$ and $L_K$ are in solar units.  The numerical constants in
Equation~(1) were supplied in Table~B1 of Zibetti \etal\ (2009), which provides
mass-to-light ratio estimates for a variety of rest-frame bands and colors.
The color term ($B-V$) in Equation~(1) accounts for variations in stellar
age (or star-formation history).  Therefore, Equation~(1) is applicable to the
full range of stellar ages and galaxy types.  Uncertainties in the above
calibration are at the $\approx$0.15~dex level, and we adopt this uncertainty
throughout the rest of this paper.  As we discuss below, the stellar masses
derived from this method for a large subset of our galaxies are in good
agreement with those derived from more detailed SED fitting techniques.

We calculated SFRs for the galaxies in our sample using UV and far-IR tracers.
The UV emission is assumed to arise from the young stellar populations, and the
portion of the UV light absorbed and re-radiated by dust is assumed to produce
the far-IR emission.  The majority of the galaxies in our sample have
photometric estimates of the continuum emission near rest-frame 2800~\AA,
thereby allowing accurate estimates of rest-frame 2800~\AA\ monochromatic
luminosities, $\nu l_\nu(2800$~\AA) from the best-fit ZEBRA-based SED templates
used by Xue \etal\ (2012) (see discussion above).  To determine the broad-band
portion of the SED that is associated with only the young UV-emitting
population, we constructed an SED that assumes a constant SF history that
spanned 100~Myr.  For each galaxy, we used this SED to calculate the total {\it
observed} UV emission for the young population following $L_{\rm UV, obs}^{\rm
young} = C(A_V) \nu l_\nu(2800$~\AA), where $C(A_V)$ scales $\nu
l_\nu(2800$~\AA) to the bolometric luminosity, given an attenuation, $A_V$, and
a Calzetti \etal\ (2000) extinction curve.  In our case, $A_V$ was calculated
for all galaxies by X12 in the SED-fitting process.  The observed UV luminosity
alone provides a poor tracer of the SFRs, since the {\it intrinsic} emission
from young stars is, in most cases, attenuated by several multiplicative
factors due to dust extinction.  Therefore, whenever possible, we also measured
galaxy IR luminosities (\hbox{8--1000$\mu$m}; $L_{\rm IR}$) to estimate
directly the levels of UV power obscured and reprocessed by dust (see, e.g.,
Kennicutt~1998 and Kennicutt \& Evans 2012 for reviews).  

To determine $L_{\rm IR}$, we cross-matched our galaxy catalog positions with
the far-IR 24--160$\mu$m GOODS-\herschel\ catalogs.\footnote{GOODS-\herschel\
catalogs, including \spitzer-MIPS 24$\mu$m sources and photometry, are
available via the Herschel Database in Marseille at
http://hedam.lam.fr/GOODS-Herschel/.}  These catalogs include data from
\spitzer-MIPS 24$\mu$m, as well as the \herschel-PACS 100$\mu$m and 160$\mu$m
fluxes of the 24$\mu$m sources; there are no unique sources at 100$\mu$m and
160$\mu$m that are not detected by \spitzer\ 24$\mu$m imaging (Elbaz \etal\
2011).  We identified 931 far-IR counterparts (using a constant matching radius
of 1\arcsec) to the 24,941 galaxies in our base sample (i.e., 3.8\%).  Using
the IR SED template presented by Chary \& Elbaz~(2001), we estimated $L_{\rm
IR}$ for a given IR-detected galaxy by (1) normalizing the template SED to the
$l_\nu$ value derived from the flux of the galaxy in the reddest far-IR
detection band, and (2) integrating the normalized template SED across the
8--1000~$\mu$m band.  We tested for systematic differences between $L_{\rm IR}$
values derived using one band versus another, but found no significant offsets
or any trends with redshift. The corresponding mean ratios and 1$\sigma$
scatter between bands were $\log [L_{\rm IR}$(160$\mu$m)/$L_{\rm
IR}$(100$\mu$m)]~=~$0.08 \pm 0.16$ and $\log [L_{\rm IR}$(100$\mu$m)/$L_{\rm
IR}$(24$\mu$m)]~=~$-0.07 \pm 0.20$, implying the expected error on $L_{\rm IR}$
estimates is on the order of $\approx$0.2~dex.  Of the 931 far-IR detected
galaxies in our sample, we derived $L_{\rm IR}$ using 250, 172, and 509
detections at 24$\mu$m, 100$\mu$m, and 160$\mu$m, respectively.

%
%
\begin{figure}
\figurenum{1}
\centerline{
\includegraphics[width=8.9cm]{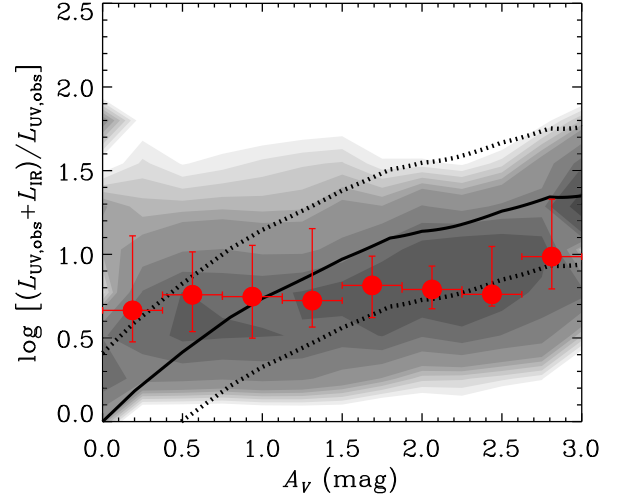}
}
\vspace{0.1in}
\caption{
Ratio of observed UV plus IR luminosity to observed UV luminosity $(L_{\rm UV,
obs}^{\rm young} + L_{\rm IR})/L_{\rm UV, obs}^{\rm young}$ versus measured
$V$-band attenuation $A_V$ (in magnitudes) for the 931 galaxies with IR
detections ({\it smooth contours\/}).  The darkest portions of the contours
indicate the highest concentration of sources.  Median values and 1$\sigma$
intervals of $(L_{\rm UV, obs}^{\rm young} + L_{\rm IR})/L_{\rm UV, obs}^{\rm
young}$, in bins of $A_V$, are indicated as filled red circles with error bars.
The black solid and dotted curves represent the predicted Calzetti \etal\
(2000) extinction curve, $3.2/C(A_V) 10^{0.72 A_V}$, and 1$\sigma$ interval
($\approx$0.5~dex), respectively.
}
\end{figure}

%
%
\begin{figure*}
\figurenum{2}
\centerline{
\includegraphics[width=15cm]{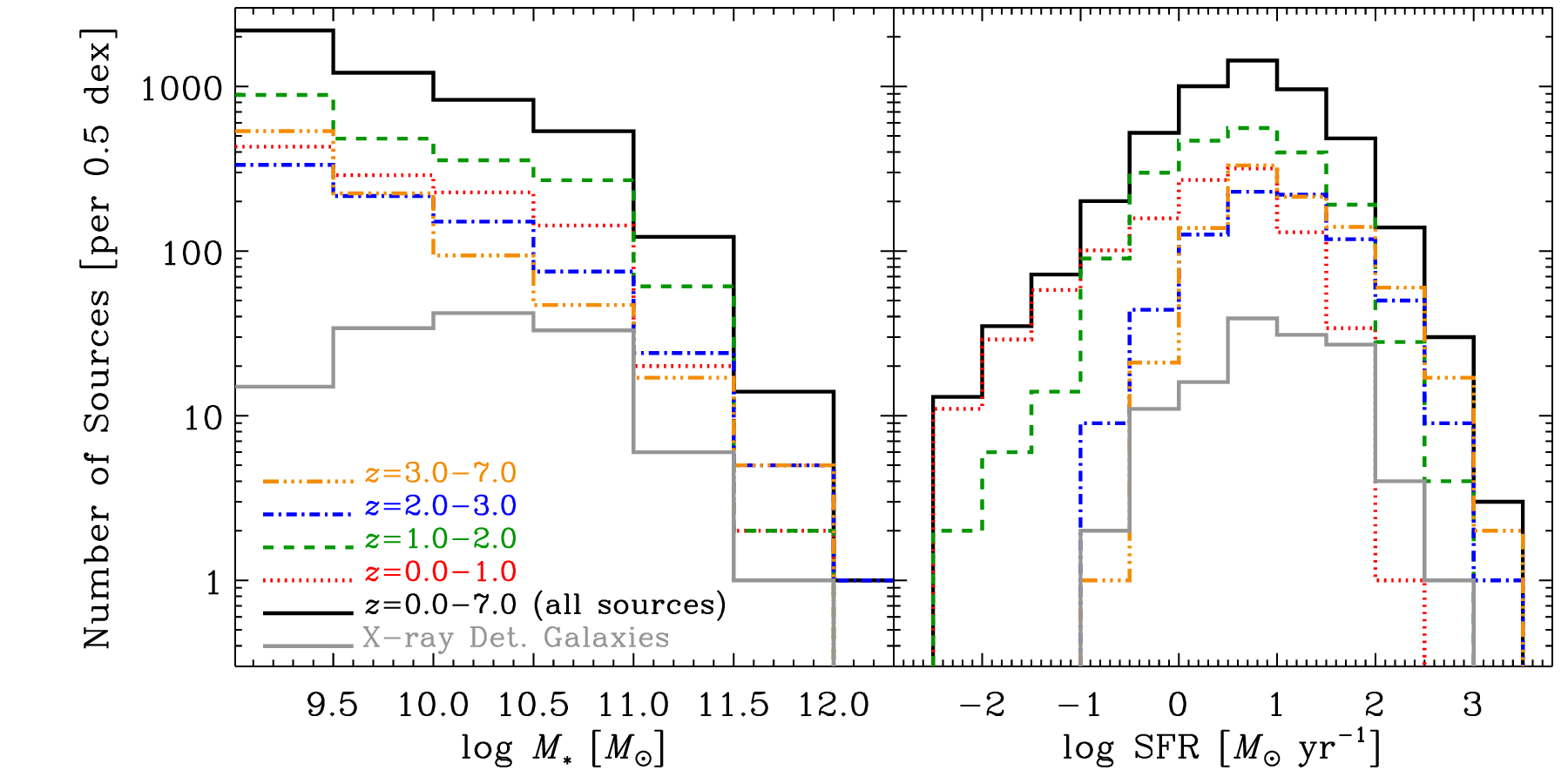}
}
\centerline{
\includegraphics[width=15cm]{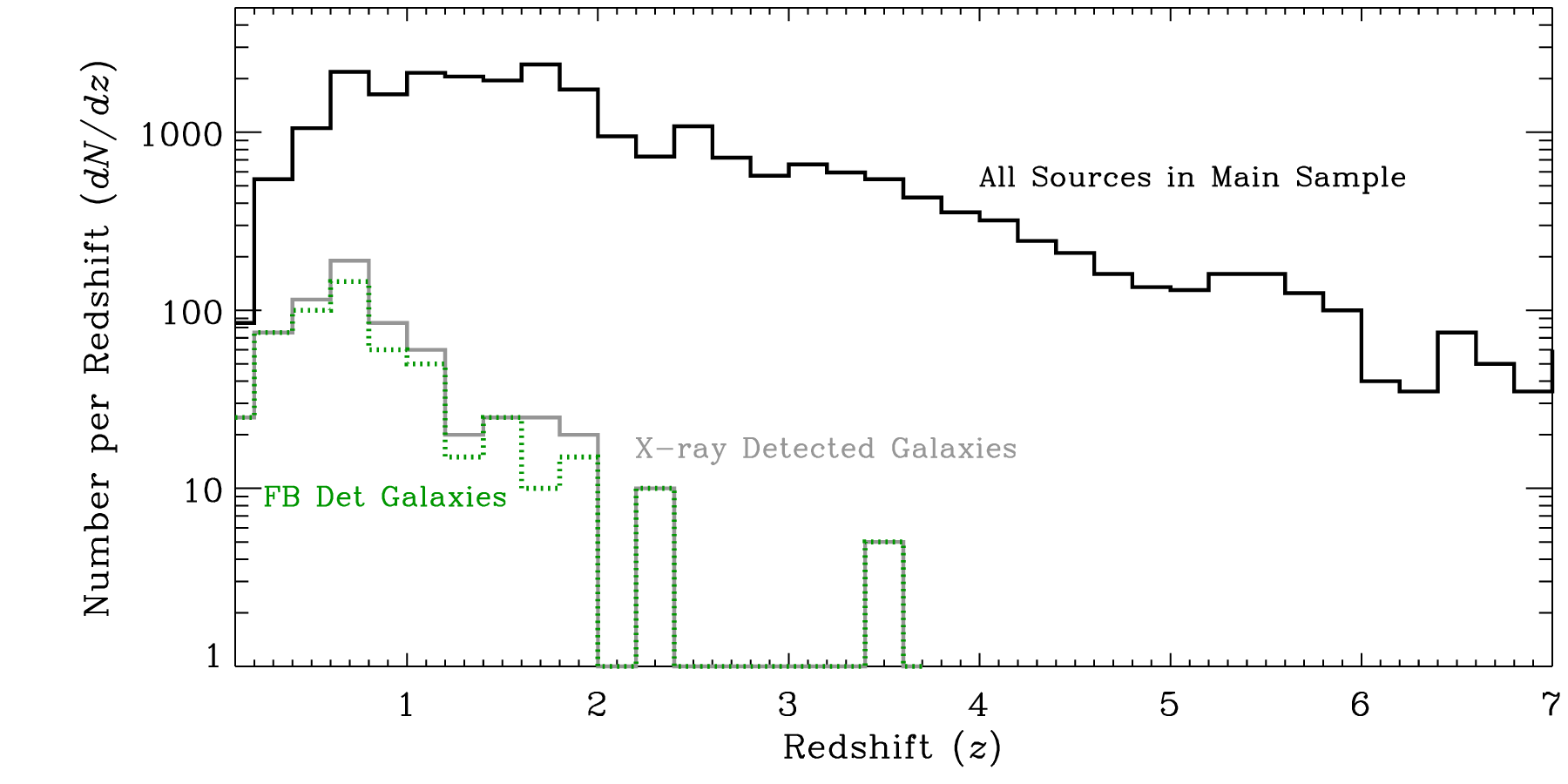}
}
\vspace{0.1in}
\caption{
Histograms showing the distributions of stellar masses ($M_\star$; $a$),
star-formation rates (SFRs; $b$), and redshifts ($z$; $c$) for the \nmcut\
galaxies that make up our main sample of $M_\star \ge 10^9$~\msol\ galaxies.
For the $M_\star$ and SFR distributions, we provide distributions of galaxies
in various redshift bins, and for all distributions, we show the distribution
of \xray\ detected sources that are classified as normal galaxies (see $\S$3
for details).  Finally, the dotted green histogram shows the redshift
distribution of normal galaxies that are detected in the \hbox{0.5--7~keV} band
(i.e., the full-band; FB).  We use the FB-detected normal galaxy sample to
estimate scaling relations based on the \xray\ detected sample in $\S$3.
}
\end{figure*}

For the 931 galaxies with both UV and far-IR detections, we utilized the
methods described in $\S$~3.2 of Bell \etal\ (2005) and estimated total galaxy
SFRs using the following equation:
$$ {\rm SFR}(M_\odot~{\rm yr}^{-1}) = 9.8 \times 10^{-11} L_{\rm bol}^{\rm
young}$$
\begin{equation}
L_{\rm bol}^{\rm young}  \approx L_{\rm UV, obs}^{\rm young} + L_{\rm IR},
\end{equation}
where all luminosities are expressed in units of solar bolometric luminosity
($L_\odot = 3.9 \times 10^{33}$~erg~s$^{-1}$) and $L_{\rm bol}^{\rm young}$ is
the bolometric luminosity associated with the 100~Myr stellar population with
constant SFR.  For the large fraction of galaxies in our sample (96.2\%) that
lack far-IR detections, we calculated $L_{\rm bol}^{\rm young}$ and SFRs using
dust-extinction corrected UV luminosities following the Calzetti \etal\ (2000)
extinction law:
\begin{equation}
L_{\rm bol}^{\rm young} \approx 10^{0.72 A_V} 3.2 \nu l_\nu (2800 \angstrom),
\end{equation}
where the factor of 3.2 provides the bolometric correction of the young
population SED (see above) to $\nu l_\nu (2800 \angstrom)$ for an $A_V = 0$
(i.e., $C(A_V=0) = 3.2$).  Figure~1 displays the distribution of $(L_{\rm UV,
obs}^{\rm young} + L_{\rm IR})/L_{\rm UV, obs}^{\rm young}$ versus $A_V$ for the 931 galaxies with
far-IR detections and overlay the formal extinction-law prediction.  It is
clear from the distribution of far-IR detected sources that there is
substantial statistical scatter in the relation, and a bias is present due to
the fact that galaxies with low $(L_{\rm UV, obs}^{\rm young} + L_{\rm IR})/L_{\rm UV,
obs}^{\rm young}$ and low $A_V$ will be excluded, since they would not be far-IR detected.
Nonetheless, these data indicate that using $A_V$ and $\nu l_\nu(2800
\angstrom)$ (without an IR measurement) to estimate $L_{\rm UV, obs}^{\rm young} + L_{\rm
IR}$ and the implied SFR for an object will result in a 1$\sigma$ uncertainty
of $\sim$0.5~dex, a factor of $\sim$2 times larger than the uncertainty of a
far-IR based measurement of $L_{\rm IR}$.  Hereafter, we adopt uncertainties of
0.2~dex and 0.5~dex for SFRs calculated using UV plus far-IR and the UV-only
data, respectively.

Of the 24,491 galaxies in our base sample, the majority are low mass ($M_\star
< 10^9$~\msol) and will not produce significant \xray\ emission per galaxy, nor
will they provide substantial contributions to the global \xray\ emission of
the Universe.  Going forward, we chose to exclude low-mass galaxies with
$M_\star < 10^9$~\msol\ from further analyses, since (1) they are more than two
orders of magnitude below the knee of the stellar mass function of galaxies,
which is at $M_\star \simgt 10^{11}$~\msol\ out to $z > 4$, and therefore not
representative of the stellar mass in the Universe (e.g., Muzzin \etal\ 2013);
(2) they are optically faint and have large uncertainties in their rest-frame
parameters; and (3) they effectively dilute \xray\ stacking signals when
included with higher-mass galaxies that have, e.g., similar SFR/$M_\star$
values.  We note that low-mass galaxies are relatively metal-poor and may
indeed have enhanced levels of XRB emission per unit stellar mass or SFR (e.g.,
Linden \etal\ 2010; Basu-Zych \etal\ 2013a, 2016; Prestwich \etal\ 2013; Brorby
\etal\ 2014, 2016; Douna \etal\ 2015); however, given their relatively
low-mass, these enhancements are not important to the overall {\itshape
average} global scaling relations.  

After removing the low-mass galaxies, our resulting {\itshape main sample}
contains \nmcut\ $M_\star \ge 10^9$~\msol\ galaxies within 7\arcmin\ of the
CDF-S aimpoint.   In Figure~2, we show the distributions of $M_\star$, SFR, and
$z$ for the main sample.  Given the low stellar mass limit of our sample
($\approx$2 orders of magnitude below the knee of the stellar mass function at
all redshifts), \xray\ scaling relations derived from this sample will be
representative of the global stellar populations in the Universe.  We tested
this assumption in the local Universe by deriving scaling relations from the
Lehmer \etal\ (2010; hereafter, L10) galaxy sample (see L10 for procedures for
deriving scaling constant) that both include and exclude $M_\star <
10^9$~\msol\ galaxies and found no material differences in derived relations.
Specifically, $[L_{\rm 2-10~keV}({\rm LMXB})/M_{\star}]_{z=0}^{\rm include} =
2.2^{+1.9}_{-1.1} \times 10^{29}$~\lum~\msol$^{-1}$ and $[L_{\rm 2-10~keV}({\rm
LMXB})/M_{\star}]_{z=0}^{\rm exclude} = 2.4^{+2.0}_{-1.2} \times
10^{29}$~\lum~\msol$^{-1}$, and $[L_{\rm 2-10}({\rm HMXB})/{\rm
SFR}]_{z=0}^{\rm include} = 1.8^{+0.5}_{-0.4} \times
10^{39}$~\lum~(\sfr)$^{-1}$ and $[L_{\rm 2-10}({\rm HMXB})/{\rm
SFR}]_{z=0}^{\rm exclude} = (1.7 \pm 0.3) \times 10^{39}$~\lum~(\sfr)$^{-1}$.
For consistency, however, when comparing stacking results with the local L10
sample, we make use of scaling relations that exclude $M_\star < 10^9$~\msol\
galaxies.

Within the main sample, 870 galaxies have far-IR based measurements of $L_{\rm
IR}$ and SFRs from $L_{\rm UV, obs}^{\rm young} + L_{\rm IR}$ (Eqn.~2), while
the remaining 4,028 galaxies have SFRs measured from $A_V$ and $\nu l_\nu(2800
\angstrom)$ (Eqns.~2 and 3).  To ensure the properties of our main sample were
robust, we compared our values of $M_\star$ and SFR to those available in the
literature.  Using the Rainbow catalog galaxy sample builder,\footnote{See
http://arcoiris.ucsc.edu/Rainbow\_navigator\_public/ for details.} we
constructed a sample of F160W-selected sources from the Cosmic Assembly
Near-infrared Deep Extragalactic Legacy Survey (CANDELS; Grogin \etal\ 2011;
Koekemoer \etal\ 2011) that included stellar mass values from Mobasher \etal\
(2015) and SFR values following Santini \etal\ (2015).  The \nmcut\ main sample
galaxies were cross-matched to the CANDELS catalog using a matching radius of
1\farcs0, and 3,284 matches were obtained.  We found that the mean ratios and
1$\sigma$ scatters between our values and those in the CANDELS catalogs were
$\log [M_\star/M_\star({\rm CANDELS})] = -0.04 \pm 0.27$ and $\log [{\rm
SFR}/{\rm SFR}({\rm CANDELS})] = -0.04 \pm 0.60$.  This result indicates that
there are no systematic differences between methods, despite there being
non-negligible scatter.  The level of scatter is comparable to our adopted
errors on these quantities, which we account for throughout the rest of this
paper.

%
\section{X-ray Detected Sources}
%

Using a {\it conservative} constant 0\farcs5 matching radius, we cross-matched
our main sample of \nmcut\ galaxies with the 613 \xray\ point-sources in the
$\approx$6~Ms CDF-S main catalog that are within 7\arcmin\ of the aimpoint.
There are 388 matches to \xray\ point sources that are detected in at least one
of the Luo \etal\ (2016, in-preparation) standard bands: 0.5--7~keV (FB),
0.5--2~keV (SB), and 2--7~keV (HB).  We note that the majority of the 225
\xray\ sources without matches within 0\farcs5 also have optical/near-IR
counterparts, but those counterparts are either (1) low-mass galaxies excluded
from our sample (see above) or (2) sources with optical offsets $>$0\farcs5.
We chose to exclude sources with offsets $>$0\farcs5 from our analysis to
strictly limit false matches.  We determined the number of likely false matches
by shifting the \xray\ source positions by 10\arcsec\ in four directions and
re-matching them to the main sample using the above criterion.  This exercise
produced 6--14 matches per shift, suggesting a false-match rate of 1.5--3.6\%.
We classified $\approx$6~Ms CDF-S sources with $z>0$ (i.e., not Galactic stars)
as either ``AGN'' or ``normal galaxies'' using the following five criteria (as
outlined in $\S$4.4 of Xue \etal\ 2011):

\begin{enumerate}

\item A source with an intrinsic \xray\ luminosity of $L_{\rm 0.5-7~keV} \ge 3
\times 10^{42}$~\lum\ was classified as an AGN.  We obtained $L_{\rm
0.5-7~keV}$ using the following procedure.  We first estimated the relationship
between the \hbox{2--7~keV} to \hbox{0.5--2~keV} count-rate ratio and intrinsic
column density and redshift for a power-law model SED that includes both
Galactic and intrinsic extinction (in {\ttfamily xspec} $wabs \times zwabs
\times zpow$) with a fixed $\Gamma_{\rm int} =  1.8$.  For each source, we
obtained the intrinsic extinction, $N_{\rm H, int}$, using the count-rate ratio
and redshift.  For some cases, the count-rate ratios were not well constrained
due to lack of detections in both the \hbox{0.5--2~keV} and \hbox{2--7~keV}
bands; for these sources an effective $\Gamma = 1.4$ was assumed.  From our SED
model, we could then obtain the absorption-corrected \hbox{0.5--7~keV} flux,
$f_{\rm 0.5-7~keV, {\rm  int}}$, and luminosity following $L_{\rm 0.5-7~keV} =
4 \pi d_L^2 f_{\rm 0.5-7~keV, int} (1 + z)^{\Gamma_{\rm int}-2}$, where $d_L$
is the luminosity distance.

\item A source with an effective photon index of $\Gamma \le 1.0$ was
classified as an obscured AGN.

\item A source with \xray-to-optical (using $R$-band as the optical reference)
flux ratio of $\log(f_{\rm X}$/$f_{\rm R}
> -1$ (where $f_{\rm X} = f_{\rm 0.5-7~keV}$, $f_{\rm 0.5-2~keV}$, or $f_{\rm
2-7~keV}$) was classified as an AGN.

\item A source with excess (i.e., a factor of $\ge$3) \xray\ emission over the
level expected from pure star formation was classified as an AGN, i.e., with
$L_{\rm 0.5-7~keV} \simgt 3 \times (8.9 \times 19^{17} L_{\rm 1.4~GHz})$, where
$L_{\rm 1.4~GHz}$ is the rest-frame 1.4~GHz monochromatic luminosity in units
of W~Hz$^{-1}$ and $8.9 \times 19^{17} L_{\rm 1.4~GHz}$ is the expected \xray\
emission level that originates from star-forming galaxies (see Alexander \etal\
2005 for details).

\item A source with optical spectroscopic features characteristic of AGN
activity (e.g., broad-line emission and/or high-excitation emission lines) was
classified as an AGN.  Using a matching radius of 0\farcs5, we cross-matched
the $\approx$6~Ms CDF-S sources with spectroscopic catalogs from Szokoly \etal\
(2004), Mignoli \etal\ (2005), and Silverman \etal\ (2010) to identify these
optical spectroscopic AGN.

\end{enumerate}

All sources with $z>0$ that were not classified as AGN from the above five
criteria, were classified as normal galaxies.  Out of the 388 \xray\
detected sources, we classified 141 as normal galaxies, with the remaining 247
sources classified as AGN.  

As described in L10), estimates of the $L_{\rm X}$(LMXB)/$M_\star$ and $L_{\rm
X}$(HMXB)/SFR scaling relations for a galaxy population can be obtained by
empirically constraining the following relation:
$$L_{\rm X}({\rm XRB}) = L_{\rm X}({\rm LMXB}) + L_{\rm X}({\rm HMXB}) = \alpha
M_\star + \beta {\rm SFR},$$ 
\begin{equation}
L_{\rm X}({\rm XRB})/{\rm SFR} = \alpha ({\rm SFR}/M_\star)^{-1} + \beta,
\end{equation}
where $L_{\rm X}({\rm XRB})$ is the total \xray\ luminosity due to the XRB
population, and \hbox{$\alpha \equiv L_{\rm X}({\rm LMXB})/M_\star$} and
\hbox{$\beta \equiv L_{\rm X}({\rm HMXB})/{\rm SFR}$} are fitting constants.
As outlined in $\S$4 below, XRBs typically dominate the total galaxy-wide
emission at energies above $\approx$1--2~keV.  Therefore, in practice,
Equation~(4) can be constrained using total galaxy-wide luminosities derived
from a hard bandpass (e.g., the \hbox{2--10~keV} band; see L10 for further
details on the $z \approx 0$ normal-galaxy population).  However, due to the
relatively high background and low \chandra\ effective area at $>$2~keV, only
23 of the 141 normal galaxies are detected in the HB, limiting our ability to
constrain $\alpha$ and $\beta$.  Fortunately, at the median redshift of the 116
normal galaxies detected in the FB, $\left < z \right
>_{\rm median} = 0.67$, the FB itself samples the rest-frame 0.8--12.8~keV
bandpass, which is expected to be dominated by XRBs (see $\S$4 below); however,
only normal galaxies at $z \simgt 2$ are likely to be entirely dominated by XRB
emission across the full 0.5--7~keV range.  Nonetheless, we can obtain
reasonable constraints on $\alpha$ and $\beta$ for the FB-detected
normal-galaxy population.

%
%
\begin{figure}
\figurenum{3}
\centerline{
\includegraphics[width=8.9cm]{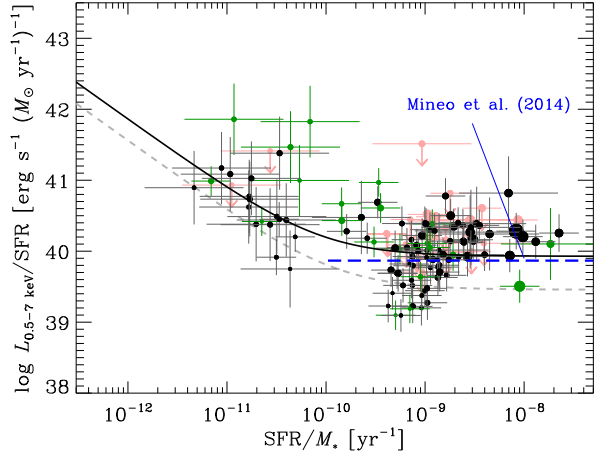}
}
\vspace{0.1in}
\caption{
Logarithm of the FB (0.5--7~keV) luminosity per unit SFR, $\log L_{\rm
0.5-7~keV}$/SFR, versus sSFR for the 141 \xray\ detected galaxies in our CDF-S
sample ({\it black filled circles with error bars\/}).  Symbol sizes increase
with source redshift; error bars are 1~$\sigma$ and represent Poisson errors on
the \xray\ counts as well as errors on the SFR measurements.  Sources that
are also detected in the HB have been distinguished using green symbols and error bars.
Upper limits ({\it red symbols\/}) are applied to sources that are detected in
either the SB or HB, but not the FB.  The solid curve represents the best-fit
solution to Equation~(4), based on the 116 FB-detected normal galaxies.  The
short-dashed gray curve represents the $z = 0$ scaling relation derived by L10
and the long-dashed blue curve represents the mean $L_{\rm 0.5-7~keV}$/SFR
value found by Mineo \etal\ (2014) using sSFR~$\simgt 10^{-10}$~yr$^{-1}$
\xray\ and radio detected galaxies in the CDF-N and CDF-S.  The best-fit
solution to the CDF-S data is likely biased towards the inclusion of galaxies
with high values of $\log L_{\rm 0.5-7~keV}$/SFR at low-sSFR due to the \xray\
selection.  
}
\end{figure}

Figure~3 displays the FB luminosity per unit SFR ($L_{\rm 0.5-7~keV}$/SFR)
versus sSFR for the 141 \xray\ detected galaxies in our sample (including 116
FB detections).  An inverse relation between $L_{\rm 0.5-7~keV}$/SFR and
SFR/$M_\star$ provides a better fit to the data than a simple constant ratio of
$L_{\rm 0.5-7~keV}$/SFR, with an anti-correlation between $L_{\rm
0.5-7~keV}$/SFR and SFR/$M_\star$ being significant at the $>$99.99\%
confidence level (based on a Spearman's $\rho$ rank correlation).  Following
the functional form presented in Equation~(4), we use these data to extract
values of $\alpha$ and $\beta$ for the FB detected sample.  The galaxy sample
has a median redshift of $\left < z \right >_{\rm median} = 0.67$ with an
interquartile range of \hbox{$\Delta z$ = 0.41--0.98}; the fits produce
\hbox{$\alpha_{\rm FB}(z \sim 0.7)$}~=~ $(7.2^{+2.5}_{-2.0}) \times
10^{29}$~erg~s$^{-1}$~$M_\odot^{-1}$ and $\beta_{\rm FB}(z \sim 0.7) =
8.5^{+0.8}_{-0.7} \times 10^{39}$~erg~s$^{-1}$~($M_\odot$~yr$^{-1}$)$^{-1}$.
This value of $\beta$ is in reasonable agreement with the mean $L_{\rm
0.5-7~keV}$/SFR~$\approx 7.4 \times
10^{39}$~erg~s$^{-1}$~($M_\odot$~yr$^{-1}$)$^{-1}$ presented by Mineo \etal\
(2014) from a sample of $z \simlt$~1.3 \xray\ and radio detected galaxies in
the CDF-N and CDF-S that have sSFR~$\simgt 10^{-10}$~yr$^{-1}$ (see Fig.~3).

If we assume a power-law \xray\ SED with photon index $\Gamma = 2$,
$\alpha_{\rm FB}$ and $\beta_{\rm FB}$ can be converted to the \hbox{2--10~keV}
bandpass equivalents by dividing by 1.64.  When compared with the $z = 0$
values of $\alpha$ and $\beta$, derived for the 2--10~keV band from L10,
$(\alpha_{z \sim 0.7}/\alpha_{z = 0})_{\rm 2-10~keV} \approx 4.7$ and
$(\beta_{z \sim 0.7}/\beta_{z = 0})_{\rm 2-10~keV} \approx 3.0$.\footnote{For
this calculation and elsewhere in this paper, we have adjusted stellar mass
values of L10 to be consistent with those calculated using Equation~(1), which
was derived by Zibetti \etal\ (2009).  L10 originally utilized prescriptions in
Bell \etal\ (2003) for calculating stellar masses.  The most notable difference
between the Zibetti \etal\ (2009) and Bell \etal\ (2003) stellar masses arises
for blue galaxies, for which the Zibetti \etal\ relation provides lower
mass-to-light ratios primarily due to the inclusion of blue-light absorption
(see $\S$2.4 of Zibetti \etal\ 2009).  For CDF-S galaxies, we found that the
Zibetti \etal\ (2009) stellar masses were in better agreement with those found
in the literature than the Bell \etal\ (2003) stellar masses, and we therefore
chose to adopt Equation~(1) here and convert L10 $M_\star$ values
appropriately.  We derive $\alpha_{z = 0} = 2.2^{+1.9}_{-1.1} \times
10^{29}$~\lum~$M_\odot^{-1}$ and $\beta_{z = 0} = 1.8^{+0.5}_{-0.4} \times
10^{39}$~\lum~(\sfr)$^{-1}$ for the 2--10~keV band L10 constraints.}  This
result suggests that both the $L_{\rm X}$(LMXB)/$M_\star$ and $L_{\rm
X}$(HMXB)/SFR scaling relations may increase significantly with redshift, with
potentially stronger evolution of the LMXB population; broadly consistent with
the F13 population-synthesis predictions.

We caution that the above result is inherently biased in nature, since the
galaxy sample includes only \xray\ detected galaxies that were identified as
normal galaxies using some dependence on \xray\ scaling relations appropriate
for the local Universe (see $\S$3.1 of Lehmer \etal\ 2012).  For example, among
other criteria, normal galaxies were discriminated from AGN by having $L_{\rm
0.5-7~keV}/L_{\rm 1.4~GHz}$ and \xray--to--optical flux ratios below some
limiting values.  These criteria effectively place a ceiling on the maximum
$L_{\rm 0.5-7~keV}$/SFR value for a normal galaxy in the \xray\ selected
sample.  Furthermore, as a result of the \xray\ selection, we are more complete
to high-SFR/high-sSFR galaxies that are \xray\ luminous compared to the
low-SFR/low-sSFR galaxy population.  This could bias the \xray\ selected sample
toward galaxies with relatively large values of $L_{\rm 0.5-7~keV}$/SFR at
low-SFR/low-sSFR, which would effectively bias $\alpha$ high.  

%
\section{Stacking Procedure}
%

As discussed in $\S$3, the vast majority of the normal galaxies in our sample
have \xray\ emission below the individual source detection limit of the
$\approx$6~Ms CDF-S.  Since scaling relations derived from the \xray\ detected
normal galaxy population are biased by both the classification of a ``normal
galaxy'' being restricted to sources that are within certain factors of $z=0$
scaling relations (see criteria in $\S$3), and the fact that the objects
include only sources that are \xray\ bright and detectable, these relations may
not be representative of the broader galaxy population.  To mitigate this
limitation, we implement \xray\ stacking analyses using complete subpopulations
of galaxies, selected by physical properties (i.e., SFR and $M_\star$).  In the
paragraphs below, we describe our stacking procedure in general terms, as it is
applied to a given subpopulation.  In the next section ($\S$5), we define the
galaxy subpopulations for which we apply the stacking procedure.  Variations of
our stacking procedure have been implemented in a variety of previous studies
of distant normal galaxies (e.g., Lehmer \etal\ 2005, 2007, 2008; Steffen
\etal\ 2007; Basu-Zych \etal\ 2013b); for completeness, we highlight here the
salient features of the procedure adopted in this paper.

Our stacking procedure begins with the extraction of on-source counts, local
background counts, and exposure times for all \nmcut\ galaxies in our main
sample using three \xray\ bandpasses: 0.5--1~keV, 1--2~keV, and 2--4~keV (see
below for justification).  Using a circular aperture with a radius of $R_{\rm
ap}$, we extracted \chandra\ source-plus-background counts $S_i$ and exposure
times $T_i$ from the \xray\ images and exposure maps, respectively.  For
galaxies with $z < 0.7$ we chose to use an aperture with $R_{\rm ap} =
2\farcs5$, and for $z \ge 0.7$, we used $R_{\rm ap} = 1\farcs5$.  These
apertures correspond to rest-frame physical radii of $\simgt$11~kpc for $z =
0.3$--0.7 where $R_{\rm ap} = 2\farcs5$, and $\simgt$10~kpc for $z =$~0.7--4
and $\approx$8--10~kpc for $z =$~4--7, where $R_{\rm ap} = 1\farcs5$.  This
choice was motivated by the goal of both encompassing the vast majority of the
\xray\ emission from XRBs within galaxies while maintaining high
signal-to-noise in our stacking (see below).  The creation of images and
exposure maps is described in Xue \etal\ (2011).  In this process, each
aperture was centered on the optical location of each galaxy.  Using background
maps, we measured the local backgrounds $B_{i, {\rm local}}$ within a circular
aperture with radius $R_{\rm local} =$~25\arcsec\ that was centered on each
source.  In practice, the background maps include \xray\ counts from the
galaxies that are not detected individually, so our local background circular
apertures will include counts from the individually undetected sources of
interest.  We estimated the expected number of background counts within each
on-source extraction aperture $B_i$ by scaling the local background counts to
the area of the extraction aperture (i.e., $B_i = B_{i, {\rm local}} \times
R_{\rm ap}^2/R_{\rm local}^2$).

%
%
\begin{figure*}
\figurenum{4}
\centerline{
\includegraphics[width=8.9cm]{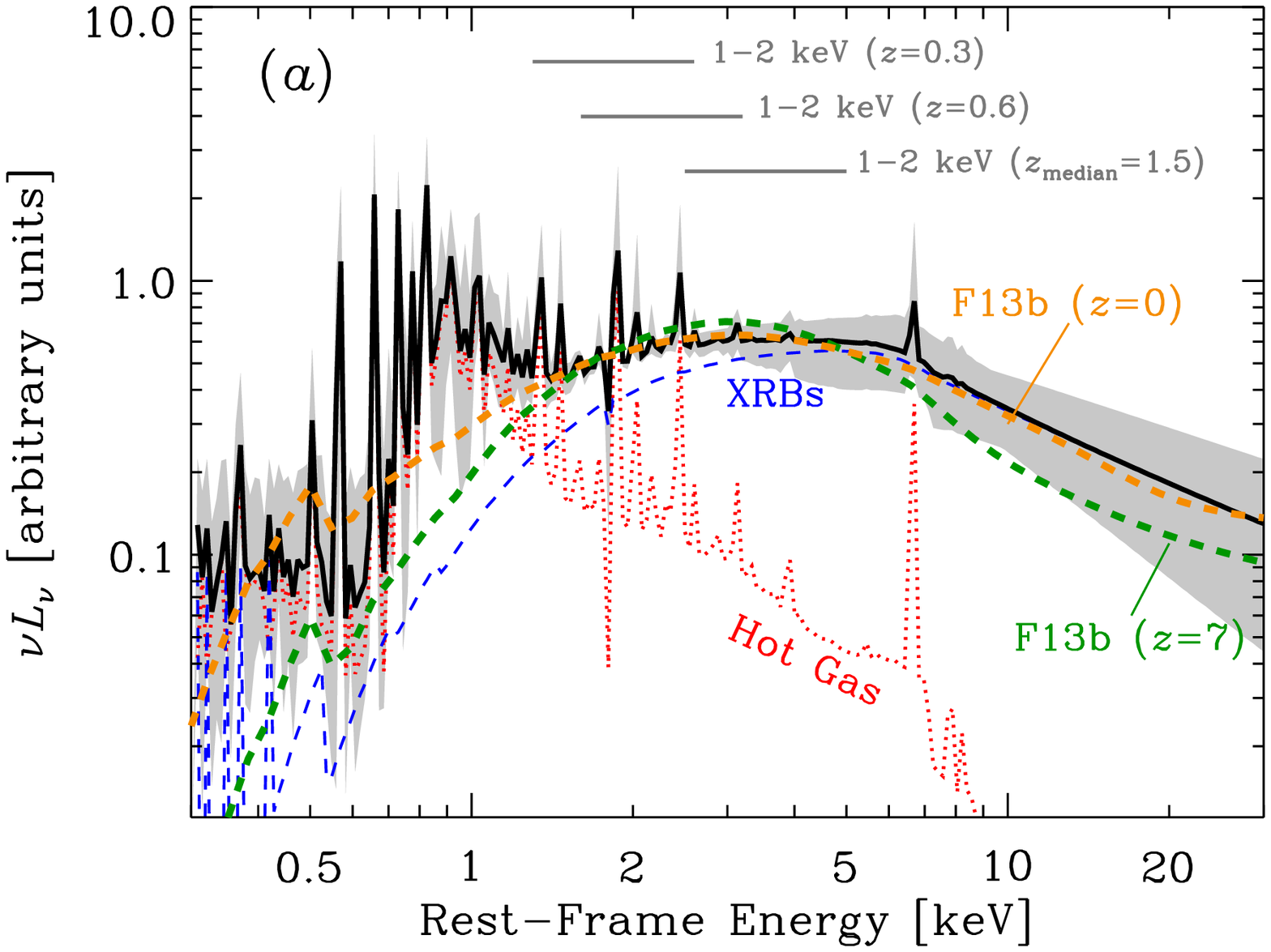}
\hfill
\includegraphics[width=8.9cm]{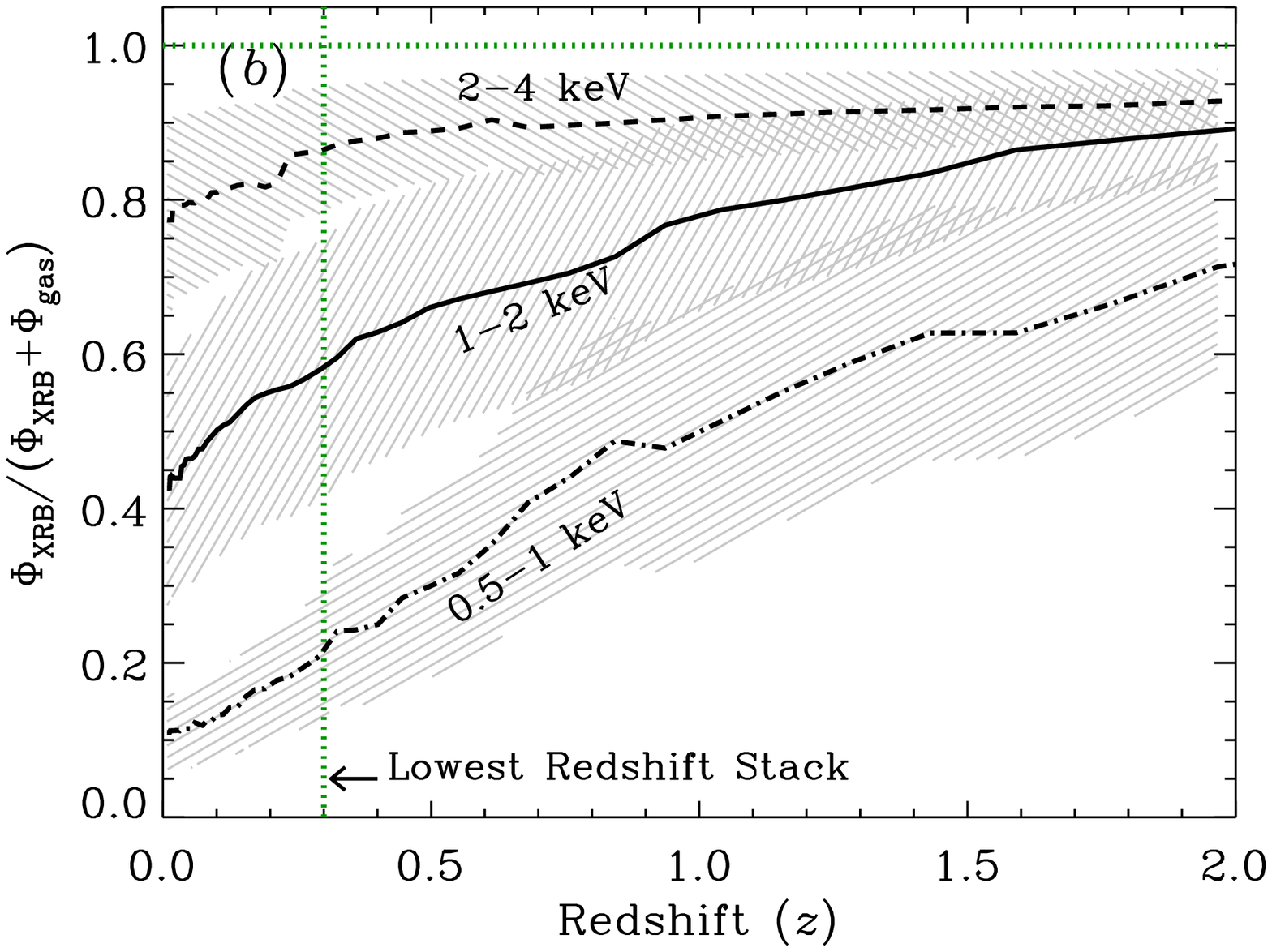}
}
\vspace{0.1in}
\caption{
($a$) Mean 0.3--30~keV SED ({\it solid curve\/}) for four local star-forming
galaxies (M83, NGC~253, NGC~3256, NGC~3310) and full range of SED shape ({\it
grey envelope\/}).  These SEDs were constrained observationally using a
combination of simultaneous \chandra/\xmm\ and \nustar\ observations (Wik
\etal\ 2014; Lehmer \etal\ 2015; Yukita \etal\ 2016).  The mean
SED was calculated by normalizing each galaxy SED to its total 0.3--30~keV
emission and averaging in energy bins.  The relative contributions of hot gas
and XRBs to the mean SED have been shown with dotted red and dashed blue
curves, respectively.  For comparison, the $z = 0$ and $z = 7$ synthetic
spectra from the Fragos \etal\ (2013b) XRB population-synthesis models are
shown as orange and green dashed curves, respectively.  The rest-frame energy
range for stacking bandpasses in the lowest redshift interval ($z = 0.3$), the
second lowest redshift interval ($z = 0.6$), and the median redshift interval
($z = 1.5$) have been annotated.
($b$) Estimated fractional contribution of XRBs to the stacked counts as a
function of redshift for the observed-frame 0.5--1~keV ({\it dot-dashed
curve\/}), 1--2~keV ({\it solid curve\/}) and 2--4~keV ({\it dashed curve\/})
bandpasses.  Throughout this paper, we utilized 0.5--1~keV and 1--2~keV stacks
and the SED presented in panel $a$ to estimate rest-frame 0.5--2~keV and
2--10~keV luminosities, respectively.
}
\end{figure*}

To avoid contamination from bright unrelated \xray\ sources, we excluded
galaxies from our stacking analyses that were located within two times the 90\%
encircled-energy fraction radius of any {\it unassociated} \xray\ detected
sources.  However, \xray\ detected sources that were classified as normal
galaxies by the criteria in Lehmer \etal\ (2012) were included in stacked
subsamples (see $\S$3).  When stacking a given galaxy subsample, we calculated
stacked source-plus-background ($S = \sum_i S_i$) and background counts ($B =
\sum_i B_i$) and used these quantities to estimate the net counts ($S-B$) of
the subsample.  For each stacked subsample, we required that the
signal-to-noise ratio (S/N~$\equiv (S-B)/\sqrt{S}$) be greater than or equal to
3 (i.e., one-sided confidence level of $\approx$99.9\%) for a detection.  Due
to the fact that our 1.5\arcsec\ radius stacking aperture encircles only a
fraction of the point-source flux\footnote{At off-axis angles $\theta \approx
3$\arcmin, our 1.5\arcsec\ radius circular aperture contains an
encircled-energy fraction of $\approx$100\% for the 0.5--7~keV band; however at
$\theta \approx 7$\arcmin, this fraction decreases to $\approx$35\%.} for
sources at relatively large off-axis angles, we calculated off-axis-dependent
aperture corrections $\xi_i$ for each stacked source $i$.  We note that the
absolute astrometric match between the optical and \xray\ frames
($\simlt$0\farcs1; Luo \etal\ 2016, in-preparation) is much smaller than our
stacking aperture size, and therefore no corrections related to errors in the
astrometric alignment need to be implemented.  Since we are calculating average
\xray\ counts from the summed emission of sources in differing backgrounds and
exposure times, we applied a single, representative exposure-weighted aperture
correction, $\xi$.  This factor, which was determined for each stacked
subsample, was calculated as follows:
\begin{equation}
\xi \equiv \frac{\sum_i \xi_i \times T_i}{T},
\end{equation}
where $T = \sum_i T_i$.  We find a range of $\xi =$~1.4--2.0 for all bandpasses
and stacking subsamples in this study.  Using these corrections, we computed
mean \xray\ count-rates $\Phi$ for each stacked subsample using the following
equation:
\begin{equation}
\Phi =  \xi \left(\frac{S-B}{T}\right).
\end{equation} For a given stacked subsample, the errors on $\Phi$ were
calculated using a bootstrap resampling technique.  For a given galaxy
subsample, we constructed 10,000 ``resampled'' galaxy lists that were stacked
using the above procedure.  Each resampled list was constructed by drawing, at
random, galaxies from the original list until the resampled list contained the
same number of sources as the original list.  A given resampled list will
typically contain multiple entries of galaxies from the original list, without
including all entries in the original list.  The 1$\sigma$ error on the count
rate of a given stacked subsample was thus obtained by calculating the scatter
in count-rates obtained for the 10,000 resampled lists. 

To convert observed count rates to fluxes, we chose to make use of the mean
\xray\ SED shape of four starburst and normal galaxies from the \nustar\ galaxy
program: NGC~253, M83, NGC~3256, and NGC~3310 (Lehmer \etal\ 2015).  These
galaxies span \hbox{sSFR~=~0.08--0.8~Gyr$^{-1}$}, SFR~=~3--30~\sfr, and
$M_\star = (0.9$--$6) \times 10^{10}$~\msol, values similar to the mean values
that we use in our stacked samples from $z \approx$~0--1.5 (see below).  The
SEDs for the four galaxies are well constrained over the \hbox{0.3--30~keV}
bandpass via simultaneous \chandra/\xmm\ and \nustar\ observations (e.g., Wik
\etal\ 2014; Lehmer \etal\ 2015; Yukita \etal\ 2016).  The galaxy-wide
\chandra/\xmm\ plus \nustar\ spectra for these galaxies were modeled using a
combination of two or three thermal plasma (hot gas) components with $kT
\approx$~\hbox{0.3--2~keV} (modeled using {\ttfamily apec} in {\ttfamily
XSPEC}) plus a component associated with XRBs (using {\ttfamily power-law} or
{\ttfamily broken power-law} models).  Both the hot gas and power-law
components had varying levels of intrinsic absorption (see references for
details); however, here we focus on observed spectra and do not attempt to
correct for intrinsic absorption.

Figure~4$a$ displays the mean \xray\ SED with relative contributions from hot
gas and XRBs indicated.  On average, hot gas and XRBs dominate below and above
$E \approx$~1.5~keV, respectively.  Below $\approx$6~keV, the mean XRB spectrum
is well characterized as an absorbed power-law with \hbox{$\Gamma \approx
1.8$--2.0,} similar to previous \chandra\ studies of the mean \xray\ spectra of
bright XRBs in star-forming galaxies (e.g., Mineo \etal\ 2012a; Pacucci \etal\
2014).  Above $\approx$6~keV, the \xray\ spectral shape develops a steeper
spectral slope ($\Gamma \approx 2.5$), a feature not widely accounted for in
XRB population studies (see, e.g., Kaaret~2014 for a discussion).  The spectral
turn over is a feature of the brightest XRBs in these galaxies, which include
black-hole binaries in ultraluminous and intermediate (or steep power-law)
accretion states (e.g., Wik \etal\ 2014).  These accretion states have
non-negligible components from both an accretion disk and a power-law component
from Comptonization (see, e.g., Gladstone \etal\ 2009; Remillard \&
McClintock~2006; Done \etal\ 2007; Bachetti \etal\ 2013; Walton \etal\ 2013,
2014; Rana \etal\ 2015).  The Fragos \etal\ (2013b) population-synthesis models
include synthetic spectra that assume both disk and power-law (Comptonization)
components.  For comparison, we show the Fragos \etal\ (2013b) synthetic XRB
spectra for the global population at $z = 0$ and $z = 7$, which brackets the
full range of spectral shapes across this redshift range.  These synthetic
spectra do not differ significantly between each other and are remarkably
consistent with the observed XRB spectrum from our local star-forming galaxy
SED at $E \simgt 1.5$~keV.  At $E \simlt 1.5$~keV, the synthetic spectra vary
somewhat as a function of redshift, primarily due to variations in the assumed
intrinsic absorption.  Fragos \etal\ (2013b) assumed that the intrinsic column
densities of XRBs varied as a function luminosity following the same
distribution as XRBs in the Milky Way.  The true intrinsic obscuration of XRBs
in high-redshift galaxies is highly uncertain.  Nonetheless, the agreement in
SED shapes at $E \simgt 1.5$~keV for the local star-forming galaxy sample and
the synthetic spectra at all redshifts suggests that there is unlikely to be
any significant variations in the XRB SED as a function of redshift.

The intensities of the hot gas and XRB components have both been observed to
broadly scale with SFR, and the hot gas temperature does not appear to vary
significantly with SFR (e.g., Mineo \etal\ 2012a,b).  Given these results, and
the predicted lack of evolution in the synthetic XRB SED from Fragos \etal\
(2013b), we do not expect the shapes of the hot gas and XRB components will
change significantly with SFR.  However, as we reach to higher redshift
galaxies, it is expected that the XRB emission components will become more
luminous per unit SFR due to declining XRB ages and metallicities (e.g., F13a).
These physical changes may have some effect on the hot gas emission per unit
SFR, as the supernova rate and mechanical energy from stellar winds are
affected by metallicity (e.g., C{\^o}t{\'e} \etal\ 2015).  The exact form of
this dependence has yet to be quantified specifically; however, some evidence
suggests that the increase in XRB emission with declining metallicity dwarfs
the hot gas emission (e.g., see the 0.5--30~keV SED of low-metallicity galaxy
NGC~3310 in Fig.~6 of Lehmer \etal\ 2015).  We therefore expect that the ratio
of hot gas to XRB emission will likely decrease with increasing redshift,
leading to a more XRB-dominant SED at high redshift.  Unfortunately, our data
are not of sufficient quality to constrain such an effect directly.

Since our primary goal is to utilize \xray\ stacking of galaxy populations to
study the underlying XRB populations, it is imperative that we perform our
stacking in a bandpass that will be dominated by XRB emission.  Clearly,
high-energy bandpasses that sample rest-frame energies $E \simgt$~2~keV would
accomplish this goal; however, since the \chandra\ effective area curve peaks
at 1--2~keV and declines rapidly at higher energies, the corresponding S/N of
stacked subsamples is prohibitively low for bandpasses above observed-frame $E
\approx$~2~keV.  Fortunately, since the {\it rest-frame} energy range for a
fixed observed-frame bandpass shifts to higher energies with increasing
redshift, we can perform stacking in bandpasses near the peak of \chandra's
response and still measure directly emission dominated by XRB populations.  

%
%
\begin{figure*}
\figurenum{5}
\centerline{
\includegraphics[width=18cm]{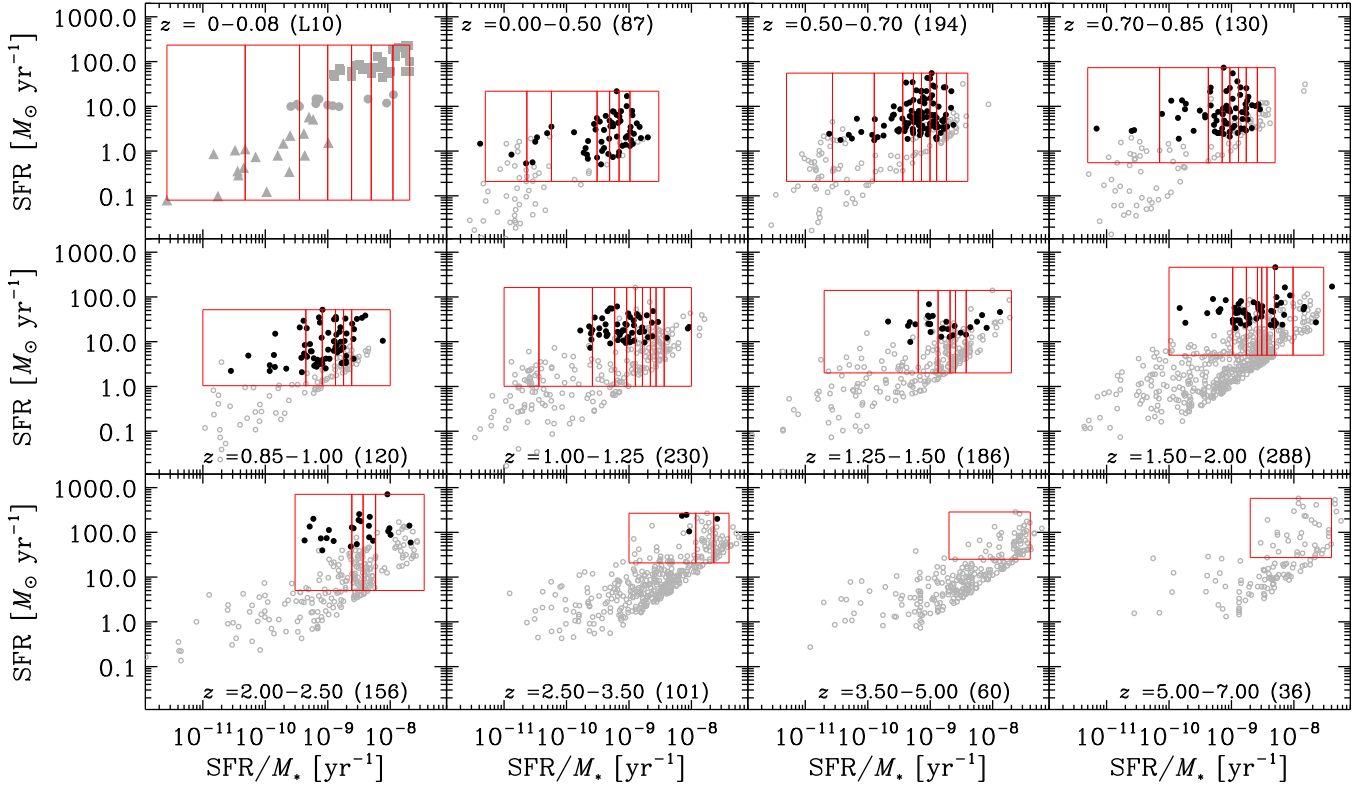}
}
\vspace{0.1in}
\caption{
SFR versus sSFR (SFR/$M_\star$) for the local galaxy sample from Lehmer \etal\
(2010; {\it upper-left panel\/}) with only $M_\star > 10^9$~\msol\
galaxies included and our main sample of \nmcut\ distant normal galaxies in
the CDF-S (known AGN are excluded) divided into 11 redshift intervals.
Galaxies with UV plus IR and UV-only estimates of SFR are indicated with black
and gray symbols, respectively.  For each panel, we outline with red
rectangles, parameter boundaries that were used for defining subsamples, for
which we obtained average \xray\ properties using averaging (L10 sample) and
stacking (CDF-S subsamples).  For reference, the numbers of galaxies that were
used in our stacking analyses are annotated in parentheses for each of the
CDF-S panels.  We note that the distributions of sources in SFR-sSFR space show
a diagonal cut-off in the lower-right regions of each panel.  This is due to
our explicit cut in stellar mass ($M_\star > 10^9$~\msol).
}
\end{figure*}

Figure~4$b$ presents the redshift-dependent fraction of the \chandra\ counts
that are provided by XRBs for observed-frame \hbox{0.5--1~keV},
\hbox{1--2~keV}, and \hbox{2--4~keV} bandpasses, assuming that the SED does not
evolve with redshift.  The \hbox{1--2~keV} and \hbox{2--4~keV} bandpasses probe
majority contributions from galaxies dominated by XRB populations beyond $z
\approx 0.15$, which corresponds to the lowest redshift galaxies in our sample
(see below).  In an attempt to measure directly emission from XRBs, while
preserving high S/N for our stacked subsamples, we perform stacking in the
\hbox{1--2~keV} bandpass and use these stacking results to measure
$k$-corrected rest-frame \hbox{2--10~keV} luminosities.  Figure~4$a$ shows the
rest-frame regions of the SED sampled by observed-frame \hbox{1--2~keV} at $z =
0.3$, $z = 0.6$, and $z = 1.5$, which are, respectively, the lowest, second
lowest, and median redshift intervals for stacked subsamples; we define
stacking subsamples explicitly in $\S$5 below.  As a check on our SED
assumptions and possible AGN contamination (see $\S$6.1 below), we also stacked
subsamples in the \hbox{0.5--1~keV} and \hbox{2--4~keV} bands.  The
\hbox{0.5--1~keV} band stacks sample softer energies that can be used to
estimate rest-frame \hbox{0.5--2~keV} luminosities by applying small
$k$-corrections.  Although our focus is on the \hbox{2--10~keV} emission, we
present, for completeness, \hbox{0.5--2~keV} constraints throughout the rest of
the paper.

For a given stacked galaxy subsample, we estimated mean \xray\ fluxes using the
following equation:
\begin{equation}
f_{E_1-E_2} =  A_{E_1-E_2} \Phi_{E_1-E_2},
\end{equation}
where $A_{E_1-E_2}$ provides the count-rate to flux conversion within the
observed-frame $E_1-E_2$ bandpass, based on our adopted SED (see Fig.~4$a$) and
the \chandra\ response function.  Errors on $f_{E_1-E_2}$ were calculated by
propagating uncertainties on $A_{E_1-E_2}$ from our starburst and normal galaxy
sample and bootstrapping errors on $\Phi_{E_1-E_2}$.  Finally, \xray\
luminosities were computed using \xray\ fluxes and $k$-corrections based on our
adopted SED following:
\begin{equation}
L_{E_1^\prime-E_2^\prime} = k_{E_1-E_2}^{E_1^\prime-E_2^\prime} 4 \pi d_L^2
f_{E_1-E_2}, 
\end{equation}
where $k_{E_1-E_2}^{E_1^\prime-E_2^\prime}$ provides the correction between
observed-frame $E_1-E_2$ band and rest-frame $E_1^\prime-E_2^\prime$ band.
Errors on $L_{E_1^\prime-E_2^\prime}$ include errors on
$k_{E_1-E_2}^{E_1^\prime-E_2^\prime}$ from our starburst and normal galaxy
sample and the errors on $f_{E_1-E_2}$.  Across the redshift range $z =$~0--4,
the $k$-correction values range from 2.1--2.7 (1.6--2.1) for correcting
observed-frame \hbox{1--2~keV} (\hbox{0.5--1~keV}) to rest-frame
\hbox{2--10~keV} (\hbox{0.5--2~keV}).

When measuring \xray\ scaling relations for stacked bins (e.g., $L_{\rm X}$
vs.~SFR), we made use of mean physical-property values $\left< A_{\rm phys}
\right>$ (where $A_{\rm phys}$ could be the SFR or $M_\star$).  Errors on these
mean quantities were calculated following a Monte Carlo approach.  For each
stacked subsample, we computed 10,000 simulated {\itshape perturbed mean
values} of a given physical quantity.  For the $k$th simulation, the perturbed
mean was computed following:
\begin{equation}
\left< A_{{\rm phys},k}^{\rm pert} \right> = \frac{1}{N_{\rm gal}}
\sum_{i=1}^{N_{\rm gal}} A_{{\rm phys},i} + N^{\rm pert}_{i, k} \sigma_{A_{{\rm
phys},i}},
\end{equation}
where $N_{\rm gal}$ is the number of galaxies in the stack and the value of
$N_{i,k}^{\rm pert}$ is a random number drawn from a Gaussian distribution,
centered at zero, with a width of unity. The value of $\sigma_{A_{{\rm
phys},i}}$ is the 1~$\sigma$ error on the physical property value $A_{{\rm
phys},i}$ measured for the $i$th galaxy in the stacked subsample.  The errors
on the SFR and $M_\star$ were presented in $\S$2.  Finally, for a stacked
subsample, the error on the mean was estimated as the standard deviation of the
perturbed mean values:
\begin{equation}
\sigma_{<A_{\rm phys}>} = \frac{1}{N_{\rm sim}} \left[ \sum_{k=0}^{N_{\rm sim}}
\left( \left< A_{{\rm phys},k}^{\rm pert} \right> - \left< A_{\rm phys} \right>
\right)^2 \right]^{1/2},
\end{equation}
where $N_{\rm sim} = 10,000$ is the number of simulations performed for a given
stacked subsample.  When computing ratios of mean quantities (e.g., $\log
L_{\rm X}$/SFR) errors on all quantities were propagated following the methods
outlined in $\S$1.7.3 of Lyons~(1991).

%
%
\begin{figure}
\figurenum{6}
\centerline{
\includegraphics[width=8.9cm]{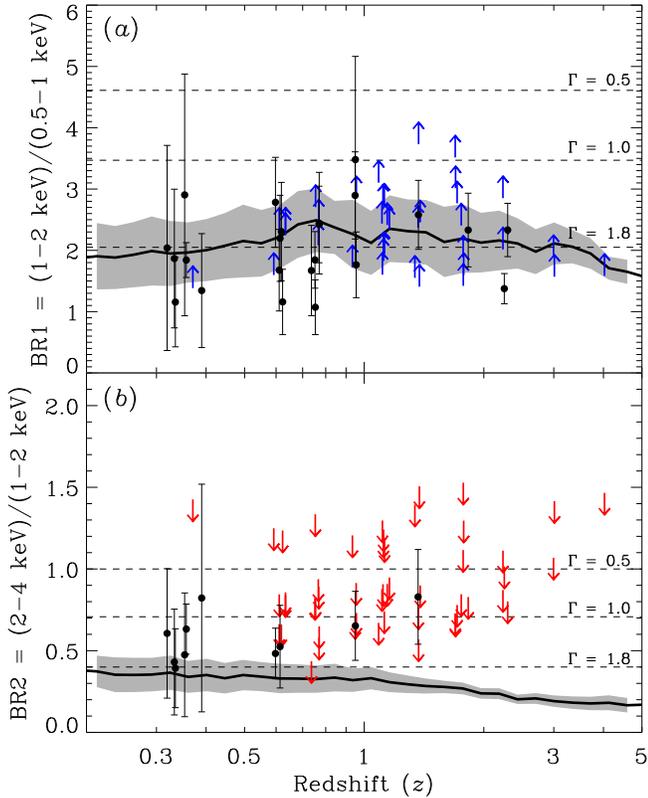}
}
\vspace{0.1in}
\caption{
Stacked average count-rate ratios BR1 ($a$) and BR2 ($b$) versus redshift for
the \nstk\ stacked subsamples ({\it filled circles and limits}).  In each panel,
the expected band-ratio from our canonical SED, presented in Figure~4, is displayed 
as a solid curve with a gray shaded band signifying the range of band-ratios expected
from our four local galaxies (NGC~253, M83, NGC~3256, and NGC~3310;
see $\S$4).  The expected band ratios for various power-law spectra ($\Gamma
=$~0.5, 1.0, and 1.8) are shown as dashed lines.  The majority of the stacked
values of BR1 and BR2 are in good agreement with our canonical SED, with the
exception of a few stacked bins at $z \approx$~1--2 that have BR1 lower limits
and measured BR2 values above the canonical SED prediction. 
}
\end{figure}

%
\section{Stacking Sample Selection and Local Comparison}
%

As discussed in $\S$1, our primary goal is to measure the redshift evolution of
the scaling factors $\alpha \equiv L_{\rm X}$(LMXB)/$M_\star$ and $\beta \equiv
L_{\rm X}$(HMXB)/SFR that were introduced in Equation~(4).  Since the ratio of
HMXB-to-LMXB emission is sensitive to sSFR, our strategy for calculating
$\alpha$ and $\beta$ involves stacking galaxy subsamples that are divided by
sSFR in a variety of redshift bins.  We began by dividing our galaxy sample
into 11 redshift intervals that were chosen to both include large numbers
($\approx$200--1000) of galaxies and span broad dynamic ranges in sSFR.
Figure~5 presents the SFR versus sSFR for each of the 11 redshift intervals
plus the $z \approx 0$ sample studied by L10.  In each panel, we have
highlighted galaxies with SFRs calculated using UV plus far-IR ({\it black
filled circles\/}) and UV-only ({\it gray open circles\/}) data.

For each redshift interval, we divided the galaxy samples into bins of sSFR for
stacking.  Similar to our choice to exclude galaxies with $M_\star <
10^9$~\msol, we further placed lower boundaries on the SFR by which galaxies
are included in stacking so that low SFR galaxies would not dilute the stacked
signals.  As displayed in Figure~5, the low-SFR bounds vary with redshift, such
that SFR$^{\rm bound}$ spans \hbox{$\approx$0.1--1~\sfr} for the majority of
the stacked subsamples ($z =$~0--1.5), and increases to $\approx$10--30~\sfr\
for the highest redshift intervals ($z \approx$~1.5--7).

To first order, obtaining average values of the \xray\ properties of galaxy
populations in bins selected by sSFR alone can be used to provide a good
characterization of the {\it average} $L_{\rm X}$/SFR as a function of sSFR;
however, finer division of the galaxy sample (e.g., by both sSFR and $M_\star$)
would allow investigation of the \xray\ emission of galaxy subsamples with
varying physical properties (e.g., metallicity and stellar age; e.g., Linden
\etal\ 2010; Basu-Zych \etal\ 2013a, 2016; Fragos \etal\ 2013b; Prestwich
\etal\ 2013; Brorby \etal\ 2014, 2016; Douna \etal\ 2015).  However, the goal
of this paper is to establish the average evolution of the scaling relations
(i.e., $\alpha$ and $\beta$ from Eqn.~4) with redshift due to changes in global
properties.  In a future paper, we will investigate how the emission from XRBs
varies with further division of the galaxy samples by physical properties.

In each redshift interval, we divided the full sSFR range into sSFR bins that
contained nearly equal numbers of galaxies per bin.  The number of galaxies per
subsample contained between 12 and 60 sources (median of 23 sources); in total,
we stacked \nstk\ galaxy subsamples.  Figure~5 shows the resulting stacking
bins for the subsamples, and Table~1 summarizes the subsample properties.  From
Figure~5, it is apparent that the sSFR bins cluster around characteristic
central sSFR values and are broader for large and small sSFR values.   This
behavior is due to the densely populated regions along the galaxy ``main
sequence,'' where the SFR and $M_\star$ are correlated (e.g., Elbaz \etal\
2007; Noeske \etal\ 2007; Karim \etal\ 2011; Whitaker \etal\ 2014).  Despite
this effect, we are able to span over 2 orders of magnitude in sSFR for stacked
subsamples in each of the redshift intervals for $z \simlt 2.5$.  Therefore, we
can place good direct constraints on $\alpha$ and $\beta$ in individual
redshift intervals (see below).

\begin{table*}
\begin{center}
\caption{Stacked Subsample Properties}
\begin{tabular}{lccccccccc}
\hline\hline
 & & & & &  SFR$_{\rm lo}$--SFR$_{\rm up}$ & $\left< {\rm SFR} \right>$ & $\log M_\star$ & sSFR$_{\rm lo}$--sSFR$_{\rm up}$ & $\left< {\rm sSFR} \right>$   \\
 \multicolumn{1}{c}{Subsample ID} & $z_{\rm lo}$--$z_{\rm up}$ & $\left< z \right>$ & $N_{\rm gal}$ & $N_{\rm det}$  & (\sfr) & (\sfr) & ($M_\odot$) & (Gyr$^{-1}$) & (Gyr$^{-1}$)  \\
 \multicolumn{1}{c}{(1)}  & (2) & (3) & (4) & (5) & (6) & (7) & (8) & (9) & (10) \\
\hline\hline
  1\ldots\ldots\ldots\dotfill &     0.0--0.5 & 0.39 &       12 &       2 &                0.2--21.7 &               0.8$\pm$0.4 &              10.7$\pm$0.1 &                0.01--0.02 &        0.02$\pm$0.01 \\
  2\ldots\ldots\ldots\dotfill &              & 0.35 &       12 &       4 &                0.2--21.7 &               1.3$\pm$0.5 &              10.5$\pm$0.1 &                0.02--0.06 &        0.04$\pm$0.02 \\
  3\ldots\ldots\ldots\dotfill &              & 0.37 &       12 &       1 &                0.2--21.7 &               1.8$\pm$0.4 &               9.9$\pm$0.1 &                0.06--0.31 &        0.22$\pm$0.07 \\
  4\ldots\ldots\ldots\dotfill &              & 0.32 &       12 &       3 &                0.2--21.7 &               1.9$\pm$0.4 &               9.7$\pm$0.1 &                0.31--0.49 &        0.40$\pm$0.12 \\
\smallskip
  5\ldots\ldots\ldots\dotfill &              & 0.33 &       13 &       3 &                0.2--21.7 &               5.5$\pm$1.1 &              10.0$\pm$0.1 &                0.49--0.69 &        0.58$\pm$0.08 \\
  6\ldots\ldots\ldots\dotfill &              & 0.33 &       12 &       6 &                0.2--21.7 &               4.6$\pm$1.0 &               9.7$\pm$0.1 &                0.69--1.03 &        0.81$\pm$0.22 \\
  7\ldots\ldots\ldots\dotfill &              & 0.36 &       14 &       6 &                0.2--21.7 &               3.0$\pm$0.6 &               9.4$\pm$0.1 &                1.03--3.00 &        1.30$\pm$0.36 \\
  8\ldots\ldots\ldots\dotfill &     0.5--0.7 & 0.60 &       21 &       6 &                0.2--55.1 &               0.9$\pm$0.4 &              10.7$\pm$0.1 &                0.01--0.03 &        0.02$\pm$0.01 \\
  9\ldots\ldots\ldots\dotfill &              & 0.62 &       21 &       5 &                0.2--55.1 &               1.6$\pm$0.6 &              10.6$\pm$0.1 &                0.03--0.13 &        0.05$\pm$0.02 \\
 10\ldots\ldots\ldots\dotfill &              & 0.59 &       23 &       2 &                0.2--55.1 &               3.1$\pm$0.5 &              10.1$\pm$0.1 &                0.13--0.36 &        0.24$\pm$0.07 \\
\hline
\end{tabular}
\end{center}
NOTE.---Col.(1): Unique stacked subsample identification number.  The subsample~ID has been assigned to each subsample and is ordered based on ascending redshift bin, and within each redshift bin, ascending sSFR. Col.(2): Lower and upper redshift boundaries of the subsample. Col.(3): Mean redshift. Col.(4): Number of galaxies stacked in the given bin. Col.(5): Number of sources detected individually in each bin. Col.(6): Lower and upper SFR boundaries for the subsample in units of \sfr.  Col.(7): Mean SFR and 1$\sigma$ error on the mean. Col.(8): Logarithm of the mean stellar mass $M_\star$ in units of $M_\odot$ and 1$\sigma$ error on the mean.  The stellar mass is bounded by limits on SFR and sSFR, as well as a hard lower bound of $M_\star^{\rm lim} = 10^9$~$M_\odot$. Col.(9): Lower and upper sSFR boundaries for the subsample in units of Gyr$^{-1}$.  Col.(10): Mean sSFR and 1$\sigma$ error on the mean.
{\it [Only a portion of Table~1 is printed here.  All \nstk\ entries can be found in the electronic version of the paper]}
\vspace{0.14in}
\end{table*}

\begin{table*}
\begin{center}
\caption{Galaxy Stacking Results}
\begin{tabular}{lcccccccccccc}
\hline\hline
 &  \multicolumn{3}{c}{S/N} & \multicolumn{3}{c}{Net Counts}  & & & & & & \\
 &  \multicolumn{3}{c}{\rule{1.2in}{0.01in}} & \multicolumn{3}{c}{\rule{1.8in}{0.01in}}  & & & $\log L_{\rm 0.5-2~keV}$ & $\log L_{\rm 2-10~keV}$ & $\log L_{\rm 2-10~keV}$/SFR &  $f_{\rm AGN}^{\rm 2-10~keV}$ \\
 \multicolumn{1}{c}{ID} &  0.5--1~keV & 1--2~keV & 2--4~keV &  0.5--1~keV & 1--2~keV & 2--4~keV & BR$_1$ & BR$_2$ & (\lum) & (\lum) & (\lum~(\sfr)$^{-1}$) & (\%) \\
 \multicolumn{1}{c}{(1)}  & (2) & (3) & (4) & (5) & (6) & (7) & (8) & (9) & (10) & (11) & (12) & (13)  \\
\hline\hline
  1 &                  4.5 &                  5.0 &                  3.3 &        55.1$\pm$31.1 &        77.2$\pm$53.3 &        60.8$\pm$51.5 &          1.3$\pm$3.2 &          0.8$\pm$2.8 &         40.2$\pm$0.2 &         40.1$\pm$0.2 &         40.2$\pm$0.3 &          0.00$^{+0.00}_{0.00}$ \\
  2 &                  9.4 &                 17.8 &                  9.1 &       146.3$\pm$71.5 &      437.4$\pm$296.9 &      199.5$\pm$159.3 &          2.9$\pm$6.0 &          0.5$\pm$1.5 &         40.5$\pm$0.2 &         40.7$\pm$0.2 &         40.6$\pm$0.3 &          0.00$^{+0.00}_{0.00}$ \\
  3 &                  2.4 &                  3.2 &                  0.3 &              $<$32.6 &        46.7$\pm$28.3 &              $<$50.4 &               $>$1.4 &               $<$1.1 &              $<$39.9 &         39.8$\pm$0.2 &         39.5$\pm$0.2 &          0.00$^{+0.00}_{0.00}$ \\
  4 &                  6.7 &                 10.1 &                  5.6 &        90.6$\pm$58.0 &      190.3$\pm$156.2 &       111.0$\pm$72.6 &          2.0$\pm$5.9 &          0.6$\pm$3.4 &         40.2$\pm$0.2 &         40.2$\pm$0.3 &         40.0$\pm$0.3 &          0.00$^{+0.00}_{0.00}$ \\
\smallskip
  5 &                  9.6 &                 13.7 &                  5.9 &      153.1$\pm$115.3 &      297.1$\pm$180.1 &       122.4$\pm$92.0 &          1.9$\pm$7.6 &          0.4$\pm$1.1 &         40.4$\pm$0.2 &         40.5$\pm$0.2 &         39.7$\pm$0.2 &          0.00$^{+0.15}_{0.00}$ \\
  6 &                 14.1 &                 14.4 &                  6.0 &      265.6$\pm$179.5 &      315.4$\pm$199.3 &       119.5$\pm$73.4 &          1.2$\pm$3.6 &          0.4$\pm$1.1 &         40.7$\pm$0.2 &         40.5$\pm$0.2 &         39.8$\pm$0.2 &          0.00$^{+0.00}_{0.00}$ \\
  7 &                  5.6 &                  7.9 &                  4.3 &        76.3$\pm$18.4 &       144.6$\pm$22.3 &        88.0$\pm$21.3 &          1.8$\pm$2.4 &          0.6$\pm$0.8 &         40.1$\pm$0.1 &         40.1$\pm$0.1 &         39.7$\pm$0.1 &          0.00$^{+0.00}_{0.00}$ \\
  8 &                  4.7 &                  9.4 &                  4.0 &        72.3$\pm$19.7 &       207.0$\pm$54.6 &        96.3$\pm$30.9 &          2.8$\pm$3.9 &          0.5$\pm$0.7 &         40.5$\pm$0.1 &         40.7$\pm$0.1 &         40.7$\pm$0.2 &          0.00$^{+0.00}_{0.00}$ \\
  9 &                  3.7 &                  6.4 &                  2.7 &        54.4$\pm$16.6 &       128.7$\pm$44.8 &              $<$70.6 &          2.3$\pm$3.4 &               $<$0.6 &         40.4$\pm$0.1 &         40.5$\pm$0.1 &         40.3$\pm$0.2 &          0.00$^{+0.01}_{0.00}$ \\
 10 &                  2.8 &                  3.7 &                  1.8 &              $<$44.2 &        72.8$\pm$29.3 &              $<$72.3 &               $>$1.6 &               $<$1.0 &              $<$40.2 &         40.2$\pm$0.2 &         39.7$\pm$0.2 &          0.01$^{+0.29}_{0.01}$ \\
\hline
\end{tabular}
\end{center}
NOTE.---Col.(1): Subsample identification as defined in Table~1. Col.(2)--(4): Signal-to-noise ratio obtained for each stacked subsample in the 0.5--1~keV, 1--2~keV, and 2--4~keV bands, respectively, following the procedure outlined in $\S$4.  Col.(5)--(7): Background-subtracted net counts and 1$\sigma$ errors obtained for the 0.5--1~keV, 1--2~keV, and 2--4~keV bandpasses, respectively. Col.(8) and (9): Band ratios and 1$\sigma$ errors for BR1~$\equiv$~(1--2~keV)/(0.5--1~keV) and BR2~$\equiv$~(2--4~keV)/(1--2~keV), respectively. Col.(10) and (11): Logarithm of the mean stacked 0.5--2~keV and 2--10~keV luminosities, respectively, in units of erg~s$^{-1}$ with 1$\sigma$ errors on the mean values. Col.(12) Logarithm of the mean 2--10~keV luminosity per mean SFR ratio in units of erg~s$^{-1}$~(\sfr)$^{-1}$ and 1$\sigma$ error. Col.(13) Estimate of the fraction of 2--10~keV emission, in percentage terms, due to undetected AGN; error bars are 1$\sigma$. These estimates are computed following the methodology outlined in $\S$6.1.
{\it [Only a portion of Table~2 is printed here.  All \nstk\ entries can be found in the electronic version of the paper]}
\vspace{0.14in}
\end{table*}

For the purpose of comparing our CDF-S stacking results with constraints from
local galaxies, we calculated average \xray\ luminosities for the local galaxy
sample culled by L10 using sSFR bins selected following the same basic binning
strategy defined above.  The L10 local galaxies sample contains data for 66
nearby ($D <$~400~Mpc; median $D \approx 60$~Mpc) galaxies that were all
observed by \chandra, and have SFR and $M_\star$ values calculated following
the same procedures outlined in $\S$2 (see footnote~23).  The full sample of 66
nearby galaxies includes subsamples of normal galaxies from Colbert \etal\
(2004), as well as LIRGs and ULIRGs from L10 and Iwasawa \etal\ (2009), and
spans a broad range of SFR (SFR~=~\hbox{0.08--200~\sfr}) and stellar mass
($M_\star = 10^{8-11}$~\msol).

The upper left-hand corner of Figure~5 presents the SFR versus sSFR
distribution of galaxies in the L10 sample.  A series of seven sSFR ranges were
chosen, in which we calculate average properties to be compared with the CDF-S
stacking results.  As with the CDF-S data, these sSFR ranges were defined to
contain nearly the same number of sources per bin; however, mean \xray\
luminosities for each bin were calculated by averaging the luminosities
obtained for each galaxy.  This operation should yield equivalent results to
those obtained by our CDF-S stacking procedure, and throughout the rest of this
paper we utilize these mean values, in conjunction with our \xray\ stacking
results, as constraints on the $z = 0$ population properties.  We emphasize,
however, that the L10 compilation includes only 2--10~keV galaxy-wide
luminosities, which probe directly the XRB emission.  In order to compare our
CDF-S stacked \hbox{0.5--2~keV} luminosities, we estimated comparable mean
0.5--2~keV XRB luminosities for each L10 galaxy subsample by applying a
bandpass correction to the 2--10~keV XRB emission and adding an estimate of the
\hbox{0.5--2~keV} hot gas emission.  The XRB bandpass correction was calculated
assuming a power-law spectrum with intrinsic $N_{\rm H} = 3 \times
10^{21}$~cm$^{-2}$ and $\Gamma = 2.0$ (see $\S$3.1 of Mineo \etal\ 2012a for
justification), and the \hbox{0.5--2~keV} hot gas emission was estimated for
each sample using the Mineo \etal\ (2012b) relation: $L_{\rm 0.5-2~keV}^{\rm
hot\; gas}$/(\lum)~$\approx 1.5 \times 10^{39}$~SFR/(\sfr).

%
\section{Results}
%

\subsection{Spectral Properties and AGN Contamination}

Applying the stacking procedure described in $\S$4 to the CDF-S galaxy
subsamples defined in $\S$5, we obtained mean \xray\ properties for the
subsamples, which are summarized in Table~2.  Out of the \nstk\ subsamples, we
obtained \nsbs, \nsbh, and \nhbs\ significant ($>$3$\sigma$) stacked detections
in the \hbox{0.5--1~keV}, \hbox{1--2~keV}, and \hbox{2--4~keV} bands,
respectively; \nundet\ subsamples were not detected in any of the three bands.
Figure~6 displays the (\hbox{1--2~keV})/(\hbox{0.5--1~keV}) and
(\hbox{2--4~keV})/(\hbox{1--2~keV}) mean count-rate ratios (hereafter, BR1 and
BR2, respectively) versus redshift for each of the stacked subsamples and show
the expected band-ratios for our adopted SED.  For the majority of subsamples,
BR1 and BR2 appear to be in agreement with our adopted SED for almost all
stacked subsamples, with a few exceptions where 3$\sigma$ lower limits on BR1
lie above the SED error envelope (at $z \approx 2$) and BR2 measurements are
somewhat elevated (at the $\approx$1$\sigma$ level) at \hbox{$z \approx$~1--2}.
This broad agreement provides some confidence that our stacked subsamples are
not strongly contaminated by obscured AGN and low-luminosity AGN below the
detection limit.  

To estimate quantitatively the level by which individually undetected AGN
contribute to the stacked subsamples, we employed an approach in which (1) the
supermassive black hole accretion distribution (in terms of the Eddington
accretion rate) is estimated, using the known AGN population, to very low
Eddington fractions, and (2) this distribution is used to estimate the expected
total contributions to stacked subsamples from AGN with luminosities that fall
below the individual source detection threshold.

%
%
\begin{figure}
\figurenum{7}
\centerline{
\includegraphics[width=8.7cm]{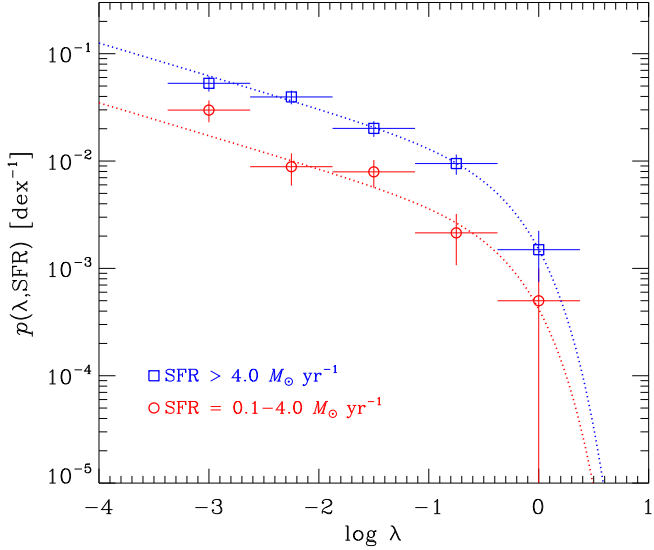}
}
\vspace{0.1in}
\caption{
Probability of a galaxy hosting an AGN with an Eddington fraction, $\lambda
\equiv L/L_{\rm Edd}$, for two different SFR regimes: SFR = 0.1--4~\sfr\ ({\it
red circles\/}) and SFR~$>$~4~\sfr\ ({\it blue squares\/}).  The curves
represent the best-fit parameterization of the $\lambda$ and SFR dependent
probability $p(\lambda, {\rm SFR})$, provided in Equation~(11).
}
\end{figure}

First, for each of the \nmcut\ galaxies in our main sample, we estimated the
central black-hole mass using the following relation: $\log M_{\rm BH} \approx
8.95 + 1.40 \log (M_\star/M_\odot)$ (Reines \& Volonteri~2015), where $M_\star$
is the stellar mass of the galaxy.  For AGN with HB detections, we calculated
rest-frame \hbox{2--10~keV} luminosities in terms of Eddington fraction,
$\lambda \equiv L_{\rm HB}/L_{\rm HB, Edd}$, where $L_{\rm HB, Edd} = (1.26
\times 10^{38}$~\lum)/$C_{\rm bol}$ ($M_{\rm BH}/M_\odot$), and $C_{\rm bol}$
is the luminosity-dependent bolometric correction as presented in Equation~(2)
of Hopkins \etal\ (2007).  For each galaxy in the main galaxy sample, we
derived an Eddington fraction limit, $\lambda_{\rm lim} = L_{\rm HB,
lim}/L_{\rm HB, Edd}$, below which we would not be able to detect an AGN if it
were present.  For a given source $L_{\rm HB, lim} = 4 \pi d_L^2 f_{\rm HB,
lim}$, where $f_{\rm HB, lim}$ is the HB flux limit, which we extracted from
the spatially-dependent HB sensitivity map constructed by Luo \etal\ (2016,
in-preparation).  

Using the above information, Eddington accretion fraction probability
distributions, $p(\lambda)$, were calculated by extracting the number of AGN
within a bin of $\lambda$ divided by the number of galaxies by which we could
have detected such an AGN (using $\lambda_{\rm lim}$).  Past studies (e.g.,
Rafferty \etal\ 2009; Aird \etal\ 2012; Mullaney \etal\ 2012; Wang \etal\ 2013;
Hickox \etal\ 2014) have suggested that such a distribution is likely to be SFR
dependent (and close to linear) due to the overall correlation between $M_{\rm
BH}$ and $M_\star$.  We therefore derived $p(\lambda)$ for two SFR regimes: SFR
= 0.1--4~\sfr and SFR $>$~4~\sfr\ to infer the SFR dependence.  Figure~7 shows
our estimates of $p(\lambda)$ for the two different SFR regimes.  We find the
distributions to have similar $\lambda$ dependencies with normalization
increasing with SFR.  We found that the following parameterization of
$p(\lambda, {\rm SFR})$ fits well the observed $\lambda$ and SFR dependencies:
\begin{equation}
p(\lambda, {\rm SFR}) = \xi (\lambda/\lambda_{\rm c})^{-\gamma_E}
\exp(-\lambda/\lambda_{\rm c}) ({\rm SFR}/1 M_\odot~{\rm yr^{-1}})^{\gamma_s},
\end{equation}
where $\xi = 0.002$~dex$^{-1}$, $\lambda_{\rm c} = 0.21$, $\gamma_E = 0.31$,
and $\gamma_s = 0.57$ are fitting constants.  This functional form was
motivated by previous estimates of similar distributions presented in the
literature (e.g., Aird \etal\ 2012; Hickox \etal\ 2014).  The curves in
Figure~7 show the predicted curves from Equation~(11) fixed to the median SFR
of each of the two SFR-regimes. 

To estimate \xray\ undetected AGN contributions for a given stacked subsample,
we used a Monte Carlo approach.  For each galaxy in a stacked subsample that
was not detected in the \xray\ band, we first drew a value of $\lambda_i$
probabilistically following the distribution in Equation~(11).  We then
converted $\lambda_i$ to a HB luminosity following $L_{{\rm HB}, i}({\rm AGN})
= \lambda_i L_{{\rm HB, Edd},i}$, where $L_{{\rm HB, Edd},i}$ is uniquely
defined for a galaxy by its black-hole mass (see above).  Occasionally a random
draw will predict a value of $L_{{\rm HB}, i}$(AGN) above the detection limit.
In such cases a new estimate of $L_{{\rm HB}, i}$(AGN) is chosen until it falls
below the detection limit.  This procedure allows us to estimate the total
contribution that AGN below the detection limit provide to the stacked emission
$f_{\rm AGN}^{E_1^\prime-E_2^\prime} = \sum_i L_{{\rm HB}, i}$(AGN)
$k^{E_1^\prime-E_2^\prime}_{\rm HB}$(AGN)/$L_{\rm E_1^\prime-E_2^\prime}$,
where $L_{\rm E_1^\prime-E_2^\prime}$ is the total stacked luminosity (see
Eqn.~8) and $k^{E_1^\prime-E_2^\prime}_{\rm HB}$(AGN) provides a bandpass
correction from the HB to the $E_1^\prime-E_2^\prime$ band for the AGN
contribution.  

To account for random errors, we computed $f_{\rm AGN}^{E_1^\prime-E_2^\prime}$
for a given stacked subsample using 1000 Monte Carlo trials, allowing us to
estimate its most probable value and 1$\sigma$ range; values of $f_{\rm
AGN}^{\rm 2-10~keV}$ are provided in Column~(13) of Table~2.  All values of
$f_{\rm AGN}^{\rm 2-10~keV}$ are less than 0.1 (i.e., 10\% contribution from
AGN), with a median value of 0.007.  These values of $f_{\rm AGN}^{\rm
2-10~keV}$ are much smaller than the errors on the stacked luminosities, and we
therefore conclude that AGN below the detection threshold do not significantly
impact our results.  

%
%
\begin{figure*}
\figurenum{8}
\centerline{
\includegraphics[width=8.9cm]{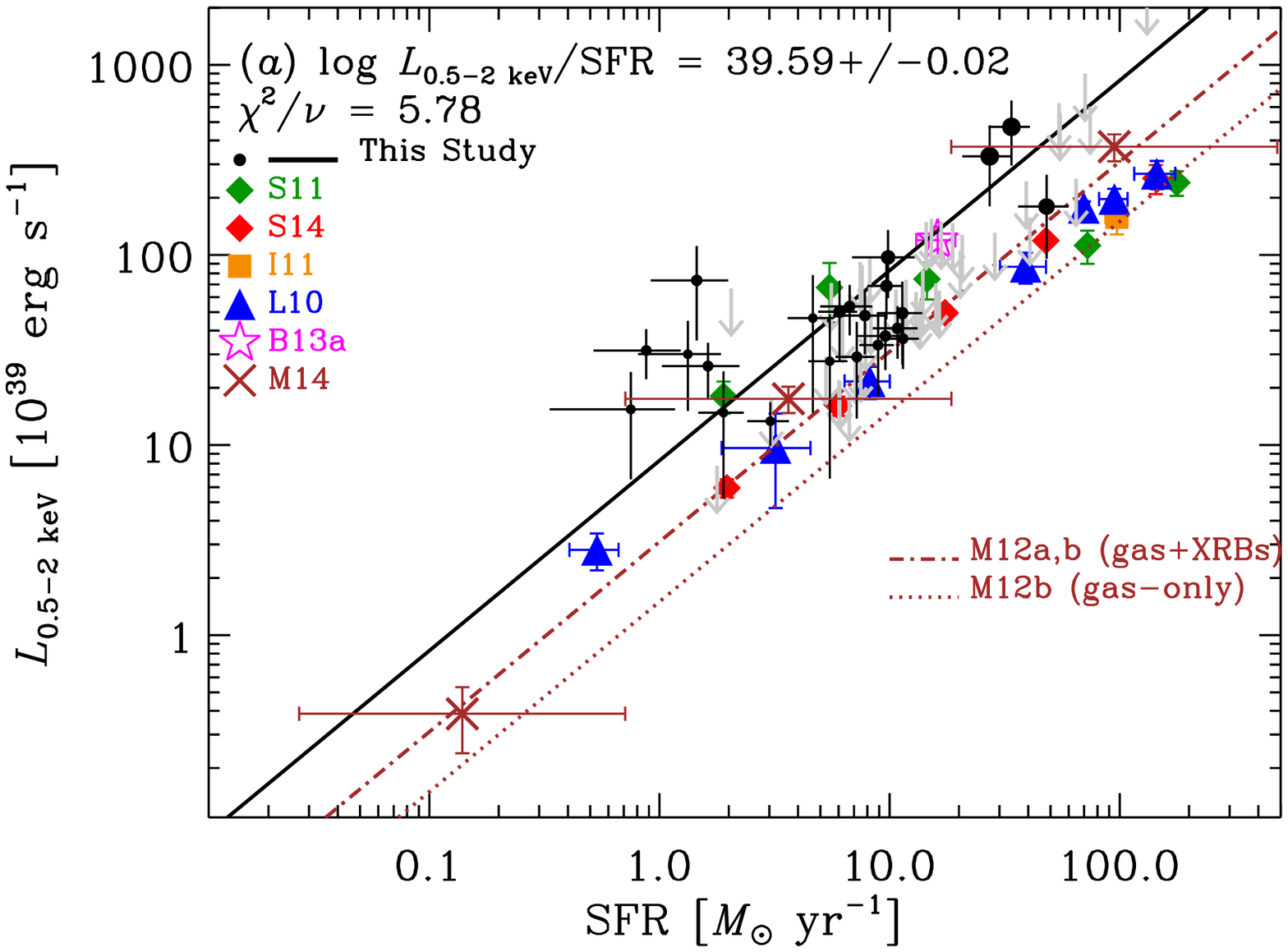}
\hfill
\includegraphics[width=8.9cm]{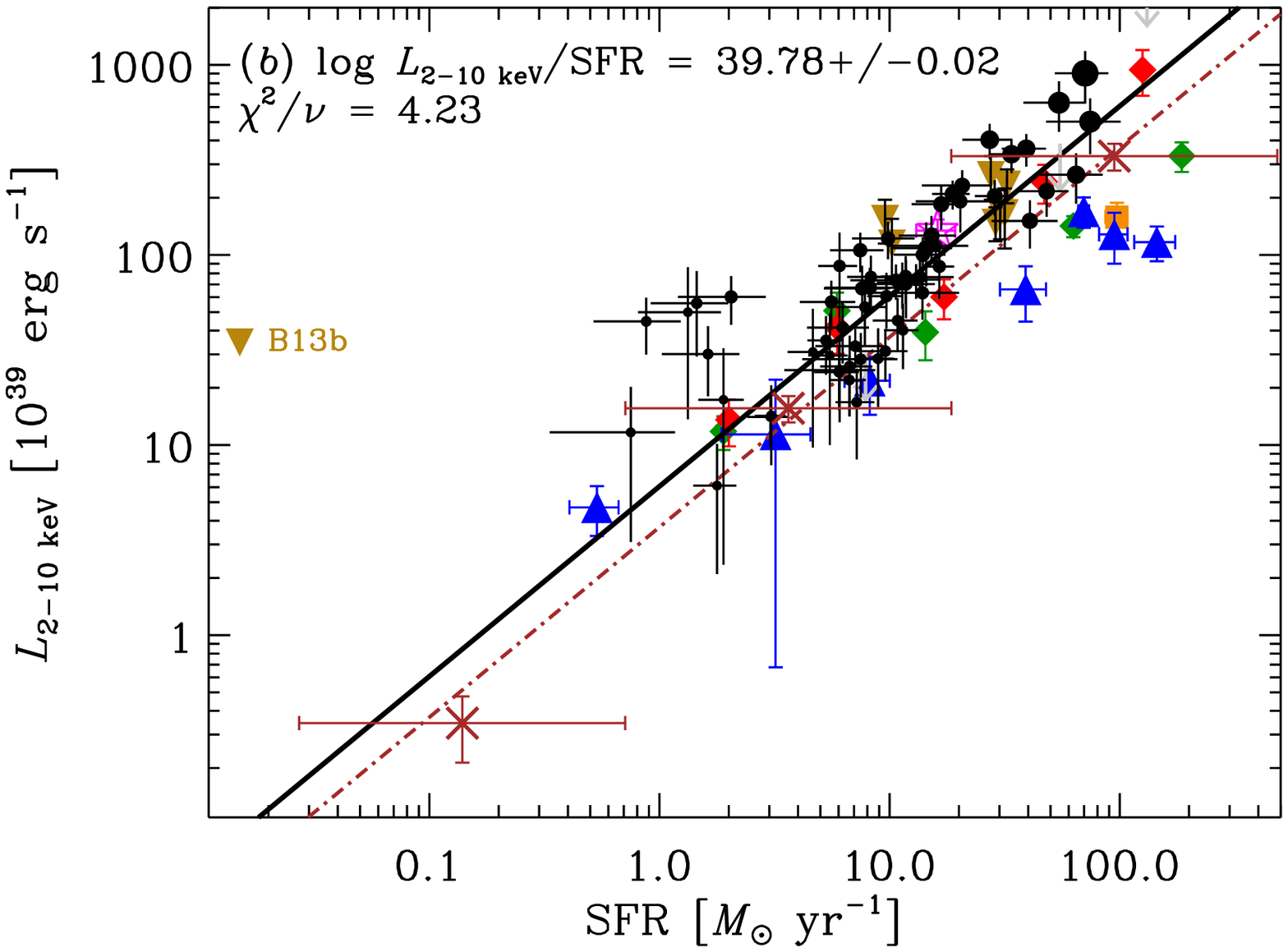}
}
\caption{
0.5--2~keV and 2--10~keV luminosity versus SFR for the local L10 sample ({\it
blue triangles\/}) and CDF-S stacked galaxy subsamples ({\it filled circles
with error bars\/} and {\it gray upper limits} [1$\sigma$]).  CDF-S stacking
symbol sizes scale with the mean redshift of the stack.  Our best-fit linear
models to the local L10 plus CDF-S stacked data following Equation~(12) are
shown as black lines.  Each data point represents the mean luminosities with
1$\sigma$ errors on the mean values.  Local galaxy comparison samples are shown
including IR-selected samples of star-forming galaxies over a broad range of
$L_{\rm IR}$ (Symeonidis \etal\ 2011; S11; {\it green diamonds\/}),
LIRGs/ULIRGs (Iwasawa \etal\ 2011; I11; {\it orange squares\/}), and
UV-selected Lyman break analogs (LBAs; Basu-Zych \etal\ 2013a; B13a; {\it
magenta stars\/}).  The local scaling relations from Mineo \etal\ (2012a,b;
M12a,b), which are based on high-sSFR galaxies, are indicated as dot-dashed
brown curves ({\it dotted curves\/}; based on M12b) for \xray\ emission due to
XRBs plus hot gas (hot gas only).  Additionally, distant galaxy samples are
displayed, including $z \approx$~1.5--4 LBGs in the CDF-S (Basu-Zych \etal\
2013b; B13b; {\it gold upside-down triangles\/}; 2--10~keV only), $z \simlt
1.3$ \xray\ and radio-detected galaxies in the CDF-N and CDF-S (Mineo \etal\
2014; M14; {\it brown crosses\/}), and $z \simlt 1.5$ \herschel\ selected
star-forming galaxies (Symeonidis \etal\ 2014; S14; {\it red diamonds\/}).  The
combined L10 ({\it blue triangles\/}) and stacked CDF-S subsamples ({\it black
points\/}) have best-fit linear scaling relations ($L_{\rm X} \propto$~SFR)
with $\chi^2/\nu = 4.09$ and 4.34 for the 0.5--2~keV and 2--10~keV bands,
repectively, thus indicating that redshift-independent linear scaling relations
do not adequately fit all data.
}
\end{figure*}

\subsection{The X-ray/SFR Correlation}

Although our primary goal is to measure $\alpha$ and $\beta$ as a function of
redshift, it is worth exploring first how basic empirical scaling relations
that have been studied and widely used for local and distant star-forming
galaxies (e.g., $L_{\rm X} \propto$~SFR) compare with those inferred from our
stacked subsamples and L10 local comparison.

Since our galaxy subsample selections were based primarily on intervals of sSFR
with strict lower limits on $M_\star$ and SFR, the mean SFRs for our galaxy
subsamples span a modest range of SFR ($\approx$2~dex; see Fig.~5), which
allows direct measurement of the $L_{\rm X}$/SFR correlations for our data.
Figure~8 presents $L_{\rm 0.5-2~keV}$ and $L_{\rm 2-10~keV}$ versus SFR for the
stacked galaxy subsamples ({\it filled black circles\/}) and local L10
comparison subsamples ({\it filled blue triangles\/}), and provides results
from additional previous studies for comparison (see discussion below).
Consistent with previous investigations, we find strong correlations between
$L_{\rm 0.5-2~keV}$ and $L_{\rm 2-10~keV}$ with SFR (e.g., Bauer \etal\ 2002;
Grimm \etal\ 2002; Ranalli \etal\ 2003, 2012; Persic \etal\ 2004; Persic \&
Rephaeli~2007; Lehmer \etal\ 2008, 2010; Iwasawa \etal\ 2009; Mineo \etal\
2012a,b, 2014; Symeonidis \etal\ 2011, 2014; Vattakunnel \etal\ 2012).  We
performed linear fits to the CDF-S stacked data and local L10 comparison sample
to derive best-fit values of the following linear model:
\begin{equation}
\log L_{\rm X}  = A_1 + \log {\rm SFR}, 
\end{equation}
where $L_{\rm X}$ is in units of \lum\ and SFR is in units of \sfr.  The
best-fit parameters for the 0.5--2~keV and 2--10~keV bands are listed in
Table~3 and plotted as solid lines in Figure~8.  Despite the strong $L_{\rm
X}$/SFR correlation for our data, the linear fits to all stacked bins do not
yield statistically acceptable fits; $\chi^2/\nu$~=~5.78 and 4.23 for the
0.5--2~keV and 2--10~keV bands, respectively, with resulting residual scatter
of 0.37 and 0.32~dex.  To test whether the poor fits were due to some
non-linearity in the $L_{\rm X}$--SFR relation, we further performed non-linear
fits to our data following the form:
\begin{equation}
\log L_{\rm X}  = A_2 + B_2 \log {\rm SFR}.
\end{equation}
The best-fit parameter $B_2$ has some small variation from unity for both the
0.5--2~keV and 2--10~keV bandpasses; however, the fit does not yield a
statistically robust characterization of the data ($\chi^2/\nu = 2.75$ and
3.38, respectively).

In addition to our local L10 sample, Figure~8 also displays the results derived
from multiple local and distant-galaxy samples.\footnote{For several of the
comparison relations, we have had to make conversions from IMFs and bandpasses
to be consistent with those adopted in our study.  For the Mineo \etal\
samples, we translated SFR values from their adopted Salpeter IMF to our Kroupa
IMF, and have converted their \hbox{0.5--8~keV} bandpass (for XRBs only) to
\hbox{0.5--2~keV} and \hbox{2--10~keV} bandpasses using their average XRB
observed SED (i.e., a power-law with intrinsic $N_{\rm H} = 3 \times
10^{21}$~cm$^{-2}$ and $\Gamma = 2.0$).  For the Symeonidis \etal\ and Iwasawa
\etal\ values, we converted $L_{\rm IR}$ directly to SFR following Equation~(2)
of this paper, assuming $L_{\rm IR} \gg L_{\rm UV}$.}  For local comparisons,
we have chosen results from IR-selected galaxies, which span normal
star-forming galaxies ($L_{\rm IR} \approx$~$10^9$--$10^{11}$~$L_\odot$) to
luminous/ultraluminous IR galaxies (LIRGs/ULIRGs; $L_{\rm IR}
\approx$~$10^{11}$--$10^{12}$~$L_\odot$; Iwasawa \etal\ 2011; Symeonidis \etal\
2011), normal star-forming galaxy samples selected to have sSFRs~$\simgt
10^{-10}$~yr$^{-1}$ (Mineo \etal\ 2012a,b), and UV-selected Lyman break analogs
(LBAs; Basu-Zych \etal\ 2013a), which are rare galaxies that have properties
similar to $z \sim$~2--3 Lyman break galaxies (LBGs, e.g., relatively low
metallicity and compact UV morphologies).  For distant-galaxy comparisons, we
have included results from Basu-Zych \etal\ (2013b), Mineo \etal\ (2014), and
Symeonidis \etal\ (2014), which presented results from \xray\ stacking of $z
\approx$1.5--4 LBGs in the CDF-S, correlating \xray\ and radio detected CDF-N
and CDF-S sources at $z \simlt 1.3$, and \xray\ stacking of \herschel\ selected
star-forming galaxies in the CDF-N and CDF-S at $z \simgt 1.5$, respectively.  

There is basic agreement in the trends found between these comparison samples
and our data.  Our data do have higher $L_{\rm X}$/SFR values on average, but
the scatter encompasses the majority of the comparison samples.  We suspect
that the heterogeneity in the comparison sample selections introduces
significant scatter in even the local $L_{\rm X}$--SFR relations.  For the
local galaxies, the scatter may be explained due to sample differences in (1)
the relative contributions of LMXBs and HMXBs (e.g., Mineo \etal\ 2012a,b use
only high-sSFR galaxies); (2) metallicity, which results in varying levels in
HMXB formation per unit SFR (e.g., Basu-Zych \etal\ 2013a study low-metallicity
galaxies); and (3) \xray\ absorption due to galaxy selection (e.g., Iwasawa
\etal\ 2011 and Symeonidis \etal\ 2011 study IR selected galaxies that may be
influenced by absorption; see also Luangtip \etal\ 2015).  

As the typical sSFR, stellar age, and metallicity of the galaxies in the
Universe evolve with redshift, it is expected that $L_{\rm X}$/SFR will change
as the HMXB and LMXB populations respond accordingly (F13a), so
redshift-related scatter is expected to be introduced by combining constraints
from various redshifts.  This behavior is evident in the scatter for the
distant-galaxy comparison samples (i.e., the Basu-Zych \etal\ 2013b, Mineo
\etal\ 2014, and Symeonidis \etal\ 2014 samples).  In fact, for
SFR~$\simgt$~10~\sfr, it is apparent from Figure~8 that the relatively large
scatter in $L_{\rm X}$/SFR for our stacked CDF-S data arises primarily due to a
redshift effect (symbol sizes are proportional to redshift).  For a given SFR,
$L_{\rm X}$/SFR seems to increase with redshift, an indication that the XRB
emission per unit SFR is indeed increasing with redshift.  For
SFR~$\simlt$~10~\sfr, the large scatter in $L_{\rm X}$/SFR is likely to be due
to varying contributions from LMXBs (see below).

\subsection{Redshift-Dependent Evolution of XRB Scaling Relations}

As discussed above, direct redshift independent scaling relations of \xray\
luminosity with SFR are not statistically robust across all galaxy subsamples
at all redshifts.  From Figure~8, there are qualitative indications for
significant redshift evolution in the scaling relations for
SFR~$\simgt$~10~\sfr.  In this section, we investigate how redshift and galaxy
physical properties influence $L_{\rm X}$.  Hereafter, we focus our discussion
on results from the rest-frame 2--10~keV band (probed by observed-frame
1--2~keV), since this bandpass probes directly XRB emission and provides the
best constraints on the redshift evolution of \xray\ emission due to the
relatively large number of significant stacked detections: (\nsbh\ in the
observed-frame 1--2~keV band versus \nsbs\ in the observed-frame 0.5--1~keV
band).  For completeness, however, we provide equivalent measurements and
results for the rest-frame \hbox{0.5--2~keV} emission throughout the rest of
this paper.

%
%
\begin{figure}
\figurenum{9}
\centerline{
\includegraphics[width=8.9cm]{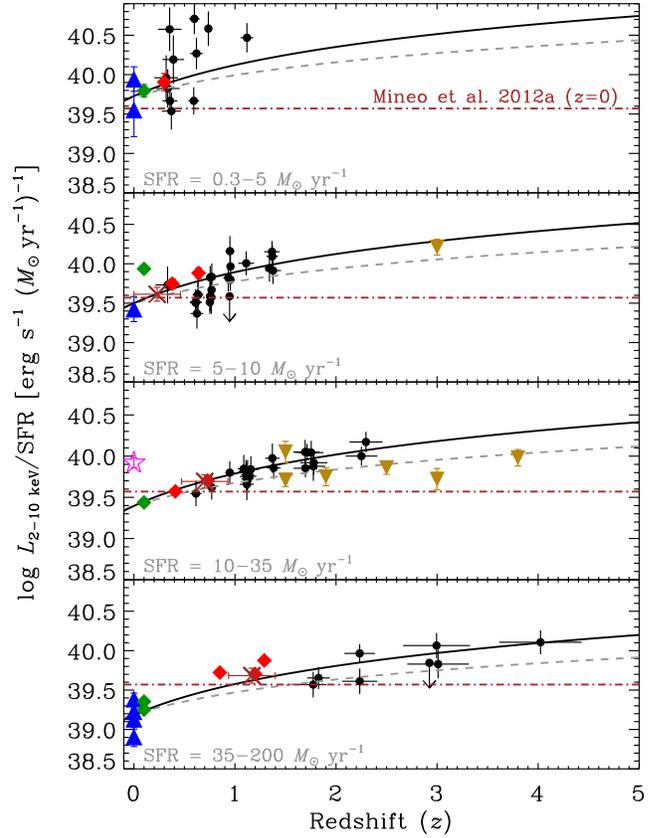}
}
\vspace{0.1in}
\caption{
Stacked 2--10~keV luminosity per unit SFR ($L_{\rm 2-10~keV}$/SFR) versus
redshift for sSFR-selected galaxy subsamples divided into four SFR intervals.
Symbols have the same meaning as in Figure~8, except that the sizes of each
symbol are constant.  In each panel, the solid curve indicates our best-fit
redshift and SFR dependent model, $\log L_{\rm 2-10~keV} = A_3 + B_3 \log {\rm
SFR} + C_3 \log (1+z)$, which was fit to the combined L10 and CDF-S stacked
data.  For comparison the Mineo \etal\ (2012a) relation for high-sSFR galaxies
has been shown as a brown dashed-dot curve.  The curves are plotted using the
median SFR of the L10 and CDF-S data points in that panel.  The dashed curves
show the equivalent relation, derived by Basu-Zych \etal\ (2013b), using the
same functional form but with different stacked data.  The redshift-dependence
of $L_{\rm 2-10~keV}$/SFR is obvious in each panel and the inclusion of
redshift as a model parameter provides a substantive statistical improvement
over a constant $L_{\rm 2-10~keV}$/SFR ratio model.
}
\end{figure}

\subsubsection{SFR and Redshift Dependence}

Figure~9 displays the $L_{\rm 2-10~keV}$/SFR ratio of our stacked subsamples
versus redshift in four SFR intervals; we include the several comparison
samples presented in Figure~8$b$.  This representation reveals that much of the
scatter in the $L_{\rm 2-10~keV}$/SFR relation and apparent discrepancies
between other studies can be reconciled by positive redshift and negative SFR
dependence on the $L_{\rm 2-10~keV}$/SFR ratio.  Some previous investigations
of \xray\ emission from distant galaxy samples (e.g., Lehmer \etal\ 2008;
Ranalli \etal\ 2012; Vattakunnel \etal\ 2012; Symeonidis \etal\ 2014) concluded
that $L_{\rm X}$/SFR at $z \approx 1$ is consistent with local scaling
relations, and that the relation does not evolve with redshift.  Such a
conclusion is dependent on both (1) the choice of the local comparison sample,
which varies for different galaxy samples (see Fig.~8), and (2) the redshift of
the sample.  For example, comparison of the $z \approx 1$ data from Symeonidis
\etal\ (2014) with the Mineo \etal\ (2012a) relation at $z=0$ ({\it brown
dot-dashed lines} in Fig.~9) indicates that the two samples have similar
$L_{\rm 2-10~keV}$/SFR.  However, the Symeonidis \etal\ (2014) values are
somewhat larger than the $z = 0$ values from L10 and Iwasawa \etal\ (2011),
which would imply that there is positive evolution of the $L_{\rm
2-10~keV}$/SFR relation with redshift.  In this study, we find that at $z
\simgt 2$, $L_{\rm 2-10~keV}$/SFR is higher than both the L10 and Mineo \etal\
(2012a) relations, indicating that the redshift evolution is robust.

Using stacked samples of $z \approx$~1--4 Lyman break galaxies (LBGs) in the
$\approx$4~Ms CDF-S survey, stacking results from star-forming galaxies in the
$\approx$2~Ms CDF-S/CDF-N surveys from Lehmer \etal\ (2008), and the local L10
galaxy sample, Basu-Zych \etal\ (2013b) found that the mean \xray\ luminosity
of star-forming galaxy samples could be characterized well using the following
relation:
\begin{equation}
\log L_{\rm X} = A_3 + B_3 \log {\rm SFR} + C_3 \log (1+z).
\end{equation}
Figure~9 indicates the Basu-Zych \etal\ (2013b) best-fit relation for
comparison ({\it dashed curves\/}), and the Mineo \etal\ (2012a) best-fit local
relation.  Using the stacking results from this study and the compiled local
galaxy samples from L10, we derived fitting constants $A_3$~=~39.82~$\pm$~0.05,
$B_3$~=~0.63~$\pm$~0.04, and $C_3$~=~1.31~$\pm$~0.11 ({\it solid curves} in
Fig.~9).  These values predict somewhat more rapid redshift evolution than
those from Basu-Zych \etal\ (2013b; $A_3 = 39.8$, $B_3 = 0.65$, and $C_3 =
0.89$), with significant divergence between fits at $z \simgt$~2.5--3.
Differences in the best-fit function are largely driven by 2--3 $z \simgt 2$
stacked samples with SFR~$\approx$~10--35~\sfr\ from Basu-Zych \etal\ (2013b;
{\it inverted triangles} in Fig.~9), and our stacked subsamples in the
low-redshift ($z \simlt 2$), low-SFR (SFR~$\approx$~\hbox{0.3--10}~\sfr)
regime, which contain significant scatter.  Our model fit to the data produced
a best-fit $\chi^2/\nu = 1.32$, which is a substantial improvement over a
single $L_{\rm 2-10~keV}$--SFR scaling relation that does not include redshift
evolution ($\chi^2/\nu \approx 4.23$), but is only marginally acceptable.  For
$\nu = 64$ degrees of freedom, there is a 4.4\% probability of obtaining
$\chi^2/\nu \ge 1.32$.

%
%
\begin{figure*}
\figurenum{10}
\centerline{
\includegraphics[width=18cm]{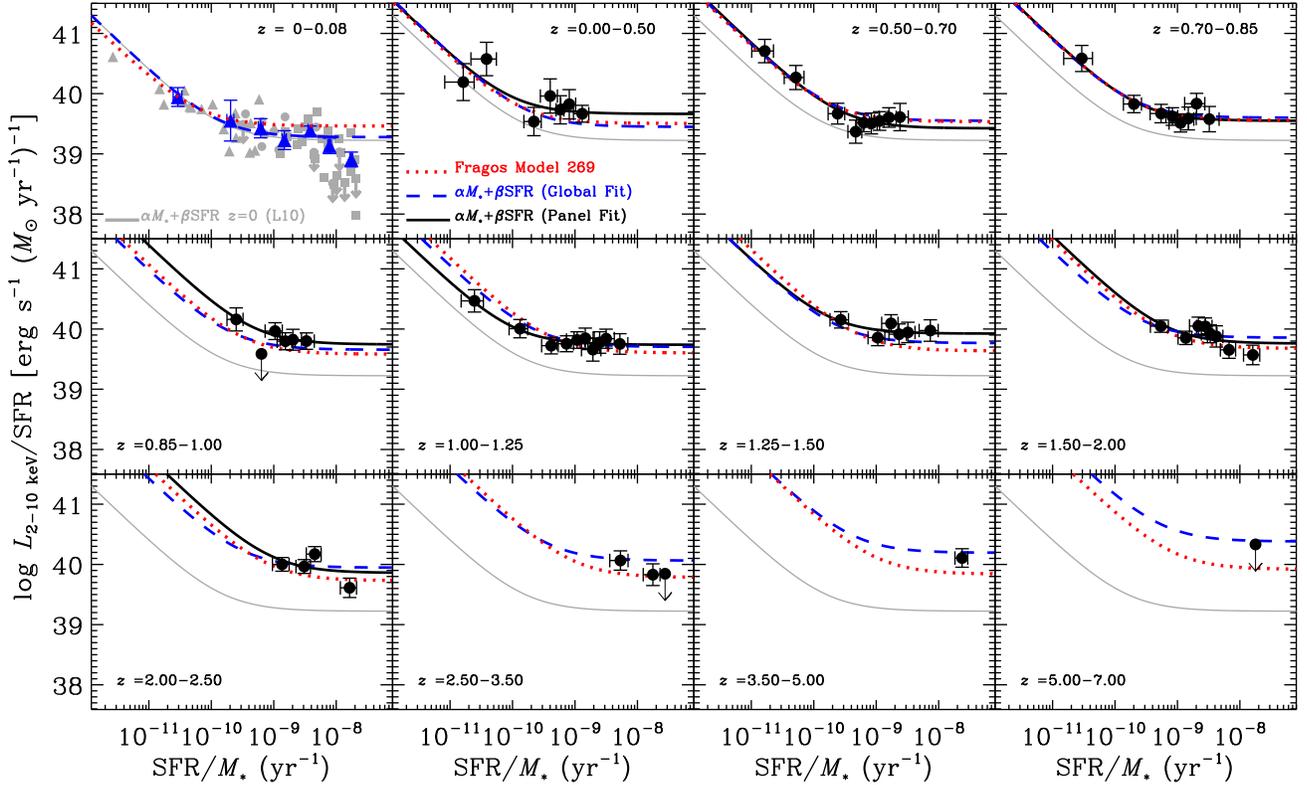}
}
\vspace{0.1in}
\caption{
Mean 2--10~keV luminosity per SFR ($L_{\rm 2-10~keV}$/SFR) versus sSFR for
local galaxies ({\it upper left panel\/}) and distant normal galaxies in the
CDF-S.  Each data point represents mean quantities.  Our best-fit model to the
function provided by Equation~(4) for each CDF-S panel is shown as a solid
curve (for $z \simlt 2.5$ panels) and our best-fit global model from
Equation~(15) is the blue dashed curve in each panel.  The red dotted curves
are those predicted by Model~269, the best fit XRB population-synthesis model
from F13a.  For comparison, we have plotted the local Universe parameterization
from L10 in each of the panels ({\it gray curves\/}).
}
\end{figure*}

Despite the statistical limitations, Equation~(14) provides a first-order
approach for estimating galaxy-wide $L_{\rm 2-10~keV}$, given values of $z$ and
SFR; the resulting best-fit relation has a statistical residual scatter of
$\approx$0.23~dex.  We caution, however, that the SFR~$\simlt$~10~\sfr\
galaxies are predicted by XRB population-synthesis models to have $L_{\rm
2-10~keV}$/SFR that flatten above $z \approx 1.5$, just above the current
limits of our survey (see Fig.~8 of Basu-Zych \etal\ 2013b).  It is therefore
likely that Equation~(14) significantly overpredicts $L_{\rm 2-10~keV}$/SFR for
$z \approx 1.5$ galaxies with SFR~$\simlt$~10~\sfr, and is not appropriate in
this regime.

Embedded within the above empirical parameterization is information about the
evolution of the underlying physical properties of galaxies (e.g., stellar age
and metallicity).  For example, Basu-Zych \etal\ (2013b) made use of the XRB
population-synthesis models of F13a, which predicted similar behavior to that
of Equation~(14), to interpret these trends as being due to the combined
evolution of LMXB and HMXB populations.  They reported that for a given
redshift interval, the low-SFR populations were predicted to have strong
contributions from both HMXBs and LMXBs.  This conclusion leads to relatively
large values of $L_{\rm 2-10~keV}$/SFR for low-SFR compared with those of
high-SFR galaxies, which are expected to be dominated by HMXBs alone.  With
increasing redshift, declining stellar ages and metallicities produce brighter
populations of LMXBs and HMXBs, respectively, thereby leading to a
corresponding increase in $L_{\rm 2-10~keV}$/SFR.  

%
%
\begin{figure*}
\figurenum{11}
\centerline{
\includegraphics[width=8.9cm]{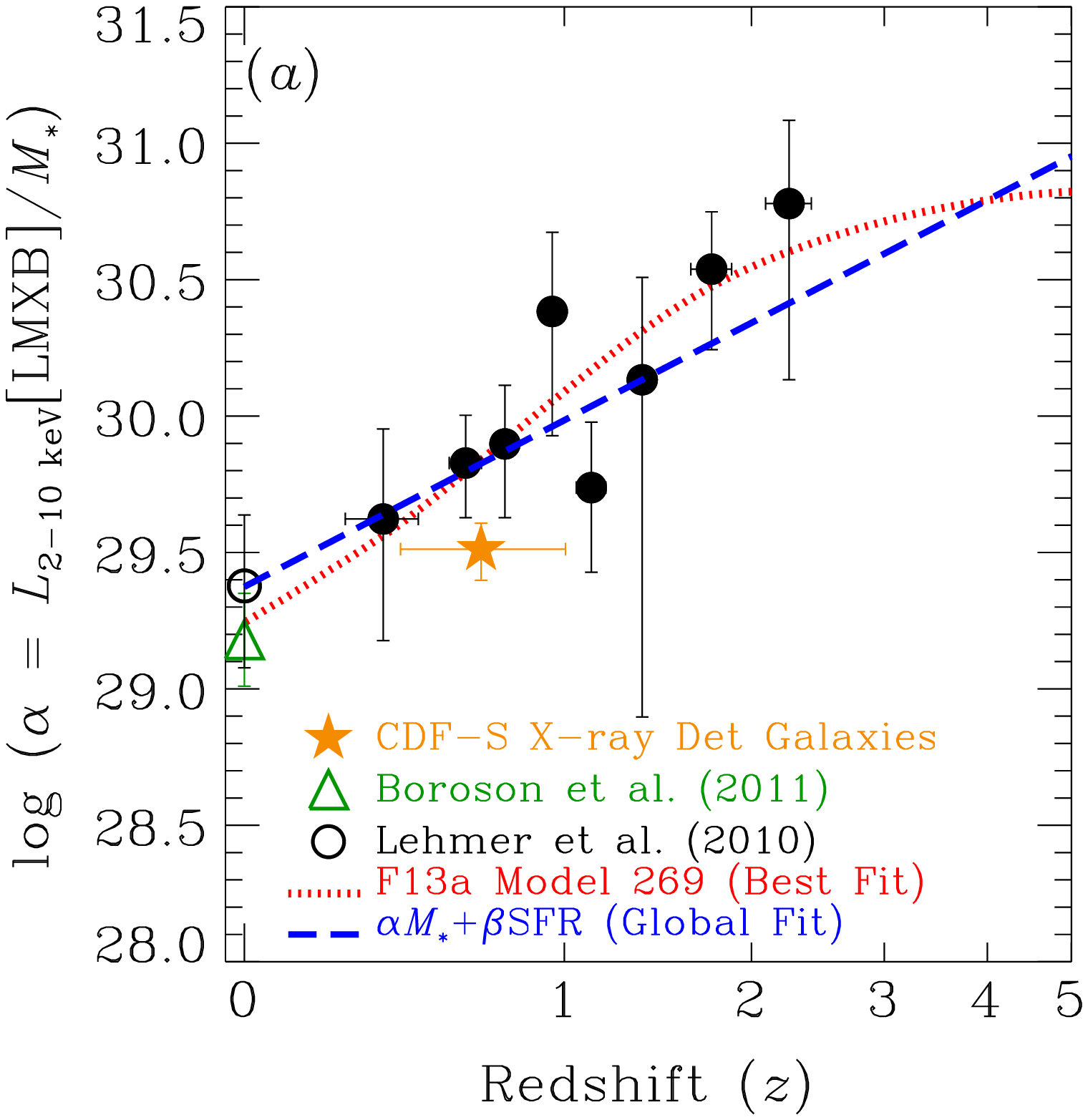}
\hfill
\includegraphics[width=8.9cm]{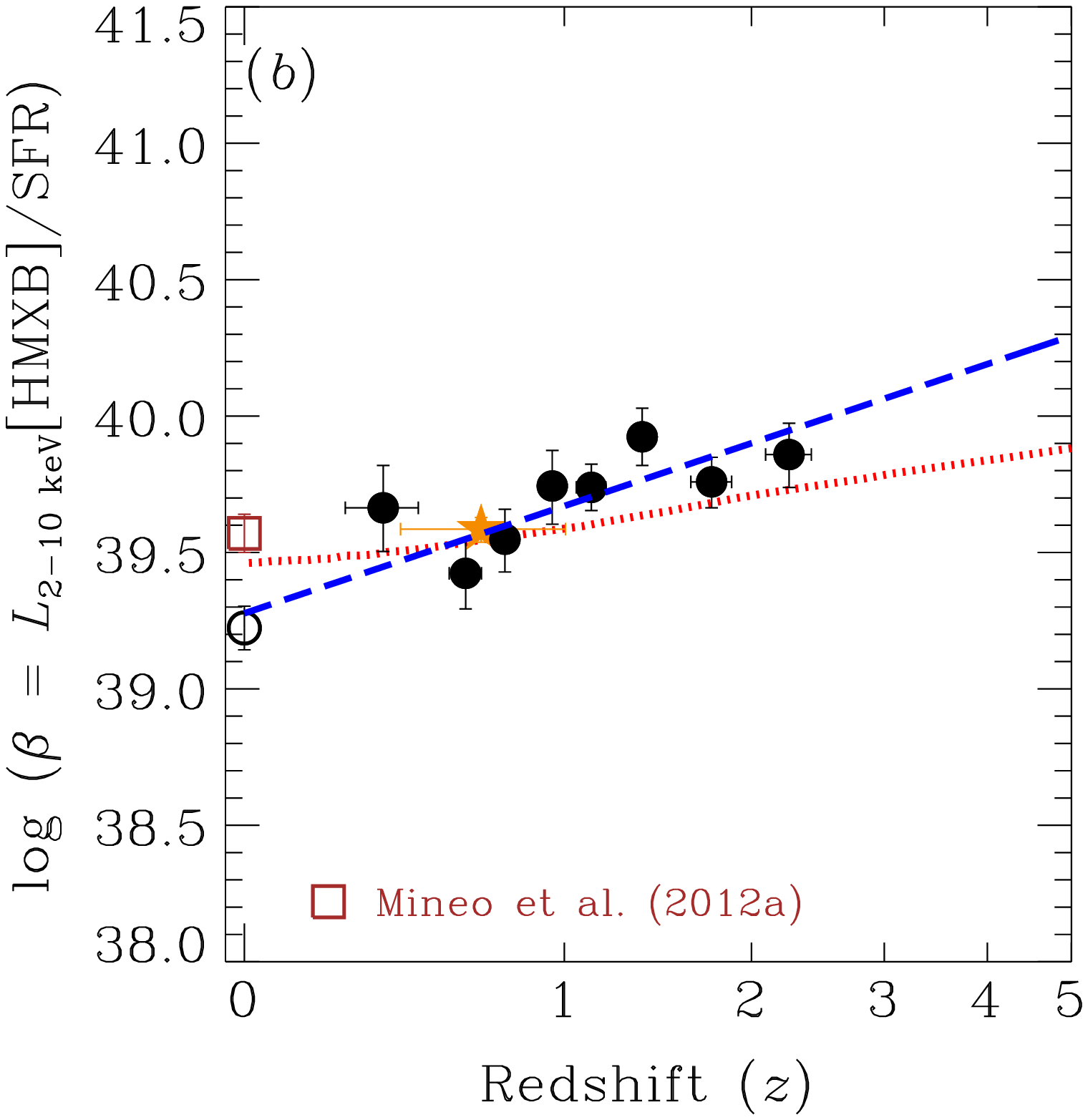}
}
\caption{
Best-fit parameterization values of $\alpha$ (panel $a$) and $\beta$ (panel $b$) versus
redshift.  Each value of $\alpha$ and $\beta$ ({\it filled circles\/} with
error bars) was obtained by fitting the $\approx$6~Ms CDF-S data in each corresponding redshift
panel of Figure~10 for $z \simlt 2.5$ intervals.  Error bars and upper limits are
1$\sigma$.  The orange stars with error bars are our estimates of $\alpha$ and
$\beta$ obtained using the \xray\ detected sample in the CDF-S (see $\S$3).
Local ($z = 0$) estimates are provided for $\alpha$ from L10 ({\it open
circle\/}) and Boroson \etal\ (2011; {\it open green triangle\/}) and for
$\beta$ from L10 and Mineo \etal\ (2012a; {\it open brown square\/}).  Results
from our best-fit global parameterization, based on Equation~(15), are shown as
dashed blue curves.  Finally, the red dotted curve shows the predicted
evolution of $\alpha$ and $\beta$ from Model~269, the best-fit F13a XRB
population-synthesis model.
\vspace{0.13in}
}
\end{figure*}

As noted above, there is substantial remaining scatter in the best-fit
parameterization from Equation~(14).  In particular, $L_{\rm 2-10~keV}$/SFR
values for stacked subsamples with SFR~$\simlt$~10~\sfr, which we argue are
expected to have contributions from both HMXBs and LMXBs, have significant
scatter for a given redshift.  We expect that our selection of subsamples by
sSFR broadens the range of LMXB contributions to these stacked subsamples,
e.g., compared to a selection by SFR alone.  

\subsubsection{Cosmic Evolution of LMXB and HMXB Populations}

As described in $\S$~5, our choice to select galaxy subsamples in bins of sSFR
was motivated by the expected scaling of LMXB and HMXB emission with $M_\star$
and SFR, respectively---i.e., to obtain $\alpha$ and $\beta$ values for a range
of redshifts.  Following Equation~(4), we expect $L_{\rm 2-10~keV}$/SFR will be
inversely proportional to sSFR.  Figure~10 presents $L_{\rm 2-10~keV}$/SFR
versus sSFR for the stacked subsamples in each redshift interval.   For the
$z < 2.0$ redshift intervals, it is apparent that $L_{\rm 2-10~keV}$/SFR
declines with increasing sSFR, as expected from Equation~(4).  Each panel of
Figure~10 displays the L10 relation for $z \approx 0$ galaxies ({\it gray
curves\/}) with the L10 data provided in the first panel of Figure~10.  With
increasing redshift, the $L_{\rm 2-10~keV}$/SFR values become increasingly
offset above the L10 relation.  In particular, the low-sSFR slopes of the
$L_{\rm 2-10~keV}$/SFR versus sSFR curves become steeper with increasing
redshift, indicating that the LMXB luminosity per unit mass (i.e., $\alpha$)
increases with increasing redshift.

For the eight redshift intervals that contained $\ge$4 detected subsamples
(i.e., all redshift intervals at $z < 2.5$), we fit the data using both a
constant model, $\log L_{\rm 2-10~keV}$/SFR~$= A_1$, as well as our canonical
model from Equation~(4), from which we extract best-fit values of $\alpha$ and
$\beta$.  Table~4 provides the best-fit values for $A_1$ and $\alpha$ and
$\beta$ in each of the eight redshift intervals, and compares the
goodness-of-fit for both models.  With the exception of the $z =$~2.0--2.5 bin,
our canonical model provides a statistically improved characterization of the
stacked data over the constant model at $\approx$83--99.99\% confidence levels
(based on an $F$-test; see Col.~10 in Table~4).  The resulting $\chi^2$ values
for these intervals indicate that the canonical model is generally a good fit
to the data ($\chi^2/\nu =$~0.2--1.2; median $\chi^2/\nu =$~0.47).  Figure~10
shows the resulting best-fit relations for the $z < 2.5$ redshift intervals as
solid curves, and Figure~11 presents the corresponding best-fit values of
$\alpha$ and $\beta$ versus redshift for the $z < 2.5$ redshift intervals as
filled black circles with error bars.  Figure~11 also indicates the values of
$\alpha$ and $\beta$ measured for the \xray\ detected sample (measured in
$\S$3; {\it orange stars\/}) and L10 ({\it open circles\/}), as well as the
value of $\alpha$ derived from LMXBs in local early-type galaxies ({\it open
green triangle\/}; Boroson \etal\ 2011) and $\beta$ derived from HMXBs in local
high-sSFR galaxies ({\it open brown square\/}; Mineo \etal\ 2012b).  

From Figure~11 it is clear that $\alpha$, the LMXB emission per unit stellar mass,
evolves rapidly with redshift out to $z \approx 2.5$, while $\beta$, the HMXB
emission per unit SFR, remains roughly constant over this redshift range,
albeit with some evidence for a mild increase with redshift.  As discussed in
$\S$1, and presented in F13a, this result is consistent with the basic
expectations from XRB population-synthesis models, which predict that the
decline in age and metallicity with increasing redshift would yield changes in
respective LMXB and HMXB scaling relations.  In the next section, we make
direct comparisons of our measurements with the XRB population-synthesis
predictions from F13a.

The constraints on $\alpha$ and $\beta$ discussed above utilize only data with
$z \simlt 2.5$ on a per-redshift interval basis; however, our stacking analyses
provide detections for galaxy subsamples out to $z \approx 4.0$.  To better
characterize the redshift evolution of $\alpha$ and $\beta$, we made use of the
following global parameterization using the full set of \nsbh\ detections out
to $z \approx 4$:
$$L_{\rm X}({\rm XRB})(z) = L_{\rm X}({\rm LMXB})(z) +  L_{\rm X}({\rm
HMXB})(z),$$
\begin{equation}
L_{\rm X}({\rm XRB})(z) = \alpha_0 (1+z)^\gamma M_\star + \beta_0 (1+z)^\delta
{\rm SFR},
\end{equation}
where $\alpha_0$, $\beta_0$, $\gamma$, and $\delta$ are fitting constants.  As
discussed above, $\alpha_0$ and $\beta_0$ have already been constrained by L10,
Boroson \etal\ (2011), and Mineo \etal\ (2012a); however, these values are
somewhat discrepant due to differences in galaxy sample properties (see
discussion in $\S$6.1).  We therefore chose to obtain values of $\alpha_0$ and
$\beta_0$ independently by fitting Equation~(15) to the CDF-S stacked data
alone.  We also combine our local L10 galaxy sample (i.e., the average values
for seven sSFR bins) with the CDF-S data to improve overall constraints on the
global evolution of the \hbox{2--10~keV} emission from galaxies.  By fitting
both CDF-S data alone and L10-plus-CDF-S data, we can show how constraints on
local galaxies influence the global redshift-dependent solution to
Equation~(15).

When using the CDF-S data alone, fitting our data to the
parameterization provided in Equation~(15) provides good overall fits to the
average data ($\chi^2/\nu = 0.91$, for $\nu = 56$ degrees of freedom) and
a significant reduction in the resulting spread (0.17~dex; or 48\%).  
Our fits characterize variations in the stacked, population-averaged
emission; however, such averaging masks galaxy-to-galaxy variations, which can
be significant.  The true intrinsic galaxy-to-galaxy spread in $L_{\rm
2-10~keV}$, within each subsample, is expected to be larger (on the order of
$\approx$0.2--0.4~dex; see, e.g., L10 and Mineo \etal\ 2012a) and sensitive to
variations in metallicity, stellar age, and statistical variations in the XRB
populations themselves.  We obtain best-fit values of $\log \alpha_0 = 29.30
\pm 0.28$, $\log \beta_0 = 39.40 \pm 0.08$, $\gamma = 2.19 \pm 0.99$, and
$\delta = 1.02 \pm 0.22$.  Interestingly, these values of $\alpha_0$ and
$\beta_0$, which are based solely on the $z \simgt 0.3$ CDF-S data, are in excellent
agreement with the range of respective values obtained for $z = 0$: $\log
\alpha_0 =$~29.1--29.2 (L10; Boroson \etal\ 2011) and $\log \beta_0
=$~39.2--39.6 (L10; Mineo \etal\ 2012a).  

\begin{table*}
\begin{center}
\caption{Summary of Global Fits to Stacked Data Sets}
\begin{tabular}{llcccccccc}
\hline\hline
  &   & \multicolumn{4}{c}{0.5--2~keV}  &   \multicolumn{4}{c}{2--10~keV} \\
  &   & \multicolumn{4}{c}{\rule{1.2in}{0.01in}}  &   \multicolumn{4}{c}{\rule{1.2in}{0.01in}} \\
 \multicolumn{1}{c}{Model Description} &  \multicolumn{1}{c}{Parameter} & \multicolumn{1}{c}{Param Value}  & $\chi^2/\nu$ & $\nu$ & $\sigma$(dex) & \multicolumn{1}{c}{Param Value}  & $\chi^2/\nu$ & $\nu$ & $\sigma$(dex) \\
\hline\hline
$\log L_{\rm X} = A_1 + \log$~SFR \dotfill   &  $A_1$ & 39.59 $\pm$ 0.02                         & 5.78        & 29 & 0.37 & 39.78 $\pm$ 0.02                         & 4.23        & 66 & 0.32     \\
\\
$\log L_{\rm X} = A_2 + B_2 \log$~SFR  \dotfill  &  $A_2$ & 40.06 $\pm$ 0.05 & & & & 40.12 $\pm$ 0.05 & & & \\
 & $B_2$ & 0.65 $\pm$ 0.04  & 2.75 & 28 & 0.29 & 0.71 $\pm$ 0.04  & 3.38 & 65 & 0.29 \\
\\
$\log L_{\rm X} = A_3 + B_3 \log {\rm SFR} + C_3 \log (1+z)$  \dotfill  &  $A_3$ & 39.83 $\pm$ 0.07 & & & & 39.82 $\pm$ 0.05 & & & \\
 & $B_3$ & 0.74 $\pm$ 0.04 & & & & 0.63 $\pm$ 0.04  & & & \\ 
 & $C_3$ & 0.97 $\pm$ 0.18  & 1.77 & 27 & 0.24 & 1.31 $\pm$ 0.11  & 1.32 & 64 & 0.20 \\
\\
$L_{\rm X} = \alpha_0 (1+z)^\gamma M_\star + \beta_0 (1+z)^\delta {\rm SFR}$ \hspace{0.1in}   {\it [CDF-S only]} \dotfill   &  $\log \alpha_0$ & 28.87 $\pm$ 0.24 & & & & 29.30 $\pm$ 0.28 & & & \\
 & $\log \beta_0$ & 39.66 $\pm$ 0.17 & & & & 39.40 $\pm$ 0.08  & & & \\ 
 & $\gamma$ & 4.59 $\pm$ 0.80 & & & & 2.19 $\pm$ 0.99  & & & \\ 
 & $\delta$ & -0.10 $\pm$ 0.72  & 0.84 & 19 & 0.16 & 1.02 $\pm$ 0.22  & 0.91 & 56 & 0.17 \\
\\
$L_{\rm X} = \alpha_0 (1+z)^\gamma M_\star + \beta_0 (1+z)^\delta {\rm SFR}$  \hspace{0.1in}   {\it [L10 plus CDF-S]} \ldots\ldots   &  $\log \alpha_0$ & 29.04 $\pm$ 0.17 & & & & 29.37 $\pm$ 0.15 & & & \\
 & $\log \beta_0$ & 39.38 $\pm$ 0.03 & & & & 39.28 $\pm$ 0.05  & & & \\ 
 & $\gamma$ & 3.78 $\pm$ 0.82 & & & & 2.03 $\pm$ 0.60  & & & \\ 
 & $\delta$ & 0.99 $\pm$ 0.26  & 0.79 & 26 & 0.16 & 1.31 $\pm$ 0.13  & 1.05 & 63 & 0.17 \\
\\
F13a Population Synthesis \dotfill & Model 245 &  & & & & & 2.71 & 67 & 0.20 \\
\\
F13a Population Synthesis \dotfill & Model 269 & & & & & &  1.56 & 67 & 0.19 \\
\hline
\end{tabular}
\end{center}
NOTE.---All global models were fit to a combination of local ($z = 0$) galaxy subsamples plus stacked high-redshift subsamples from the CDF-S (derived in this work).  For the 0.5--2~keV band, we utilized seven local galaxy subsamples compiled from L10 plus \nsbs\ stacked subsamples in the CDF-S that were significantly detected in the observed-frame \hbox{0.5--1~keV} band.  For the 2--10~keV band, we utilized seven local galaxy subsamples compiled by L10 plus the \nsbh\ stacked subsamples that were significantly detected in the CDF-S in the observed-frame 1--2~keV band.  Only detected stacked subsamples (i.e., S/N~$\simgt 3\sigma$) were used in the global fits; upper limits were excluded.
\vspace{0.14in}
\end{table*}

\begin{table*}
\begin{center}
\caption{2--10~keV Fits to Stacked Data Sets In Each Redshift Bin}
\begin{tabular}{lccccccccc}
\hline\hline
 \multicolumn{1}{c}{ } &  & \multicolumn{3}{c}{$\log L_{\rm 2-10~keV} = A_1 + \log {\rm SFR}$}   & \multicolumn{4}{c}{$L_{\rm 2-10~keV} = \alpha M_\star + \beta {\rm SFR}$ }  & \\
 \multicolumn{1}{c}{ } &  & \multicolumn{3}{c}{\rule{1.5in}{0.01in} }   & \multicolumn{4}{c}{\rule{2.4in}{0.01in} }  & \\
 \multicolumn{1}{c}{ } &  & $\log A_1$ &  &   & $\log \alpha$ & $\log \beta$ &  &  &  \\
 \multicolumn{1}{c}{$z_{\rm lo}$--$z_{\rm up}$} & $N_{\rm det}$ & ($\log$ \lum\ (\sfr)$^{-1}$) & $\chi^2/\nu$ & $\nu$  & ($\log$ \lum\ (\sfr)$^{-1}$) & ($\log$ \lum\ $M_\odot^{-1}$) & $\chi^2/\nu$ & $\nu$ & $F_{\rm prob}$ \\
 \multicolumn{1}{c}{(1)} &  (2) & (3) & (4) & (5) & (6) & (7) & (8) & (9) & (10) \\
\hline\hline
                 0.0--0.5 &   7 &   39.83$^{+0.08}_{-0.09}$ &  1.99 &   6 &   29.62$^{+0.33}_{-0.45}$ &   39.66$^{+0.16}_{-0.16}$ &  1.00 &   5 & 0.9536 \\
                 0.5--0.7 &   9 &   39.71$^{+0.06}_{-0.06}$ &  5.28 &   8 &   29.83$^{+0.18}_{-0.20}$ &   39.42$^{+0.13}_{-0.13}$ &  0.45 &   7 & 1.0000 \\
                 0.7--0.9 &   8 &   39.73$^{+0.06}_{-0.06}$ &  2.88 &   7 &   29.90$^{+0.22}_{-0.27}$ &   39.55$^{+0.11}_{-0.12}$ &  0.49 &   6 & 0.9990 \\
\smallskip
                 0.9--1.0 &   5 &   39.89$^{+0.07}_{-0.07}$ &  0.83 &   4 &   30.38$^{+0.29}_{-0.46}$ &   39.74$^{+0.13}_{-0.14}$ &  0.14 &   3 & 0.9797 \\
                 1.0--1.2 &  10 &   39.85$^{+0.05}_{-0.05}$ &  1.67 &   9 &   29.74$^{+0.24}_{-0.31}$ &   39.74$^{+0.09}_{-0.09}$ &  0.21 &   8 & 1.0000 \\
                 1.2--1.5 &   6 &   39.99$^{+0.06}_{-0.06}$ &  0.64 &   5 &   30.13$^{+0.38}_{-1.24}$ &   39.92$^{+0.11}_{-0.11}$ &  0.47 &   4 & 0.8293 \\
                 1.5--2.0 &   8 &   39.90$^{+0.05}_{-0.05}$ &  1.66 &   7 &   30.54$^{+0.21}_{-0.30}$ &   39.76$^{+0.09}_{-0.10}$ &  1.18 &   6 & 0.9012 \\
                 2.0--2.5 &   4 &   39.97$^{+0.06}_{-0.06}$ &  2.55 &   3 &   30.78$^{+0.31}_{-0.65}$ &   39.86$^{+0.12}_{-0.12}$ &  3.41 &   2 & 0.3311 \\
\hline
\end{tabular}
\end{center}
NOTE.---Model fits to stacked galaxy subsample 2--10~keV luminosities include only stacked detections with observed-frame 1--2~keV S/N~$\ge 3$.  Only redshift bins with more than five detected galaxy subsamples are included in this table.  The redshift range and number of detected galaxy subsamples are provided in Columns~(1) and (2), respectively.  A constant model ($\log L_{\rm 2-10~keV} = A_1 + \log {\rm SFR}$; Col.~(3)--(5)) is tested against our canonical model ($L_{\rm 2-10~keV} = \alpha M_\star + \beta {\rm SFR}$; Col.~(6)--(9)) for each redshift range.  In all cases, except perhaps, for the last bin (i.e., at $z\approx2.5$), the canonical model provides a better goodness of fit ($\chi^2$; Col.~(4) and (8)), and the significance of improvement in terms of the $F$-test is provided in Column~(10).
\vspace{0.14in}
\end{table*}

%
%
\begin{figure*}
\figurenum{12}
\centerline{
\includegraphics[width=19cm]{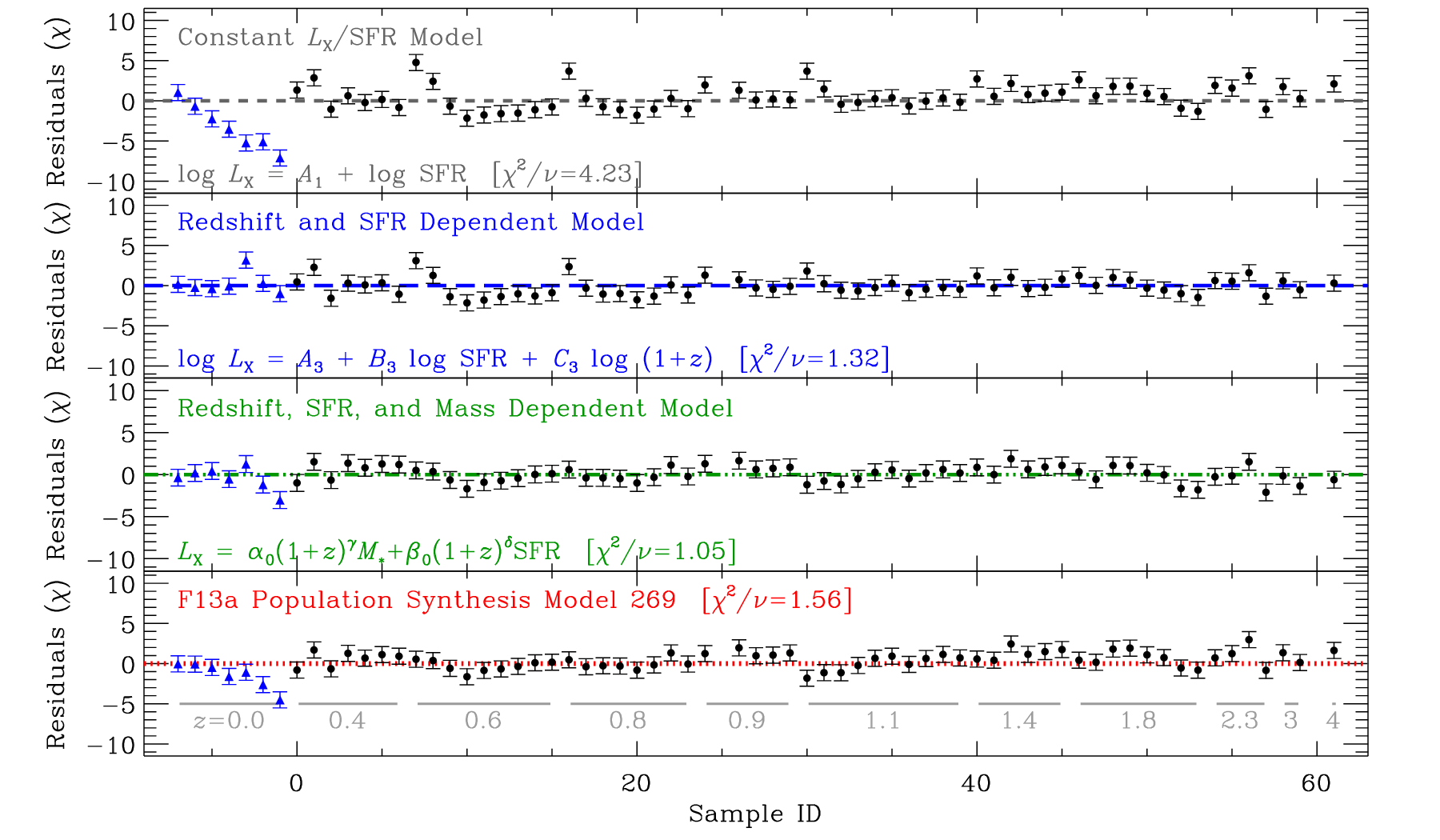}
}
\caption{
Summary of residuals to model fits to the L10 local sample ({\it blue
triangles\/}) and \nstk\ stacked galaxy subsamples in the CDF-S ({\it black filled
circles\/}).  In order of descending panels, our models include a constant
$L_{\rm 2-10~keV}$/SFR ratio ({\it top panel}; Eqn.~12), a redshift and SFR
dependent model ({\it second panel}; Eqn.~14), a redshift and sSFR model ({\it
third panel}; Eqn.~15), and the best-fitting XRB population-synthesis model
from F13a ({\it bottom panel}; Model~269 from F13a).  Both the redshift and
sSFR model, as well as the XRB population-synthesis Model~269, provide the best
characterization of the global \xray\ emission of galaxy populations.
}
\end{figure*}

When we combine the L10 local comparison values with the CDF-S data and re-fit
the ensemble data set to Equation~(15), we obtain $\chi^2/\nu \approx 1.05$ for
$\nu = 63$ degrees of freedom, and values consistent with the CDF-S only fits:
$\log \alpha_0 = 29.37 \pm 0.15$, $\log \beta_0 = 39.28 \pm 0.05$, $\gamma =
2.03 \pm 0.60$, and $\delta = 1.31 \pm 0.13$.  These parameters indicate that
the redshift increases of $\alpha$ and $\beta$ (as noted above from the fits in
each of the $z \simlt 2.5$ redshift intervals) are significant at the
$\approx$3.4~$\sigma$ and $\approx$10.1~$\sigma$ levels, respectively.
However, the significances of evolution in both $\alpha$ and $\beta$ are
dependent on the inclusion of the L10 local galaxy constraints.  From the CDF-S
data alone, the respective redshift evolution of LMXB and HMXB scaling
relations is significant at the $\approx$2.2~$\sigma$ and $\approx$4.6~$\sigma$
levels, respectively.  The L10, Boroson \etal\ (2011), and Mineo \etal\ (2012a)
measurements of $\alpha_0$ and $\beta_0$ are based on galaxy samples drawn from
the archive, which have selection biases that are different from those in our
CDF-S galaxy subsamples; therefore, future investigations to measure $\alpha_0$
and $\beta_0$ in a local galaxy sample with similar selection to the CDF-S
galaxy subsample would help clarify the results presented in this paper.

%
\section{Discussion}
%

The above results indicate that the $L_{\rm X}$--SFR correlation is not
universal, but rather depends critically on both SFR and $M_\star$ and evolves
with redshift.  The top three panels of Figure~12 show the 2--10~keV residuals
to the best-fit global relations presented throughout $\S$6, which include: (1)
a constant ``universal'' relation $\log L_{\rm 2-10~keV}$/SFR~=~$A_1$
(Eqn.~12); (2) a SFR and redshift-dependent relation $\log L_{\rm
2-10~keV}$/SFR~=$A_3 + B_3 \log {\rm SFR} + C_3 \log (1+z)$ (Eqn.~14); and (3)
a SFR, $M_\star$, and redshift-dependent relation $L_{\rm 2-10~keV} = \alpha_0
(1+z)^\gamma M_\star + \beta_0 (1+z)^\delta {\rm SFR}$ (Eqn.~15).  This
succession of global models produces significant improvement in goodness of fit
($\chi^2/\nu \approx 4.2$, 1.3, and 1.1, respectively), as well as significant
reductions in residual scatter ($\sigma =$~0.32, 0.20, and 0.17~dex,
respectively).  The key results from this study can be summarized as (1) the
LMXB emission per unit stellar mass, $\alpha \equiv L_{\rm X}$(LMXB)/$M_\star$,
increases rapidly with increasing redshift $\propto (1+z)^{2-3}$; and (2) the
HMXB emission per unit SFR, $\beta \equiv L_{\rm X}$(HMXB)/SFR, increases
mildly with increasing redshift $\propto (1+z)$.  These proportionalities
appear to be valid out to at least $z \approx$~2--3; however, as we will
discuss below, there are physical reasons to expect that these
proportionalities will not extend to even higher redshifts.  In this section,
we discuss our results in the context of the XRB population-synthesis models
from F13a, which have made similar predictions.

\subsection{What Drives the Evolution of LMXB and HMXB Populations?}

The F13a XRB population-synthesis study constructed 288 unique XRB models,
which predict the evolution of $\alpha$ and $\beta$ over the history of the
Universe (based on the {\ttfamily StarTrack} population-synthesis code;
Belczynski \etal\ 2002, 2008), accounting for evolution of the star-formation
history and metallicity using the {\ttfamily Millenium~II} cosmological
simulation with a semi-analytic galaxy evolution prescription (Guo \etal\
2011).  These models include prescriptions for XRBs that are formed through
stellar evolutionary channels only and do not include XRBs that may form due to
dynamical interactions in high stellar density environments like globular
clusters (see, e.g., Benacquista \& Downing~2013 for a review).  As noted in
F13a, dynamically-formed XRBs are significant to the overall \xray\ emission in
the most massive elliptical galaxies, but are minority populations globally
(see \S2.1 of F13a for further details).

The six parameters that were varied in the 288 unique scenarios in F13a
included common-envelope efficiency (two parameters), wind prescriptions,
binary mass ratio distribution, kicks from SNe, and the stellar IMF (see F13a
for details).   Six out of the 288 models (Models~245, 229, 269, 205, 249, and
273, ordered by decreasing likelihood; see F13a for details on each of these
models) provided acceptable predictions of $z=0$ observational constraints
(from Tzanavaris \& Georgantopolous~2008; L10; Boroson \etal\ 2011; Mineo
\etal\ 2012a), with the remaining 282 were deemed to be likely unrealistic.
The six models with the highest probability had the same prescriptions for
common-envelope efficiency and binary mass-ratio distribution, but varied in
prescriptions for stellar-wind strength, SNe kick amplitudes, and stellar IMFs.
One of the limitations in constraining the F13a models was the lack of strong
observational constraints at $z > 0$, a limitation that is mitigated
significantly by the current study.

We compared each of the 288 F13a model predictions for the redshift evolution
of $\alpha$ and $\beta$ with our combined local L10 and CDF-S stacked sample
measurements.  Our comparisons were limited to the 2--10~keV constraints, which
are expected to directly probe the XRB populations with negligible
contamination from hot gas.  Each of the 288 models provides a unique redshift,
SFR, and $M_\star$ dependent prediction for every subsample with no free
parameters.  Although these models are not strictly fit to the data, a
goodness-of-fit parameter can be assigned to each model and compared across
models.  

Table~5 provides a list of the $\chi^2/\nu$ values for the top ten F13a models,
sorted by ascending $\chi^2$.  The highest probability model, Model~269, has a
$\chi^2/\nu = 1.56$ for $\nu=67$ degrees of freedom.  Given the number of
degrees of freedom, the probability of obtaining $\chi^2/\nu \ge 1.56$ is
0.22\%, indicating that this model is formally unacceptable; however, given the
limited ranges and numbers of parameters that were varied in the F13a models,
it is likely that minor tweaks, and the inclusion of additional parameters not
varied by F13a (e.g., the fraction of stars in binaries), could yield
statistically robust fits to our data.   Future generations of XRB
population-synthesis models will explore these issues.  For further comparison,
Table~5 lists the relative rankings of the F13a models from the Tzanavaris
\etal\ (2013) study of XRB luminosity functions in local galaxies and the
Tremmel \etal\ (2013) study of the evolution of normal galaxy \xray\ luminosity
functions.  These investigations, along with the F13a original study, rank
Model~269 highly (within the top 12 models).  Notably, Model~269 provides a
better global prediction to the data than the best-fit constant $L_{\rm
2-10~keV}$/SFR (see Fig.~11 and Table~3) and a comparable fit to the data as
the redshift and SFR dependent (Eqn.~14) models.  

From a physical point of view, Model~269 is the same in all six parameters as
the previous best-fit model from F13a, Model~245, except that Model~269 allows
for binary systems to emerge after a common-envelope phase involving the donor
stars going through the Hertzsprung gap, while Model~245 assumes that such a
common-envelope phase will automatically lead to the merging of the two stars
and a termination of any possible subsequent XRB phase.  This has the effect of
Model~269 having mildly elevated emission from HMXBs (i.e., elevated values of
$\beta$) over Model~245 across the redshift range studied here.  The variation
of this parameter is motivated on theoretical grounds and is currently
unconstrained by observations.  Theoretically, a Hertzsprung gap star does not
have a clear entropy jump at the core-envelope transistion (Ivanova \&
Taam~2004), so once a companion is engulfed within such a star during the
common-envelope phase, there is no clear boundary where an inspiral would
cease, and consequently a merger is expected (see, e.g., Taam \&
Sandquist~2000).  However, on energetic grounds, it is possible for the system
to successfully exit the common-envelope phase without merging.  If binaries
indeed survive a common-envelope phase where donor stars are in the Hertzsprung
gap (i.e., Model~269), then the predicted numbers of gravitational wave sources
from double black hole mergers is predicted to be much larger (by up to a
factor of $\sim$500) than if they are destroyed (i.e., the Model~245 case; see
F13a and Belczynski \etal\ 2007 for details).  Future studies from Advanced
LIGO and Advanced Virgo gravitational wave detectors will likely constrain this
effect independently (Belczynski \etal\ 2015, 2016; Abbott \etal\ 2016).

Figure~10 displays the Model~269 predicted $L_{\rm 2-10~keV}$/SFR versus sSFR
for all redshift intervals, spanning $z \approx$~0--7 ({\it dotted red
curves\/}).  The predictions for Model~269 are remarkably similar to those from
our best-fit global model (Eqn.~15; {\it dashed blue curves\/} in Fig.~10) over
the redshift range \hbox{$z \approx$~0--2.5}.  In general, Model~269 and our
best-fit global model are in good agreement; however, differences in the
high-sSFR predictions are clear at $z \simgt$~1.5--2, with Model~269 generally
predicting lower values of $L_{\rm X}$/SFR.

Figure~11 shows the redshift-dependent trajectories of $\alpha$ (Fig.~11$a$)
and $\beta$ (Fig.~11$b$) for Model~269.  As before, Model~269 and the best-fit
global model are in general agreement in terms of the predicted evolution of
$\alpha$ and $\beta$ with redshift, with two key exceptions: (1) Model~269
predicts that $\alpha$ flattens above $z \simgt 2$, a regime not well
constrained by our data; and (2) Model~269 predicts that $\beta$ increases more
slowly with redshift than our data reveal.

%
%
\begin{figure*}
\figurenum{13}
\centerline{
\includegraphics[width=9cm]{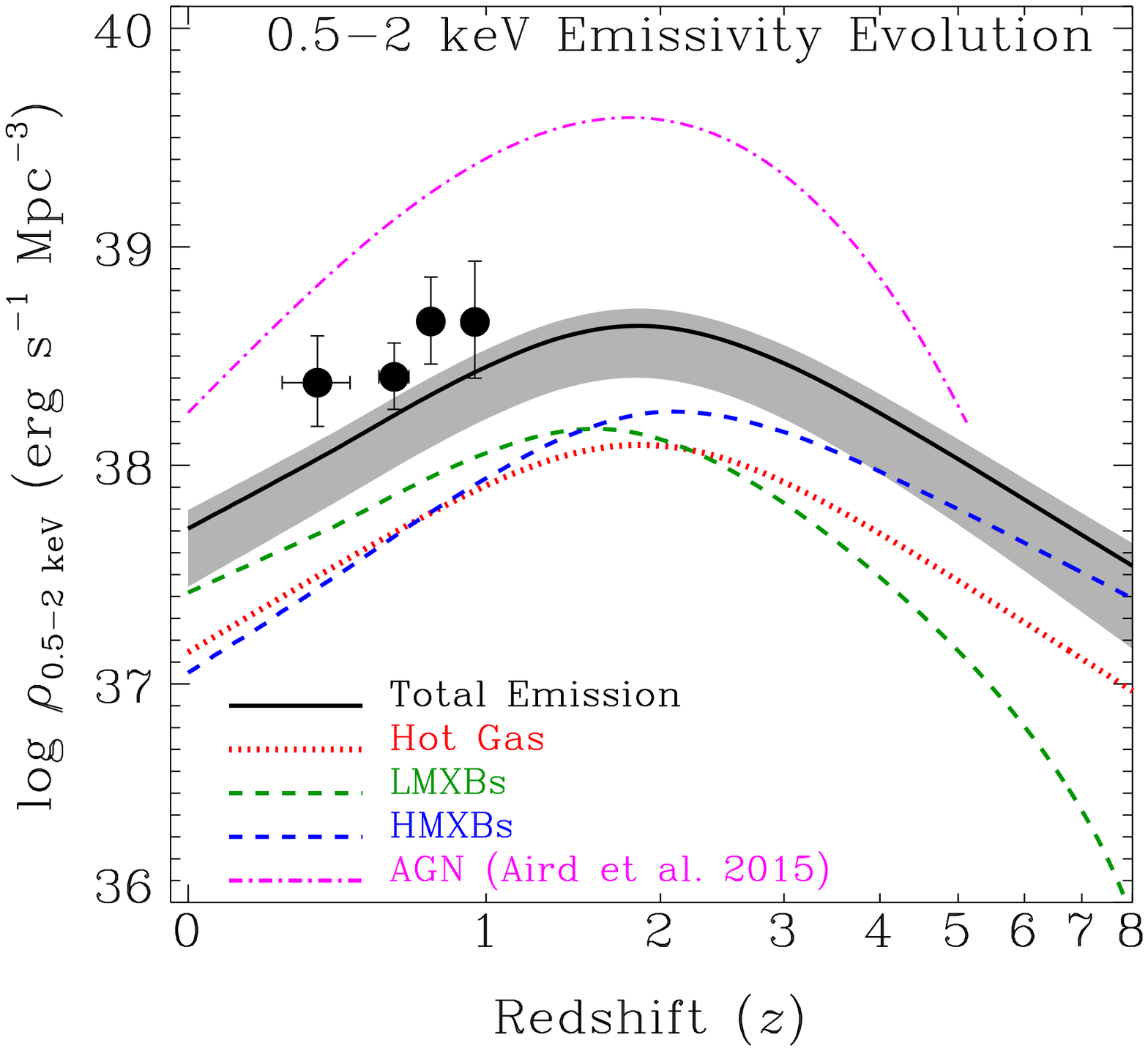}
\hfill
\includegraphics[width=9cm]{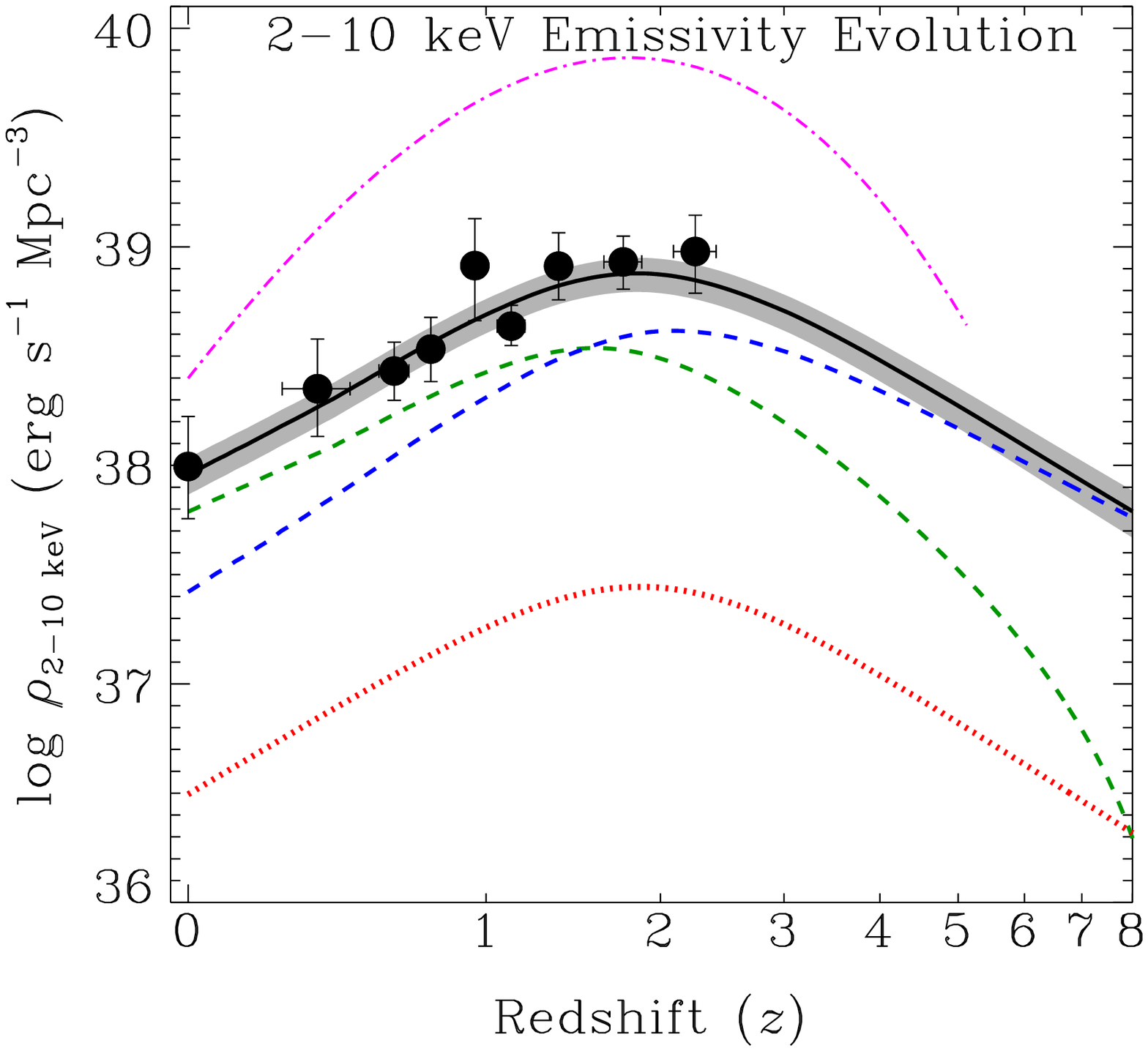}
}
\vspace{0.1in}
\caption{
Estimated normal-galaxy \xray\ emissivity of the Universe versus redshift for
the 0.5--2~keV ($a$) and 2--10~keV ($b$) bands.  For comparison, the AGN
emissivity, as computed by Aird \etal\ (2015), are shown for both bandpasses as
magenta dot-dashed curves.  Data points with 1$\sigma$ error bars correspond to
scaling the best-fit values of $\alpha$ and $\beta$ provided in Table~4 to the
Madau \& Dickinson~(2014) analytic estimates of stellar-mass and SFR density,
respectively.  The solid curves and shaded 1$\sigma$ error regions represent
the theoretical best-estimates for the evolution of \xray\ scaling relations
with cosmic time, including typical uncertainties in measurements of the SFR
densities $\psi$ and stellar mass densities $\rho_\star$, as well as
uncertainties in the XRB SED due to absorption (for the 0.5--2~keV band only).
The LMXB and HMXB contributions to these curves were computed by scaling,
respectively, $\alpha$ and $\beta$ from XRB population-synthesis Model~269 to
the analytic stellar-mass and SFR densities ({\it dashed curves\/}).  We scaled
the local hot gas scaling relations with SFR to the analytic SFR density curve
to compute the hot gas contributions (see text for details).  Under these
assumptions, XRBs dominate the normal galaxy emissivity of the Universe at all
redshifts, with LMXBs dominating at $z \simlt$~1--2 and HMXBs at $z
\simgt$~1--2, and there is an indication that the normal galaxy emissivity will
exceed that of AGN at $z \simgt$~6--8.
}
\end{figure*}

The good agreement between the Model~269 and canonical model trends allows for
a physical interpretation of the rapid increase in $\alpha$ and mild increase
in $\beta$ observed in our data.  The close similarities between Model~269 and
the F13a best model, Model~245, imply that the conclusions drawn here are the
same basic conclusions drawn by F13a: the rise in $\alpha$ with redshift arises
due to the shifting to higher-mass LMXB donor stars, and more luminous LMXBs,
as the population age declines with increasing redshift.  The predicted turn
over in $\alpha$ at $z \simgt 2.5$ occurs when the Universe was
$\approx$2.5~Gyr old, and the stellar populations were on average
$\approx$1~Gyr in age, which is the peak stellar age for LMXB emission.
Therefore, the turn over in $\alpha$ around $z \approx 2.5$ corresponds to the
peak stellar age for LMXB formation.  Populations with \hbox{0.5--1.5~Gyr} age
($z =$~2--5) are expected to produce comparably bright LMXB emission per unit
$M_\star$, due to contributions from massive ($\approx$3~\msol) red giant (RG)
donors (see, e.g., Kim \etal\ 2009).  $L_{\rm X}$(LMXB)/$M_\star$ then declines
rapidly at ages above $\approx$1.5~Gyr (i.e., $z \simlt 2.5$), as the brightest
RG donor star mass shifts to lower masses ($\simlt$1--3~\msol).  Some studies
of LMXBs in local elliptical galaxies of different ages have provided initial
support for this prediction (e.g., Kim \& Fabbiano~2010; Lehmer \etal\ 2014);
however, no statistically conclusive results have been reached based on how
LMXBs evolve with stellar-population age using local galaxies (e.g., Boroson
\etal\ 2011; Zhang \etal\ 2012).

For HMXBs, mild evolution is predicted for $\beta$, continuing to high
redshifts, due to a decline in metallicity with redshift.  As argued by Linden
\etal\ (2010) and F13a, star-formation under low-metallicity conditions can
yield significantly larger numbers of compact objects (neutron stars and black
holes), compared with higher metallicities, due to a reduction in the
efficiency of stellar-wind mass loss over the lifetimes of massive stars.
Additionally, massive stars at lower metallicity tend to expand to large radii
later in their evolution compared to higher metallicity ones. Hence, lower
metallicity stars form better defined and more massive cores before entering
the common envelope.  This in turns allows for the easier ejection of the
common-envelope of stars that will produce black holes (Linden \etal\ 2010;
Justham \etal\ 2015).  This increased pool of black holes in low-metallicity
environments results in an excess of HMXBs (and thus HMXB emission) per unit
SFR compared to higher metallicity environments.  Furthermore, the most massive
black holes formed in low metallicity environments can be quite large,
increasing the potential for the \xray\ emission from very luminous \xray\
sources to vastly exceed the collective emission from much more numerous
low-luminosity XRBs.  These sources are thought to provide important
contributions to heating the intergalactic medium at $z \simgt 10$ (e.g.,
Mirabel \etal\ 2011; Fragos \etal\ 2013b; Kaaret~2014).  Studies of the
metallicity dependence of ultraluminous \xray\ source (ULX) formation and
$L_{\rm X}$/SFR have demonstrated this effect empirically, and find dependences
similar to that expected from the F13a population-synthesis predictions (e.g.,
Mapelli \etal\ 2009, 2010; Basu-Zych \etal\ 2013a; Brorby \etal\ 2014; Douna
\etal\ 2015).  These findings, along with the results presented in this paper,
therefore support the idea that XRB emission was enhanced in the primordial $z
\simgt 10$ Universe, and could provide a non-negligible contribution to the
heating of the intergalactic medium. 

\subsection{Implications for the Cosmic X-ray Emissivity and X-ray Background Contribution}

Given the potential for normal-galaxy populations to be substantial contibutors
to heating of the intergalactic medium at high redshifts, we use our
measurements of the evolution of $\alpha$ and $\beta$, along with estimates of
galaxy stellar mass and SFR density evolution, to provide updated constraints
on the evolution of the \xray\ emissivity.  The stellar mass and SFR density of
the Universe has been constrained by numerous past studies using
radio--to--far-UV emission.  The recent review by Madau \& Dickinson~(2014)
provides a coherent census of these constraints and presents analytic formulae
describing the redshift evolution of the stellar mass density, $\rho_\star$,
and SFR density, $\psi$, of the Universe that are valid out to $z \approx 8$.
Using the Madau \& Dickinson~(2014) formalism, corrected to our adopted
Kroupa~(2001) IMF, we estimated the redshift-dependent LMXB and HMXB emissivity
as a function of redshift following:
$$\rho_{\rm X}^{\rm LMXB} = \alpha(z) \rho_{\star}(z),$$
\begin{equation}
\rho_{\rm X}^{\rm HMXB} = \beta(z) \psi(z).
\end{equation}
%

\begin{table}
\begin{center}
\caption{Fragos Model Goodness of Fit}
\begin{tabular}{lccccc}
\hline\hline
 \multicolumn{1}{c}{F13a} & \multicolumn{1}{c}{Rank} & \multicolumn{1}{c}{Rank} & \multicolumn{1}{c}{Rank} &\multicolumn{1}{c}{Rank}  \\
 \multicolumn{1}{c}{Model} & \multicolumn{1}{c}{(This Study)} & \multicolumn{1}{c}{(F13a$^\ast$)} & (Tz13$^\dagger$) & (Tr13$^\ddagger$) & $\chi^2/\nu$ \\
\hline\hline
269 \ldots\ldots\ldots\ldots &   1 &   3 &   5 &  12 &  1.56 \\
273 \dotfill &   2 &   6 &  12 &   6 &  1.64 \\
 77 \dotfill &   3 &  21 &  19 &  27 &  1.82 \\
201 \dotfill &   4 &  10 &   9 &  19 &  1.98 \\
 81 \dotfill &   5 &  26 &  24 &  20 &  2.14 \\
 53 \dotfill &   6 &  20 &  11 &  22 &  2.16 \\
246 \dotfill &   7 &  25 &  23 &  23 &  2.24 \\
232 \dotfill &   8 &  35 &  60 &  28 &  2.30 \\
230 \dotfill &   9 &  17 &  35 &  16 &  2.33 \\
231 \dotfill &  10 &  37 &  34 &  31 &  2.43 \\
\hline
\end{tabular}
\end{center}
$^\ast$Rank of likelihood based on F13a.\\
$^\dagger$Rank from Tzanavaris \etal\ (2013) study of local SINGS galaxies.\\
$^\ddagger$Rank from Tremmel \etal\ (2013) study of normal-galaxy \xray\ luminosity function evolution.
\end{table}

Equation~(16) can be applied using $\alpha$ and $\beta$ values constrained
either in individual redshift intervals or via global parameterizations.
Figure~13 presents constraints from the individual redshift intervals for the
0.5--2~keV and 2--10~keV bands as filled circles with 1$\sigma$ error bars.
These values were compared with the LMXB and HMXB emissivity predictions from
Model~269, extended from $z =$~0--8 ({\it dashed curves\/} in Fig.~13).  We
must also account for hot gas emission, in particular at 0.5--2~keV, to compare
with the data.  We computed a first-order estimate of the hot gas emissivity of
the Universe at all redshifts by assuming a universal hot gas scaling with SFR,
of the form:
\begin{equation}
\rho_{\rm 0.5-2~keV}^{\rm hot\; gas}/{\rm (erg~s^{-1}~Mpc^{-3})} = 1.5 \times
10^{39} \psi(z)/{\rm(}M_\odot~{\rm yr^{-1}~Mpc^{-3})},
\end{equation}
which is based on the Mineo \etal\ (2012b) relation.  We then utilized the hot
gas component of our canonical SED (see Fig.~4$a$) to scale Equation~(17) to
the 2--10~keV band; we find,
\begin{equation}
\rho_{\rm 2-10~keV}^{\rm hot\; gas}/{\rm (erg~s^{-1}~Mpc^{-3})} = 3.4 \times
10^{38} \psi(z)/{\rm(}M_\odot~{\rm yr^{-1}~Mpc^{-3})}.
\end{equation}
Figure~13 indicates the breakdown of LMXB, HMXB, and hot gas contributions to
the \xray\ luminosity density of normal galaxies.  All three components provide
substantial contributions to the 0.5--2~keV normal galaxy emissivity, with XRBs
dominating at all redshifts for both bands.  Similar to the results from F13a,
LMXBs dominate the \xray\ emissivity at $z \simlt$~1--2, with HMXBs dominating
at $z \simgt$~1--2, as the stellar mass density of the Universe drops
precipitously.  For comparison, Figure~13 displays the AGN emissivity derived
in Aird \etal\ (2015); these curves are shown to $z \approx 5$, where direct
observational constraints are available.  Reasonable extrapolations of the AGN
curves to $z > 5$ indicate that the normal galaxy emissivity should overpower
AGN at \hbox{$z \simgt$~6--8}, during the epoch of reionization.  This
conclusion was also reached by Fragos \etal\ (2013b), based on Model~245, and
has important implications for the role of ionizing photons from XRBs in
heating the neutral intergalactic medium during the reionization epoch (see,
e.g., discussions in Mirabel \etal\ 2011; McQuinn~2012; Pacucci \etal\ 2014;
Artale \etal\ 2015).

As noted by Dijkstra \etal\ (2012), the measured unresolved cosmic \xray\
background (CXB) provides a firm upper limit on the possible evolution of
\xray\ scaling relations.  From the above estimates of the X-ray emissivity
evolution of the Universe, it is straightforward to calculate the contributions
to the observed CXB from normal galaxy populations throughout the Universe. The
normal galaxy CXB intensity in bandpass $E_1-E_2$, $\Omega^{\rm CXB,
gal}_{E_1-E_2}$, can be calculated following:
\begin{equation}
\Omega_{E_1-E_2}^{\rm CXB, gal} =  \int_{0}^{z_{\rm max}}
k_{E_1^\prime-E_2^\prime}^{E_1-E_2} \frac{\rho_{E_1^\prime-E_2^\prime}}{4\pi
d_L^2} \frac{dV}{dz d\Omega}  dz,
\end{equation}
where $k^{E_1-E_2}_{E_1^\prime-E_2^\prime}$ provides a $k$-correction from
rest-frame $E_1^\prime-E_2^\prime$ to observed-frame $E_1-E_2$ following our
canonical SED (defined in $\S$4), $\frac{dV}{dz d\Omega}$ is the
cosmology-dependent differential volume element (comoving volume per unit
redshift per unit solid angle), and we integrate to $z_{\rm max} = 8$.  

As reported in Lehmer \etal\ (2012), the total unresolved \xray\ background
plus resolved normal-galaxy emission from the $\approx$4~Ms CDF-S survey has
sky intensities of $\approx$$2.2 \times
10^{-12}$~erg~cm$^{-2}$~s$^{-1}$~deg$^{-2}$ and $\approx$$3.2 \times
10^{-12}$~erg~cm$^{-2}$~s$^{-1}$~deg$^{-2}$ for the observed-frame 0.5--2~keV
and 2--8~keV bands, respectively.  These values are, respectively, 27\% and
18\% of the total CXB intensity and constitute the maximum intensities that
normal galaxies could contribute to the CXB in these bands.  Using
Equation~(19) with our estimates of the redshift evolution of $\rho_{\rm
0.5-2~keV}$ and $\rho_{\rm 2-10~keV}$ (\hbox{Eqns.~16--18}), we derive
normal-galaxy intensities of $\Omega_{\rm 0.5-2~keV}^{\rm CXB, gal} \approx 1.0
\times 10^{-12}$~erg~cm$^{-2}$~s$^{-1}$~deg$^{-2}$ and $\Omega_{\rm
2-8~keV}^{\rm CXB, gal} \approx 1.1 \times
10^{-12}$~erg~cm$^{-2}$~s$^{-1}$~deg$^{-2}$, which are $\approx$46\% and
$\approx$35\% of the maximum possible.  We therefore find that the redshift
evolution of \xray\ scaling relations are fully consistent with constraints set
by the CXB intensity.

%
\section{Summary and Future Direction}
%

Using the $\approx$6~Ms depth data in the \chandra\ Deep Field-South, we have
studied the evolution of \xray\ emission in normal galaxy populations since $z
\approx 7$.  Our findings can be summarized as follows:

\begin{itemize}

\item Scaling relations involving not only star-formation rate (SFR), but also
stellar mass ($M_\star$) and redshift ($z$) provide a significantly better
characterization of normal-galaxy \xray\ luminosity ($L_{\rm X}$) than a
single, ``universal'' $L_{\rm X}$/SFR relation, which has been widely assumed
in the literature.  

\item We deduce that emission from low-mass \xray\ binary (LMXB) and high-mass
\xray\ binary (HMXB) populations drive correlations involving $M_\star$ and
SFR, respectively.  We find that out to at least $z \approx 2.5$, the 2--10~keV
emission, which probes directly emission from \xray\ binary populations,
evolves as $L_{\rm 2-10~keV}({\rm LMXB})/M_\star \propto (1+z)^{2-3}$ and
$L_{\rm 2-10~keV}({\rm HMXB})/{\rm SFR} \propto (1+z)$.  This evolution is
consistent with basic predictions from \xray\ binary population-synthesis
models, which indicate that the increases in LMXB and HMXB scaling relations
with redshift are primarily due to effects related to declining stellar ages
and metallicities, respectively (see $\S$7.1 and F13a for details).  However,
the best-fit XRB population synthesis model is only marginally consistent with
our data (at the $\approx$7\% confidence level), calling for minor revisions of
the current suite of models.

\item When convolving the redshift-dependent scaling relations $L_{\rm
2-10~keV}({\rm LMXB})/M_\star$ and $L_{\rm 2-10~keV}({\rm HMXB})/{\rm SFR}$
with respective redshift-dependent stellar mass and SFR density curves from the
literature, we find that LMXBs are likely to dominate the \xray\ emissivity of
normal galaxies out to $z \approx$~1--2, with HMXBs dominating at higher
redshifts.  We find that \xray\ binary populations dominate the \xray\
emissivity of normal galaxies in both the \hbox{0.5--2~keV} and
\hbox{2--10~keV} bands; however, hot gas is inferred to provide significant
contributions to the \hbox{0.5--2~keV} emissivity over the majority of cosmic
history.

\item The total \xray\ emissivity from normal galaxies peaks around $z
\approx$~1.5--3, similar to the SFR density of the Universe.  However, the
\xray\ emissivity of normal galaxies declines more slowly at $z \simgt 3$ than
the SFR density due to the continued rise in $L_{\rm X}$/SFR scaling relation
with redshift.  Extrapolation of our results suggests that normal galaxies will
provide an \xray\ emissivity that exceeds that of AGN at $z \simgt$~6--8,
consistent with \xray\ binary population-synthesis predictions.  

\end{itemize}

The results presented in this paper could be significantly expanded upon by
future investigations with \chandra\ and planned observatories.  In particular,
some of the results and interpretations provided in this paper are in need of
verification using data for local galaxies.  For example, the rise in  $\alpha
\equiv L_{\rm 2-10~keV}({\rm LMXB})/M_\star$ and $\beta \equiv L_{\rm
2-10~keV}({\rm HMXB})/{\rm SFR}$ scaling relations with cosmic time is thought
to be associated with changes in the mean stellar age and metallicity of the
Universe.  However, attempts to prove that $\alpha$ and $\beta$ vary with
stellar age and metallicity in local galaxies, e.g., using \chandra\ resolved
\xray\ binary populations, have not yet yielded statistically robust results
(see $\S$7.1 for discussion of such studies).  This situation is due to
challenges in (1) measuring accurate stellar ages and metallicities for
galaxies, and (2) obtaining \chandra\ observations for statistically
significant samples of galaxies that span broad ranges of these parameters.
Future observations with \chandra\ (and to a lesser extent \xmm) are needed to
target significant numbers of galaxies with stellar ages and metallicities that
are well constrained by optical/near-IR spectroscopic data and span broad
ranges of these parameters. 

In addition to local galaxy studies, future stacking investigations of distant
galaxy populations separated by all of the most relevant {\it physical}
parameters (i.e., SFR, $M_\star$, metallicity, and stellar ages) would allow
for direct measurements of how physical properties influence \xray\ binary
formation, independent of redshift.  Statistically significant samples of
galaxies could be drawn from the combination of wide and deep \chandra\ surveys
(see, e.g., Brandt \& Alexander~2015 for a review).  

Future \xray\ missions will provide significant gains in our knowledge of
\xray\ binary populations in extragalactic environments for both the nearby and
distant Universe.  For example, \athena\ (e.g., Nandra \etal\ 2013) will
provide new spectral and timing constraints for \xray\ binaries in a variety of
environments within the local galaxy population.  Wide-area \athena\ surveys
are planned that will reach depths comparable to the most sensitive regions of
a $\approx$1~Ms \chandra\ survey, but over $\approx$1--10~deg$^2$ regions (see,
e.g., Aird \etal\ 2013).  Such a survey would yield 10,000--100,000 normal
galaxy detections with most sources being at $z \simgt 0.5$ (see Lehmer \etal\
2012).  Significant clarification of the detailed evolution of \xray\ binary
populations would come from future missions with imaging capabilities that
supercede those of \chandra, and provide direct detections of \xray\ galaxies
to high redshifts.  The \surveyor\ mission concept (e.g., Weisskopf \etal\
2015) would reach depths of a few~$\times 10^{-19}$~\flux\ in the SB for a 4~Ms
depth image, and would yield normal galaxy sky densities of
$\sim$500,000~deg$^{-1}$.  Such flux levels are a factor of $\simgt$10 times
lower than the mean stacked fluxes for this study, suggesting that most
galaxies that are currently studied through stacking would be detected
individually by \surveyor.

\acknowledgements

We thank the referee for their thorough reading of the manuscript and helpful
comments, which have improved the quality of this paper.  We thank Myrto
Symeonidis and James Aird for gratiously sharing data, which have been valuable
for comparing with our results.  This work has made use of the Rainbow
Cosmological Surveys Database, which is operated by the Universidad Complutense
de Madrid (UCM), partnered with the University of California Observatories at
Santa Cruz (UCO/Lick,UCSC).

B.D.L. gratefully acknowledges financial support from \chandra\ \xray\ Center
(CXC) grant GO4-15130B and NASA ADAP grant NNX13AI48G.  A.E.H. and A.B.Z.
acknowledge funding through CXC program GO4-15130Z and NASA ADAP grant
09-ADP09-0071.  W.N.B. and B.L.~thank CXC grant GO4-15130A and NASA ADP grant
NNX10AC99G.  T.F.  acknowledges support from the Ambizione Fellowship of the
Swiss National Science Foundation (grant PZ00P2\_148123).  F.E.B. acknowledges
support from CONICYT-Chile (Basal-CATA PFB-06/2007, FONDECYT Regular 1141218,
``EMBIGGEN'' Anillo ACT1101), the Ministry of Economy, Development, and
Tourism's Millennium Science Initiative through grant IC120009, awarded to The
Millennium Institute of Astrophysics, MAS.  Y.Q.X. acknowledges support of the
Thousand Young Talents program (KJ2030220004), the 973 Program (2015CB857004),
the USTC startup funding (ZC9850290195), the National Natural Science
Foundation of China (NSFC-11473026, 11421303), the Strategic Priority Research
Program ``The Emergence of Cosmological Structures'' of the Chinese Academy of
Sciences (XDB09000000), and the Fundamental Research Funds for the Central
Universities (WK3440000001).

%

%

\end{document}